\RequirePackage{fix-cm}
\documentclass[smallcondensed]{svjour3}     % onecolumn (ditto)
\smartqed  % flush right qed marks, e.g. at end of proof
\usepackage[numbers]{natbib}
\usepackage{graphicx}
\usepackage{soul}
\usepackage{url}
\usepackage[hidelinks]{hyperref}
\usepackage[utf8]{inputenc}
\usepackage[small]{caption}
\usepackage{graphicx}
\usepackage{amsmath}
\usepackage{booktabs}
\usepackage{bm}
\usepackage{dsfont}
\makeatletter \let\cl@chapter\relax \makeatother
\usepackage{numprint}
\usepackage{makecell}
\usepackage{tabularx}
\usepackage[shortlabels]{enumitem}
\usepackage{mathrsfs} 
\usepackage{upgreek}
\usepackage{amsmath} 
\usepackage{mathtools}

\usepackage{subcaption}
\DeclareCaptionSubType * [alph]{table}
\usepackage{multirow}
\usepackage[table, dvipsnames]{xcolor}
\definecolor{color1}{RGB}{175,238,238}
\definecolor{color2}{RGB}{250,235,215}
\definecolor{color3}{RGB}{238,232,170}
\definecolor{color4}{RGB}{255,228,225}
\usepackage{color}
\usepackage[misc]{ifsym}

\usepackage{booktabs,dcolumn}
\newcolumntype{d}[1]{D..{#1}}
\usepackage[linesnumbered,ruled,vlined]{algorithm2e}

\SetCommentSty{mycommfont}

\SetKwFunction{lenth}{lenth}
\SetKwFunction{pop}{pop}
\SetKwFunction{push}{push}

\SetKwData{qc}{$Ms$}
\SetKwData{nr}{$D_N$}
\SetKwData{hr}{$D_H$}
\SetKwData{bs}{$l_r$}
\SetKwData{lh}{$l_H$}
\SetKwData{ts}{$T$}

\usepackage{cleveref}

\Crefformat{figure}{#2Fig.~#1#3}
\Crefmultiformat{figure}{Figs.~#2#1#3}{ and~#2#1#3}{, #2#1#3}{ and~#2#1#3}
\crefrangeformat{figure}{figs. #3(#1)#4 to #5(#2)#6}
\Crefformat{table}{#2Table~#1#3}
\Crefmultiformat{table}{Table~#2#1#3}{ and~#2#1#3}{, #2#1#3}{ and~#2#1#3}
\crefrangeformat{table}{table #3#1#4 to #5#2#6}
\Crefformat{equation}{#2Eq.~(#1#3)}
\crefrangeformat{equation}{eqs.~#3(#1)#4 to #5(#2)#6}
\Crefmultiformat{equation}{Eqs.~#2(#1)#3}{ and~#2(#1)#3}{, #2(#1)#3}{ and~#2(#1)#3}

\Crefformat{algorithm}{#2Algorithm ~#1#3}
\Crefmultiformat{algorithm}{Algorithms ~#2#1#3}{ and~#2#1#3}{, #2#1#3}{ and~#2#1#3}
\crefrangeformat{algorithm}{Algorithms #3(#1)#4 to #5(#2)#6}

\Crefformat{lemma}{#2Lemma ~#1#3}
\Crefmultiformat{lemma}{Lemmas ~#2#1#3}{ and~#2#1#3}{, #2#1#3}{ and~#2#1#3}
\crefrangeformat{lemma}{Lemmas #3#1#4 to #5#2#6}

\Crefformat{section}{#2Section~#1#3}
\Crefmultiformat{section}{Sections~#2#1#3}{ and~#2#1#3}{, #2#1#3}{ and~#2#1#3}
\crefrangeformat{section}{Sections #3#1#4 to #5#2#6}
\Crefformat{subsection}{#2Subsection~#1#3}
\Crefmultiformat{subsection}{Subsections~#2#1#3}{ and~#2#1#3}{, #2#1#3}{ and~#2#1#3}
\crefrangeformat{subsection}{Subsections #3#1#4 to #5#2#6}

\newcommand*{\affaddr}[1]{#1} % No op here. Customize it for different styles.
\newcommand*{\affmark}[1][*]{\textsuperscript{#1}}

\begin{document}
	
	\title{Stratified and Time-aware Sampling based Adaptive Ensemble Learning for Streaming Recommendations%\thanks{Grants or other notes
	}
	%\subtitle{STS-AEL for Streaming Recommendation}
	
	
	\author{%
		Yan Zhao \protect\affmark[1]\textsuperscript{,}\affmark[2] \and Shoujin Wang \affmark[2]
		 \and Yan Wang\affmark[2] \and Hongwei Liu\affmark[1]
	}
%	\authorrunning{Author A \and Author B\and Author C\and Author D \and Author E}
	
%	\author{Yan Zhao  \and
%		Shoujin Wang \and
%		Yan Wang \and
%		Hongwei Liu 
%	}
	
	
	\institute{
		\hspace*{1.5em}Yan Zhao \at
		\hspace*{1.5em}\email{zhaoyan@ftcl.hit.edu.cn}           %  \\
		\and
		\hspace*{1.5em}Shoujin Wang \at
		\hspace*{1.5em}\email{shoujin.wang@mq.edu.au}
		\and
		\hspace*{1.5em}Yan Wang \at
		\hspace*{1.5em}\email{yan.wang@mq.edu.au}
		\and
		\Letter  ~  Hongwei Liu \at
		\hspace*{1.6em}\email{liuhw@hit.edu.cn}\\
		\and
		\affaddr{\affmark[1] ~~~School of Computer Science and Technology, Harbin Institute of Technology, Harbin, China}\\
		\affaddr{\affmark[2] ~~Department of Computing, Macquarie University, Sydney, Australia}\\
%		\affaddr{\LaTeX\ University}%
	}

%	\institute{Yan Zhao \at
%		\email{zhaoyan@ftcl.hit.edu.cn}           %  \\
%		%             \emph{Present address:} of F. Author  %  if needed
%		\and
%		Shoujin Wang \at
%		\email{shoujin.wang@mq.edu.au}
%		\and
%		Yan Wang \at
%		\email{yan.wang@mq.edu.au}
%		\and
%		Hongwei Liu \at
%		\email{liuhw@hit.edu.cn}
%	}
	
	\date{Received: date / Accepted: date}
	
	{\let\thefootnote\relax\footnotetext{This paper was accepted by Applied Intelligence in July 2020. The final authenticated publication is available online at https://doi.org/[DOI to be inserted] \\}}
	\maketitle
	
	\begin{abstract}
		Recommender systems have played an increasingly important role in providing users with tailored suggestions based on their preferences. However, the conventional offline recommender systems cannot handle the ubiquitous data stream well. To address this issue, Streaming Recommender Systems (SRSs) have emerged in recent years, which incrementally train recommendation models on newly received data for effective real-time recommendations. Focusing on new data only benefits addressing \textit{concept drift}, i.e., the changing user preferences towards items. However, it impedes capturing \textit{long-term user preferences}. In addition, the commonly existing \textit{underload} and \textit{overload} problems should be well tackled for higher accuracy of streaming recommendations. To address these problems, we propose a \textbf{S}tratified and \textbf{T}ime-aware \textbf{S}ampling based \textbf{A}daptive \textbf{E}nsemble \textbf{L}earning framework, called STS-AEL, to improve the accuracy of streaming recommendations. In STS-AEL, we first devise stratified and time-aware sampling to extract representative data from both new data and historical data to address concept drift while capturing long-term user preferences. Also, incorporating the historical data benefits utilizing the idle resources in the underload scenario more effectively. After that, we propose adaptive ensemble learning to efficiently process the overloaded data in parallel with multiple individual recommendation models, and then effectively fuse the results of these models with a sequential adaptive mechanism. Extensive experiments conducted on three real-world datasets demonstrate that STS-AEL, in all the cases, significantly outperforms the state-of-the-art SRSs.
	\end{abstract}

	\section{Introduction}
	
	\noindent Recommender Systems (RSs) are a type of information filtering systems, which aim to assist users to make decisions more effectively and efficiently. With the ever-growing data volume, RSs are playing an increasingly important role in both academia and industry to confront the information explosion. The conventional RSs can be mainly divided into three categories based on their work mechanisms, i.e., 1) collaborative filtering based RSs~\cite{applied_intelligence_neural_variation,applied_intelligence_cf_time_windows}, which make recommendations based on the past behaviours of similar users, 2) content-based RSs~\cite{content-based_recommendation_recsys,content-based_recommendation_survey}, which make recommendations based on the item properties, and 3) hybrid RSs~\cite{hybrid-based_recommendation_Infor-Sci,hybrid-based_recommendation_WWW}, which combine the first two for a higher recommendation accuracy. Recently, focusing on the recommendation issues in different scenarios, more types of RSs, e.g., session-based RSs~\cite{session_survey,session_shoujin}, cross-domain RSs~\cite{cross_domain_1,cross_domain_2}, and social RSs~\cite{social_1,social_2}, have emerged. For comprehensively introducing the existing work on recommendations, survey papers have been published to present the development and current status of generalized RSs~\cite{RS_survey_1,RS_survey_3}, fuzzy tool based RSs~\cite{RS_survey_2}, and the deep learning based RSs~\cite{RS_survey_4}. Moreover, to improve the users' experience and increase businesses' profits, RSs have been widely deployed in various areas, including e-commerce, e.g., Amazon~\footnote{https://www.amazon.com/} and eBay~\footnote{https://www.ebay.com/}, online video services, e.g., Youtube~\footnote{https://youtube.com/} and Netflix~\footnote{https://www.netflix.com/},  and online learning, e.g., Coursera~\footnote{https://www.coursera.org/}.

	Despite the fast development and wide applications of RSs, delivering accurate recommendations in the streaming scenario remains a challenge \cite{srec}.  Data stream commonly exists in the real world, e.g., purchases, clicks, and ratings; thus, how to deliver accurate streaming recommendations is an essential problem we need to address. However, the conventional RSs periodically train the recommendation models with large-volume historical data, and thus cannot process data stream in time. To this end, a new trend has emerged 
	%in recent years 
	to train the recommendation model on new data instantly to perform real-time recommendations. The RSs following this trend are commonly referred to as \textit{online} \cite{eals} or \textit{Streaming Recommender Systems} (SRSs) \cite{2019_streaming}. 
	
	%The existing SRSs are mainly developed in two stages: 1) \textit{adaption-based SRSs}, and 2) \textit{stream-oriented SRSs}. As an earlier attempt, \textit{adaption-based SRSs}~\cite{ISGD,eals} aim to adapt the conventional offline recommender systems to the streaming setting with an online update mechanism. In recent years,  \textit{stream-oriented SRSs}~\cite{srec,spmf} have emerged, which are specifically devised for the streaming recommendation.
	
	The existing SRSs are mainly developed in two stages, i.e., 1) \textit{adaption-based SRSs}, and 2) \textit{stream-oriented SRSs}. As an earlier attempt, \textit{adaption-based SRSs} aim to adapt the conventional offline RSs to the streaming setting by incrementally training their recommendation models with new data, such as Incremental Collaborative Filtering (ICF)~\cite{ICF} and Element-wise Alternating Least Squares (eALS)~\cite{eals}. In recent years, stream-oriented SRSs have been devised specifically for the streaming scenario, such as Stream-centered Probabilistic Matrix Factorization (SPMF)~\cite{spmf} and Neural Memory Recommender Networks (NMRN)~\cite{nmrn-gan}.
	
	%For example, sRec~\cite{srec} captures temporal dynamics with streaming data, SPMF~\cite{spmf} performs efficient ranki5+^ng based recommendation in the streaming setting, and the recent OCFIF~\cite{ocfif} combines multiple SRSs to avoid the limitations of a single fixed model.
	
	%SRSs have been proposed by adapting the conventional offline recommender systems to the streaming setting (~\cite{ISGD,eals}), or devising recommendation models specifically for the streaming scenario (~\cite{srec,spmf}) to increase streaming recommendation accuracy.	Despite the extensive research, the following three challenges need to be well addressed for SRSs:
	
	%Although many solutions have been reported either adapting the conventional offline RSs to the streaming setting~\cite{ISGD} or devising new approaches specifically for the streaming setting~\cite{spmf}, the following three challenges still need to be well addressed to improve the accuracy of streaming recommendation:
	Although many solutions have been reported, the following three challenges still need to be well addressed to improve the accuracy of streaming recommendations:
	\begin{itemize}
		
		\item \textbf{CH1:} how to address \textit{concept drift}, i.e., the changing user preferences towards items over time, while capturing \textit{long-term user preferences}. The evolutional user preferences and item properties over time cause concept drift in streaming recommendations~\cite{spmf}. For example, Alice has different preferences on the styles of clothes as she grows up. Another problem is the loss of long-term user preferences. For example, Bob likes to read history books, however, the shopping website wrongly recommends him math books as he recently bought some math books just for the examination. It is a challenging task for SRSs to simultaneously handle the above two problems, i.e., concept drift and the loss of long-term user preferences.
		\item  \textbf{CH2:} how to tackle the \textit{underload} problem, i.e., the scenario where the data receiving speed is lower than the data processing speed. The low resource utilization ratio commonly exists in real-world applications~\cite{underload_book,underload_forbes}, as the systems are designed to be scalable and be prepared for the peak demand. Resource management approaches~\cite{underload_common,underload_cikm} have been studied to address the underload problem for general stream processing. However, no studies have been reported to particularly address the underload problem in streaming recommendations. 
		
		%existing SRSs have explicitly addressed the underload problem w.r.t. the features of the streaming recommendation.
		
		\item  \textbf{CH3:} how to tackle the \textit{overload} problem, i.e., the scenario where the data receiving speed is higher than the data processing speed. The overload problem has attracted much research interest in the stream processing areas~\cite{overload_survey,overload_sigmod}, including the streaming recommender systems~\cite{srec,spmf,ssrm}. As the velocity of the data stream keeps increasing, SRSs should be prepared to handle the intensive workload beyond their capacities.
	\end{itemize}

	The main idea and limitations of the existing SRSs, which are proposed to address the aforementioned three challenges, are briefly introduced as follows. Targeting \textbf{CH1}, SPMF~\cite{spmf} and NMRN~\cite{nmrn-gan} have proposed reservoir-based and neural memory network based approaches, respectively. However, SPMF has difficulties in dealing with concept drift as it does not effectively utilize new data to capture the changing user preferences, while NMRN has limited capability to capture long-term user preferences as the memory recording preferences might update frequently over the data stream. Different from the first challenge, \textbf{CH2} has not yet been discussed in the literature of SRSs. However, it is an important issue, since computation resources are wasted in the underload scenario and should be effectively utilized for improving the recommendation accuracy. As the underload scenario commonly exists, we argue that SRSs which can effectively use the idle resources should be proposed. In addition, to address \textbf{CH3}, sampling methods have been proposed to reduce the training workload of the monolithic SRS \cite{ssrm,spmf}, which employs a single model for recommendations. However, monolithic SRSs cannot tackle the overload problem well due to their limited computational capabilities.

	\bigskip
	\noindent\textbf{Our Approach and Contributions:} To address all the above three challenges, we propose a novel \textbf{\textbf{S}}tratified and \textbf{T}ime-aware  \textbf{S}ampling based \textbf{A}daptive \textbf{E}nsemble \textbf{L}earning framework, called STS-AEL, for higher accuracy of streaming recommendations. Specifically, STS-AEL contains two main components: 1) \textbf{S}tratified and \textbf{T}ime-aware \textbf{S}ampling (STS), which samples training data from both new data and historical data in a time-aware manner, i.e., assigning higher sampling probabilities to newer data in the sample space. In addition, the sample sizes of the new data and reservoir are determined based on the characteristics and receiving speed of the streaming data; and 2) \textbf{A}daptive \textbf{E}nsemble \textbf{L}earning (AEL), which first trains multiple individual models in parallel with the sampled data and then fuses the results of these trained models with a novel sequential adaptive mechanism. Note that AEL is specifically devised for the streaming setting, where the training and test are iteratively conducted over the data stream. To be specific, AEL dynamically calculates the fusion weights with a sequential adaptive mechanism based on the testing accuracy of each individual model in the last iteration to conduct more effective fusion in the current iteration.
	%., which dynamically calculates the fusion weights for multiple individual models
	% based on their performances.% where the training and test are iteratively conducted along with data stream. %To be specific, AEL calculates the fusion weights with a sequential adaptive mechanism based on the test accuracies of the individual models in the last iteration to conduct more effective fusion in the current iteration.
	%Different from the conventional offline ensemble learning, which simply averages the outputs of individual models,  
	%Thus AEL  to conduct more effective fusion in the current iteration is may degrade the ensemble performance, since different individual models may contribute differently to the final result w.r.t.\ different users and items. To this end, 
	%for higher streaming recommendation accuracy. 
	
	%, based SRS. We first propose a novel weighted positive sampling approach, called STS, to address \textbf{CH2} to better prepare the input for the individual recommendation model; and then we perform joint trainig with both sampled historical data and new data (\textbf{CH1, CH2}) when the processing speed is faster than the receiving speed; finally we ensemble multiple individual recommendation models to reduce the variance (\textbf{CH3}) and further improve the recommendation accuracy.
	
	The characteristics and contributions of our work are summarized as follows:
	\begin{itemize}
		
		%\item In this paper, we propose a novel \textbf{\textbf{S}}tratified and \textbf{T}ime-aware  \textbf{S}ampling based \textbf{A}daptive \textbf{E}nsemble \textbf{L}earning framework, called STS-AEL, for more accurate streaming recommendation, which contains STS and AEL.
		\item In this paper, we propose a novel STS-AEL framework for more accurate streaming recommendations, which contains two main components, i.e., STS and AEL.
		
		\item To address \textbf{CH1} and \textbf{CH2}, we propose STS to extract representative data from both new data and historical data while guaranteeing the proportion of new data. In this way, through elaborately incorporating both new data and historical data, STS can not only capture both short-term and long-term user preferences, but also effectively utilize the idle resources in the underload scenario to deliver higher recommendation accuracy.
		
		\item To address \textbf{CH2} and \textbf{CH3}, we propose AEL to increase recommendation accuracy in both the underload scenario and overload scenario. AEL first conducts concurrent training to address the excessive data in both the underload scenario (complemented by the sampled historical data) and the overload scenario via multiple individual models, and then fuses the prediction results of these models to deliver higher recommendation accuracy. Moreover, AEL utilizes the sequential adaptive fusion specifically devised for the streaming setting to further improve the accuracy of streaming recommendations.
	\end{itemize}
	Extensive experiments demonstrate that the proposed STS-AEL framework significantly outperforms the state-of-the-art SRSs in terms of recommendation accuracy on all three widely-used datasets w.r.t. both the underload scenario and overload scenario. 
	
	\section{Related Work}
	
	In this section, we review the existing SRSs in two groups: adaption-based SRSs and stream-oriented SRSs. Besides, we also introduce ensemble learning, based on which we propose the STS-AEL framework.

	\subsection{Adaption-Based SRSs}
	
	The early SRSs adapt conventional offline RSs to the streaming setting by devising online update mechanisms.	
	For example, ICF~\cite{ICF} enhanced user-based collaborative filtering with incremental updates of user-to-user similarities to process streaming data. Later on, the matrix factorization based approaches have been widely adapted to the streaming setting with different online update mechanisms. For example, Vinagre et al.~\cite{ISGD} devised an incremental stochastic gradient descent method to update the recommendation model dynamically, Devooght et al.~\cite{rcd} improved the randomized block coordinate descent learner to adapt the matrix factorization to the streaming scenario, and He et al.~\cite{eals} proposed a fast alternating least square approach to update the matrix factorization with newly received data. To further exploit the potential of matrix factorization in the streaming scenario, researchers have applied various optimization methods which are originally designed for the offline recommendations to the streaming scenario. For instance,  Rendle et al.~\cite{rkmf} endowed matrix factorization with nonlinearities in the streaming scenario with a regularized kernel matrix factorization model, Diaz-Aviles et al.~\cite{rmfx} employed a pairwise approach to optimize the matrix factorization model w.r.t. the personalized ranking, and Silva et al.~\cite{vbmf} performed an efficient online Bayesian inference approach to improve the matrix factorization model in the streaming scenario.\\
	%	Later on, the singular value decomposition based method was adapted to the streaming setting by~\cite{time-svd++} through incremental time-aware representation of user preferences.
	%The user-based collaborative filtering and support vector machine were adapted to the streaming setting by~\cite{ICF} and~\cite{time-svd++} through incrementally updating user-to-user similarities and devising a incremental time-aware representation method of user preferences, respectively.
	%	More recently, other applications of matrix factorization in the streaming setting have also been exploited by \cite{mf2012online,ISGD,eals} via well-designed online update mechanisms.\\
	
	\vspace{1pt}
	\noindent \textbf{Summary:} As an early attempt, adaption-based SRSs adapt the conventional offline RSs to the streaming setting for making recommendations based on the data stream. Moreover, some optimization methods have also been employed by these adaption-based SRSs to improve the recommendation accuracy. 
	
	\noindent \textbf{Gaps:} The adaption-based SRSs mainly focus on the online update mechanisms to perform online training with data stream. However, they are not devised to solve the aforementioned essential challenges in streaming recommendations, i.e., addressing concept drift while capturing long-term user preferences, the underload problem, and the overload problem.
	%The application of the widely studied approach matrix factorization~\cite{matrixFactorization} in the streaming setting has been exploited by researchers. \cite{RKMF} enhances matrix factorization with a flexible kernel for deriving models based on new data. Later on,~\cite{ISGD} proposed a fast online matrix factorization with incremental stochastic gradient descent, and the work in~\cite{eals} employed alternating least squares technique to more efficiently optimize matrix factorization with unobserved data.
	%In addition to the independent versions, parallel SRS has also been exploited.~\cite{tencentrec} enhanced item-based collaborative filtering with parallel incremental updates and real-time pruning to process the high-velocity data more efficiently. 
	
	%\vspace{-1mm}
	\subsection{Stream-Oriented SRSs}
	In recent years, SRSs specifically devised for the streaming scenario have been proposed. Compared with the adaptation-based SRSs, stream-oriented SRSs focus more on the challenges in the streaming scenario, e.g., handling concept drift, capturing long-term user preference, the underload problem, and the overload problem. 
	
	%~\cite{srec} devised a Bayesian approach to address concept drift via efficient online inference, but it fails to capture the long-term user preferences. 
	
	Targeting the concept drift issue in the streaming scenario, Chang et al. have proposed a Bayesian inference based approach, i.e., Streaming RECommender system (sRec) \cite{srec} to effectively capture the short-term user preferences. Specifically, in sRec, a random process is first used to model the dynamic creation of users as well as the concept drift, and then a variational Bayesian approach is employed to perform online predictions. However, sRec focuses on the new data only once the parameters are initialized, and thus cannot well capture the long-term user preferences embedded in the historical data. Therefore, to address concept drift while capturing long-term user preferences, Wang et al.~\cite{spmf}, Wang et al.~\cite{nmrn-gan}, and Chen et al.~\cite{2019_streaming} have proposed the reservoir-based approach,  the neural memory network based approach, and the forgetting mechanism based approach, respectively. Specifically, the work in~\cite{spmf} first maintains a reservoir which stores the representative historical data, and then trains the recommendation model with both the sampled new data and historical data. However,~\cite{spmf} has limited capability to address concept drift, as it treats new data and historical data the same when sampling the training data and thus the importance of new data is overlooked.
	Differently,~\cite{nmrn-gan} employs the neural memory network to keep both the long-term and short-term user preferences. However, it has difficulties to capture long-term user preferences effectively as the memory recording preferences may update frequently over the data stream. In~\cite{2019_streaming}, two forgetting mechanisms to filter out the outdated interactions and the outliers were proposed, respectively. However, as the approach in~\cite{2019_streaming} utilizes the matrix factorization model for the recommendations, it has limited capability to model the complex interactions and well learn the user preferences when compared with more powerful machine learning techniques, e.g., ensemble learning.
	
	As for the overload problem, sampling-based approaches have been employed to reduce the training workload when confronting the excessive amount of data. The most representative sampling approach for SRSs is ranking-based sampling, which was proposed by Wang et al. to prepare training data for SPMF~\cite{spmf}. This ranking-based sampling method was later employed by the Streaming Session-based Recommendation Machine (SSRM)~\cite{ssrm} for its effectiveness in sampling informative data. When conducting the ranking-based sampling, all available user-item interactions are first evaluated by the recommendation model, and then the user-item interactions with lower prediction accuracies will be given higher probabilities to be sampled. Through this way, the recommendation model can achieve larger improvement as it can get more knowledge from the user-item interactions which receive low recommendation accuracies. However, the excessive computation complexity restricts its effectiveness, as the ranking-based sampling methods have to evaluate all the available user-item interactions. Furthermore, SPMF and SSRM are both monolithic SRSs, which employ a single model for recommendations. However, monolithic SRSs can only partially solve the overload problem due to their limited computational capabilities. Although Online Collaborative Filtering with Implicit Feedback (OCFIF) proposed in~\cite{ocfif} employs multiple individual recommendation models for streaming recommendations, it feeds the individual models with the same data, thus cannot well address the overload problem.\\
	
	\vspace{1pt}
	\noindent \textbf{Summary:} Stream-oriented SRSs have utilized different types of techniques to address the challenges in streaming recommendations. For example, the reservoir-based approach, the neural memory network based approach, and the forgetting mechanism based approach have been employed to capture both long-term and short-term user preferences, while the sampling based approach has been employed to make more accurate recommendations in the overload scenario.
	
	\noindent \textbf{Gaps:} Despite attempts have been made by the stream-oriented SRSs, they have difficulties in making a trade-off between the new data and historical data when training the recommendation models with data stream, which 1) makes it difficult to address the concept drift issue well when the new data are not emphasized enough, and 2) causes the loss of the long-term user preferences when the historical data are overlooked. Moreover, existing SRSs are unable to perform well in both the underload scenario and overload scenario, which degrades the recommendation performance as 1) computational resources might not be fully utilized for training recommendation models in the underload scenario, and 2) valuable data might be overlooked for improving recommendation accuracies in the overload scenario.
	
	\subsection{Ensemble Learning}
	%\vspace{-1mm}
	Ensemble learning combines multiple individual models for better performance~\cite{ensemble2012book} and has been employed in various research areas including natural language process~\cite{ensemble_nlp_1,ensemble_nlp_2}, image classification~\cite{ensemble_image_1,ensemble_image_2}, and speech recognition~\cite{ensemble_speech_recognition_1,ensemble_speech_recognition_2}. The effectiveness of ensemble learning in the recommendation area has also been validated in the existing work~\cite{ensemble2019,ensemble2017,ensemble2014,ensemble2012}, which are all for the offline scenario. As one of the most promising directions to process streaming data~\cite{survey-ssl}, ensemble learning has been applied to a wide range of streaming scenarios~\cite{survey2017gomes}.  As an early attempt, Oza~\cite{oza2005online} adapted two offline ensemble learning methods, i.e., Bagging and Boosting, to the streaming setting. Then, the work in~\cite{oza2005online} was improved by Bifet~\cite{LevBag}. 
	Later on, researchers employed ensemble learning to address concept drift in the streaming scenario~\cite{Learn++.NSE,MD3}. In addition, to increase the throughput, sliding window based streaming ensemble learning approaches~\cite{window2017multi,OWE} have been devised.
	
	Despite the promising potential of ensemble learning in streaming recommendations,
	OCFIF~\cite{ocfif} is the only SRS ensembling multiple individual models to avoid the limitations of monolithic SRSs. However, OCFIF does not fully exploit the potential of ensemble learning for the following three reasons: 1) it trains multiple individual models with the same data, which limits the data processing speed. Moreover, it also reduces the diversities of the individual models, which makes them difficult to complement one another, 2) OCFIF selects only one individual model for the final prediction, which restricts its performance, and 3) OCFIF is specifically designed for ensembling the matrix factorization models, thus its generality is limited.\\
	
	\vspace{1pt}
	\noindent \textbf{Summary:} Ensemble learning is a powerful machine learning technique and performs well in multiple areas. Moreover, its effectiveness in the streaming scenario has also been verified in the literature.
	
	\noindent \textbf{Gaps:} The potential of ensemble learning in streaming recommendations is not fully exploited by the existing SRSs to further improve the accuracy of streaming recommendations, although the effectiveness of ensemble learning has been verified in other streaming areas.

	\section{STS-AEL Framework}

	\begin{table}
		%	\footnotesize{
		\begin{center}
			\begin{tabularx}{0.99\textwidth}{c|l}
				\hline
				\hline
				\textbf{Notation} & \textbf{Description} \\
				\hline
				
				$\alpha$ & the proportion of $\mathbf{SS}_{new}$ in \textbf{SS} \\\hline
				$\lambda_{new}$ and $\lambda_{res}$ &  the decay ratio for $\mathbf{N}$, and the decay ratio for $\mathbf{R}$ \\\hline 
				$a$ &  the activation function \\\hline
				$acc$ &  the recommendation accuracy \\\hline
				$bs$ &  the size of training batch \\\hline
				\multirow{2}{*}{$c_{u,v}^k$} &  the estimated confidence of  $\text{im}_k$ for the prediction of the \\  
				& interaction between user $u$ and item $v$ \\\hline	
				$fw_k$ & the fusion weights for $\text{im}_k$ \\\hline
				$im_k$ & the $k^{th}$ individual model \\\hline
				\textbf{IM} = \{im$_1$, im$_2$, ..., im$_o$\} & the set of individual models \\\hline
				$m$ &  the number of users \\\hline
				$n$ &  the number of items \\\hline
				$\mathbf{N} \subseteq \mathbf{Y}$ &  the newly received data from data stream \\\hline
				$o$ &  the number of individual models \\\hline
				$\mathbf{p_u}$ and $\mathbf{q_v}$ &  the user embedding, and the item embedding\\\hline
				$\mathbf{P}_k=\{\langle acc_{1}^k, {u_1}^k,{v_1}^k\rangle,$&	the set containing the recommendation accuracies from $\text{im}_k$ \\
				$ \cdots,\langle {acc_{g}^k,u_g}^k,{v_g}^k\rangle\}$ &  and corresponding user-item pairs in the last test iteration \\\hline
				\multirow{2}{*}{$\mathbf{R} \subseteq \mathbf{Y}$} &  reservoir, i.e., a set which contains representative historical \\ & data \\\hline 
				$sp_{p}$ and $sp_{r}$ & the data processing speed, and the data receiving speed \\\hline
				\multirow{2}{*}{$\mathbf{S}_k \subseteq \mathbf{P}_k$} &  the set which contains top $e$ tuples from $\mathbf{P}_k$ those have\\
				& most similar user-item pairs to the target user-item pair \\\hline
				\multirow{2}{*}{$\mathbf{SS}_{new} \subseteq \mathbf{N}$ and  $\mathbf{SS}_{his} \subseteq \mathbf{R}$} & the set of sampled data from $\mathbf{N}$, and the set of sampled \\ & data  from $\mathbf{R}$ \\\hline
				$\mathbf{SS} = \mathbf{SS}_{new} \cup \mathbf{SS}_{his}$ & the set of sampled data for training \\\hline
%				$\mathbf{T} = \{t_1, t_2, ..., \}$ &  the set of time steps \\\hline
				$\mathbf{U} = \{u_1, u_2, ..., u_m\}$ &  the set of users \\\hline
				$\mathbf{V} = \{v_1, v_2, ..., v_n\}$ &  the set of items \\\hline
				$\mathbf{W}$ and $\mathbf{b}$ &  the weight matrix, and the bias vector for neural network \\\hline
				\multirow{3}{*}{$y_{i,j} \in \textbf{Y}$} & the notation which indicates whether an interaction exists,   \\
				& i.e., it  is 1 if an interaction exists between user $u_i$ and item  \\ &  $v_j$, and 0 otherwise \\\hline
				$\mathbf{Y} \in R^{m*n}$ &  the matrix of interactions between $\mathbf{U}$ and $\mathbf{V}$ \\\hline
				$|*|$ &  the size of a set, e.g., $|\mathbf{U}|$ represents the number of the users\\\hline
				$||*||_1$ &  the L1 norm of a vector \\\hline
				$||*||_2$ &  the L2 norm of a vector \\\hline
				$[\mathbf{a};\mathbf{b}]$ &  the concatenation of vector $\mathbf{a}$ and vector $\mathbf{b}$ \\\hline
				\hline
			\end{tabularx}
		\end{center}
		%	}
		%	\vspace{-4mm}
		\caption{Important Notations}
		%	\vspace{-5mm}
		\label{table:notation1}
	\end{table}

	In this section, we first introduce the notations and present the problem statement. After that, we propose a novel \textbf{S}tratified and \textbf{T}ime-aware \textbf{S}ampling based \textbf{A}daptive \textbf{E}nsemble \textbf{L}earning framework, called STS-AEL, and then introduce its two key components, i.e., \textbf{S}tratified and \textbf{T}ime-aware \textbf{S}ampling (STS) and \textbf{A}daptive \textbf{E}nsemble \textbf{L}earning (AEL).
	
	\subsection{Notations and Problem Statement}

	For the readability purpose, in~\Cref{table:notation1}, we list the important notations used in the rest of this paper. Based on these notations, we introduce the streaming recommendation problem studied in this work as follows.
	
	In this paper, we focus on streaming recommendations with implicit user-item interactions, e.g., users' clicks on items. With the interaction set \textbf{Y}, user set \textbf{U}, item set \textbf{V}, let $\mathcal{Y}=\{y_{u^1,v^1}^{1}, y_{u^2,v^2}^{2}, \dots, y_{u^k,v^k}^{k},\dots\}$ be the list of currently received interactions, where $ y_{u^k,v^k} ^k \in \mathbf{Y}$ indicates an interaction between user $u^k \in \mathbf{U}$ and item $v^k \in \mathbf{V}$. In addition, the interactions in $\mathcal{Y}$ are ordered based on their receiving time, e.g., $y_{u^k,v^k}^k$ indicates the $k^\text{th}$ received interaction. Note that the adjacent interactions (e.g., $y_{u^k,v^k}^k$ and $y_{u^{k+1},v^{k+1}}^{k+1}$)  in $\mathcal{Y}$ may be related to different users (e.g., user $u^k$ is not user $u^{k+1}$) as the real-world data stream. Then, the task of the SRS is to predict the probability of a future interaction between the given user $u'$ and item $v'$ based on the currently received interactions $\mathcal{Y}$, i.e., $\hat{y} = P(y_{u',v'}|\mathcal{Y})$. Compared with conventional offline recommendations, streaming recommendations take continuous and infinite data stream, e.g., streaming clicks, as input, thus it is more challenging.

	\subsection{The STS-AEL Framework}
	
	\begin{figure}[!ht]
		\centering
		\includegraphics[width= 0.95\textwidth, keepaspectratio]{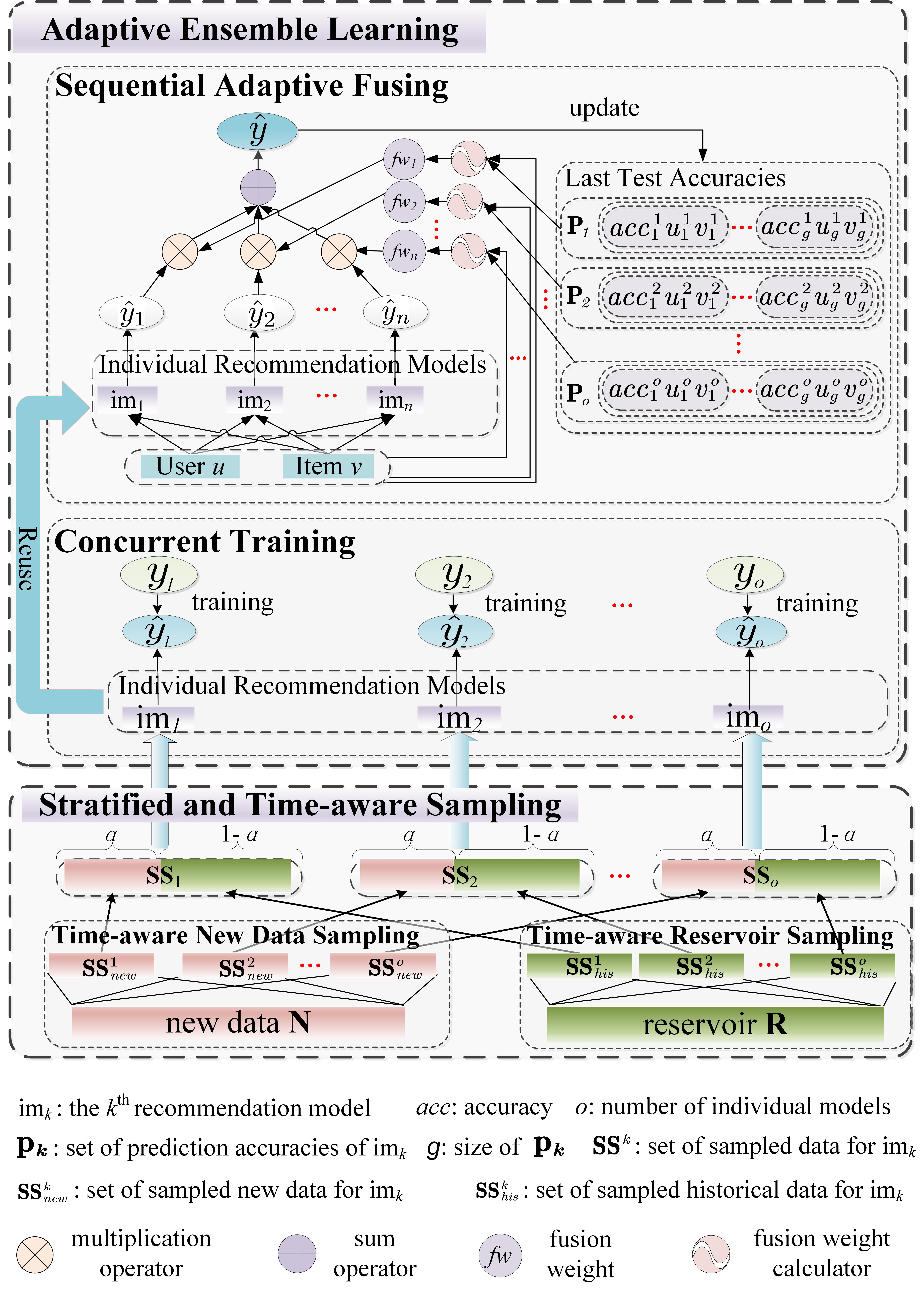}	
		%	\vspace{-6mm}
		\caption{Our Proposed STS-AEL Framework (the data flow is roughly from the bottom to top). STS-AEL contains two main components, i.e., Stratified and Time-aware Sampling (STS, with more details introduced in Section 3.3) and Adaptive Ensemble Learning (AEL, with more details introduced in Section 3.4). Specifically, STS first samples representative data from both the new data and historical data. After that, with the sampled data, AEL concurrently trains multiple individual models, and then fuse these models with our proposed sequential adaptive fusion mechanism.}
		
		\label{figure:architecture} 
	\end{figure}
	
	\begin{algorithm}[t]
		\SetKwInOut{Input}{Input}
		\SetKwInOut{Output}{Output}
		\DontPrintSemicolon 
		\SetKwBlock{DoParallel}{do in parallel for each model \textbf{im} in \textbf{IM}}{end}
		\Input{~New Data $\mathbf{N}$, Reservoir $\mathbf{R}$, Set of Individual Models \textbf{IM}}
		\Output{~Recommendations}
		%get the sample sizes of $S$ and $R$ ssn and ssp by \\
		%get the sample probabilities of $S$ and $R$ xxx and xxx by\\
		%get samples of new data xx1 based on with replacement\\
		%get samples of reservoir xx2 based on with replacement\\
		
		\SetNoFillComment
		
		\uIf{$\mathbf{N}$ is for training}{
			\tcc{Stratified and Time-aware Sampling}
			Sample $|\mathbf{IM}|$ sets of training data from $\mathbf{N}$ and $\mathbf{R}$ by~\Cref{eq:degrade_lambda,eq:freshness,eq:sample_probability,eq:merge}\\
			\tcc{Concurrent Training}
			\DoParallel{
				Update \textbf{im} with sampled training data by optimizing the loss in~\Cref{eq:loss}
			}
			
		}
		\Else{
			\tcc{Sequential Adaptive Fusing}
			\DoParallel{
				Get the prediction results of \textbf{im}, taking NeuMF as an example, by~\Cref{eq:gmf1,eq:gmf2,eq:mlp1,eq:mlp_input,eq:mlp_inner,eq:mlp_output,eq:neumf}\\
			}
			Get the fusion weights \textbf{fw} for the models in \textbf{IM}  by~\Cref{eq:pk,eq:sim,eq:pp,eq:vector,eq:weight1,eq:weight2}\\
			Get the final predictions $\mathbf{\hat{y}}^{final}$ by fusing the prediction results by~\Cref{eq:fusion_for_ael} \\
			Store the prediction accuracies\\
			\Return{$\mathbf{\hat{y}}^{final}$}
		}
		\caption{STS-AEL Framework}
		\label{alg:sts-ael}
	\end{algorithm}
	To perform accurate streaming recommendations, we propose \textbf{S}tratified and \textbf{T}ime-aware \textbf{S}ampling based \textbf{A}daptive \textbf{E}nsemble \textbf{L}earning framework, called STS-AEL. As ~\Cref{figure:architecture} shows, the proposed STS-AEL mainly contains two components, i.e., STS and AEL, which are introduced in detail in Section 3.3 and Section 3.4, respectively.
	%For clarity, we present the working process of STS-AEL in~\Cref{figure:architecture}. 
	Specifically, STS first samples representative data from both new data and the reservoir. Then, with the sampled data, AEL efficiently performs concurrent training for all individual recommendation models and, after that, effectively fuses the results of all these models with a sequential adaptive fusion approach to obtain the final recommendation result. 
	
	To better introduce the workflow of STS-AEL, we have presented its high-level procedure in~\Cref{alg:sts-ael}. Specifically, as ~\Cref{alg:sts-ael} illustrates, the training data are first prepared by STS (line 2), with which the multiple individual models are trained in parallel (lines 3 and 4). After that, when conducting the predictions, the trained individual models generate prediction results in parallel (lines 6 and 7). Then, these prediction results from multiple individual models are fused into the final one with the sequential adaptive fusion method, which elaborately calculates the fusion weights based on the recommendation accuracies of the last batch of received interactions (lines 8 and 9). Finally, the prediction accuracies are stored for calculating the fusion weights regarding the next batch of received interactions (line 10). More details about STS-AEL are presented in the following.
	%More details about STS and AEL are presented in~\Cref{ssec:sts} and~\Cref{ssec:ael}, respectively.
	
	%The more detailed process of STS-AEL is shown in~\Cref{alg:STS-AEL}. As indicated in Lines~\ref{line:alg_sampling_start} - \ref{line:alg_sampling_end}, STS-AEL first employs STS to extract representative and diversified samples from new data (in the overload scenario) or historical data (in the underload scenario). Then, the sampled data will be used to train the recommendation model by AEL (Line~\ref{line:alg_ensemble}). After that, the historical data will be adjusted based on new data $D_N$ and size limit of historical data \lh (Lines~\ref{line:alg_history_start} - \ref{line:alg_history_end}). Finally, the ensemble model is obtained (Line~\ref{line:alg_recommend}) for streaming recommendation.

	%temporary historical data will all be used, instead of being sampled, to conduct joint training when it is less than required. It should be noted that STS-AEL can also handle the overload scenario, as shown in Lines~\ref{line:alg_HEL_Rec_insu_start} - \ref{line:alg_HEL_Rec_insu_end}, by training only sampled new data instead of joint training. In addition, as mentioned in this subsection earlier, the individual models can be trained in parallel although not demonstrated in~\Cref{alg:STS-AEL} for concision.
	
	\label{ssec:num1}

	%	\begin{algorithm}[t]
	%		\footnotesize
	%		
	%		\Input{individual models: \qc, new records: \nr, whole historical records: \hr, size limit of positive records and \hr: \bs, \lh, \dfrac{discounting factor}{den}s for \nr and \hr: $\lambda_1$, $\lambda_2$
	%		}
	%		\Output{the aggregated model: $\mathcal{M}$ }
	%		
	%		
	%		\While{\nr \rm{is not empty}}
	%		{
	%			\If{$|\nr| \geq$ \bs \label{line:alg_sampling_start}}
	%			{
	%				get $T_N$ by STS $\bs$ records from $\nr$ without replacement \\
	%				$T_H$ $\leftarrow$ $\emptyset$
	%				
	%			}
	%			\Else
	%			{
	%				%	\If{$|\hr| \leq \bs - |\nr|$ \label{line:alg_HEL_Rec_insu_start}}
	%				%	{$\ts  \leftarrow \nr \cup \hr$\\
	%				%		$S \leftarrow$  $\emptyset$\label{line:alg_HEL_Rec_insu_end}
	%				%	}
	%				%	
	%				%	\Else
	%				%	{
	%				$k$ $\leftarrow$ $\bs - |\nr|$ \\
	%				get $T_H$ by STS $k$ records from $\hr$ without replacement \\
	%				$T_N$ $\leftarrow$ $N$
	%				\label{line:alg_sampling_end}
	%				%$S \leftarrow {r[i] \sim \mathcal{D}(\lambda_2, \hr), i = 1,2,...,\bs - |\nr|}$\\
	%				%	$sampled\_data$ = ${data \stackrel{sample}{\longleftarrow} \mathcal{D}(\lambda_2, |\hr|) \text{ for }l\text{ times}}$ \\
	%				%	}
	%				
	%			}
	%			{\fontsize{8.5}{10.2}\selectfont$\mathcal{M}\leftarrow$ Hybrid\_Ensemble\_ Learning($\qc$, $T_N$, $T_H$, $\bs$, $\hr$)}  \label{line:alg_ensemble} \\ 
	%			\hr.\push{\nr}  \label{line:alg_history_start} \\
	%			\While{$|\hr| \rangle \lh$}
	%			{
	%				\hr.\pop{}
	%			}\label{line:alg_history_end}
	%			\caption{STS-AEL}
	%			\label{alg:STS-AEL}
	%			\Return{$\mathcal{M}$} \label{line:alg_recommend}
	%		}
	%	\end{algorithm}
	%	
	
	\subsection{Stratified and Time-aware Sampling}
	\label{ssec:sts}
	Training the recommendation model with the entire dataset is impractical for SRSs, as streaming data is continuous and infinite.
	% while the processing time and computation resources are limited.
	To this end, we propose STS to sample representative data to reduce the training workload effectively.
	
	%Different from the existing sampling methods which commonly utilize new data only~\cite{eals} or fail to sufficiently emphasize on new data~\cite{spmf} for training recommendation models, our proposed STS 
	To capture both short-term and long-term user preferences, STS elaborately incorporates both new data and historical data while guaranteeing the proportion of new data.
	%to capture both short-term and long-term user preferences. 
	Specifically, STS contains five key steps: 1) maintain a reservoir containing representative historical data, which is a widely-used technology~\cite{reservoir-boosting,spmf} in the streaming processing area, 2) calculate the sample sizes of both this reservoir and new data, 3) calculate the probabilities to sample user-item interactions from both the reservoir and new data, 4) with the sampling sizes and sampling probabilities, obtain sample sets $\mathbf{SS}_{his}$ and $\mathbf{SS}_{new}$ from the reservoir \textbf{R} and new data \textbf{N}, respectively, and 5) merge $\mathbf{SS}_{his}$ and $\mathbf{SS}_{new}$ to form the final sample set $\mathbf{SS}$ as the input of the subsequent concurrent training.
	
	%	To ensure the recommendation accuracy, STS addresses concept drift while capturing long-term user preferences. To achieve this goal, STS incorporates both new data and historical data while guaranteeing the proportion of new data. Specifically, STS first maintains a reservoir, which contains a pre-set amount of representative historical data, and then samples data from both this reservoir and the new data. Besides, considering the additional sampled historical data also helps to increase the resource utilization ratio in the underload scenario. 
	\begin{figure}[]
		\centering
		\includegraphics[width= 0.7\textwidth, keepaspectratio]{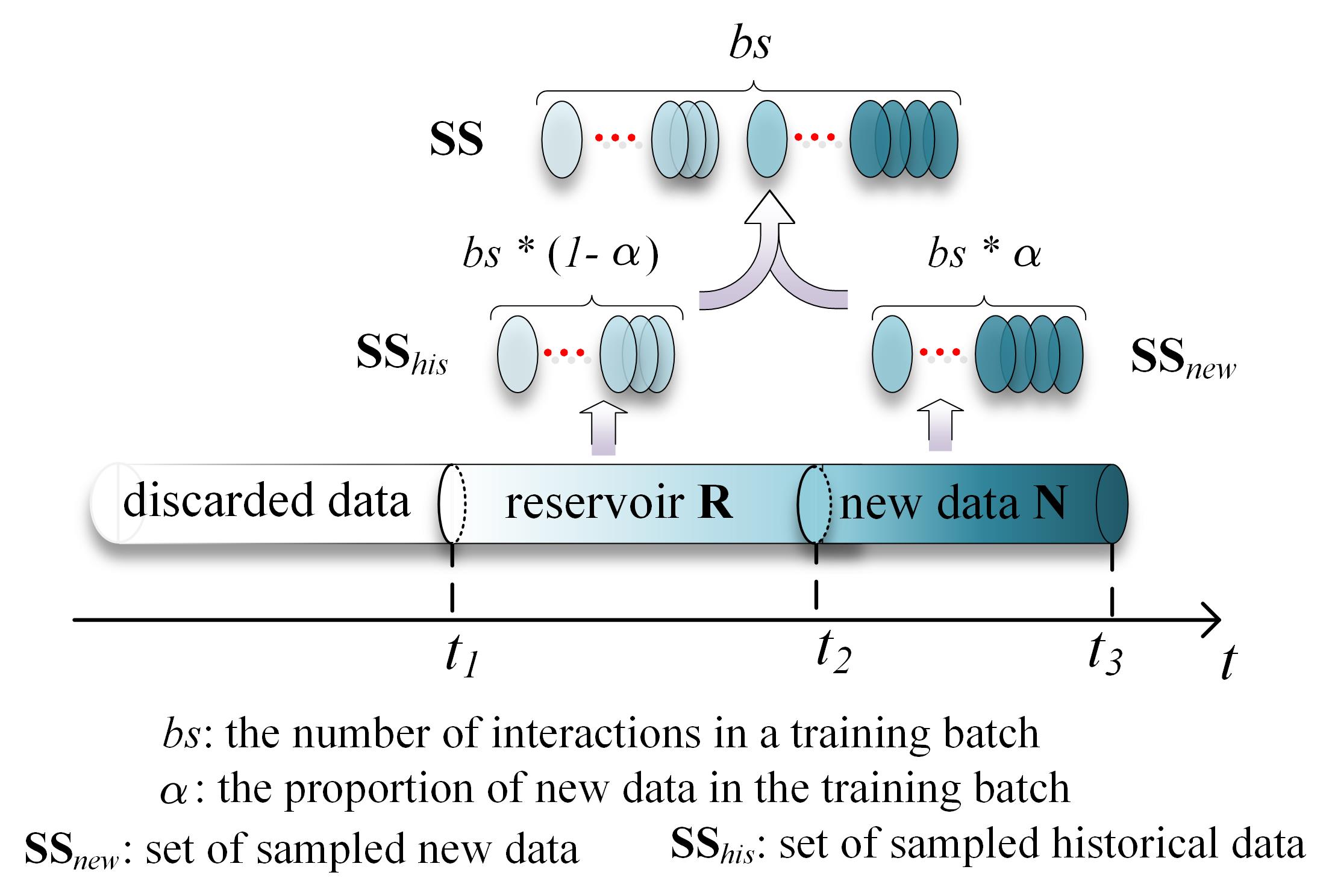}
		%	\vspace{-5mm}
		\caption{Stratified and Time-aware Sampling (STS) approach. STS first maintains a reservoir which records representative historical data, and then utilizes a time-aware method to sample data from both this reservoir and new data, respectively, to captures both long-term and short-term user preferences. Note that, for the reservoir maintenance, STS incorporates the new data and discards the oldest data when the reservoir runs out of the space as newer data commonly reflect more timely user preferences.}
		%	\vspace{-5mm}
		\label{figure:STS} 	
	\end{figure}
	To better illustrate our proposed STS approach, we present its sample process in~\Cref{figure:STS}, where the darker color indicates the newer data, i.e., the data received more recently. As shown in~\Cref{figure:STS}, the whole data can be partitioned into three parts, i.e., new data, reservoir, and discarded data, based on receiving time. During the reservoir maintenance, STS incorporates newly received data and discards the oldest data, as newer data contains more timely user preferences towards items. To guarantee the proportion of the sampled new data in the entire training data, STS adopts a stratified sampling strategy and utilizes parameter $\alpha$ to adjust the proportion of the sampled new data. With this proportion $\alpha$ and the training batch size $bs$, STS first calculates the sample size of new data: $|\mathbf{SS}_{new}| = bs * \alpha$ and the sample size of the historical data from the reservoir: $|\mathbf{SS}_{his}| = bs * (1-\alpha)$. And then,  $\mathbf{SS}_{new}$ and $\mathbf{SS}_{his}$ are sampled from new data and reservoir, respectively, both in a time-aware manner. Specifically, to assign newer data in the sample space higher sampling probability, we employ decay ratios $\lambda_{new}$ ($\lambda_{new} \geq 1$) and $\lambda_{res}$ ($\lambda_{res} \geq 1$) for the new data \textbf{N} and the reservoir \textbf{R}, respectively. Next, we present the sampling process for $\mathbf{SS}_{new}$ in detail, while the sampling process for $\mathbf{SS}_{his}$ is similar. Given the sampling probability $p_{k-1}$ of the $(k-1)^{th}$ user-item interaction, the sampling probability $p_{k}$ of the $k ^{th}$ user-item interaction can be calculated as below, 
	%	\vspace{-1 mm}
	\begin{equation}
	\label{eq:degrade_lambda}
	p_k = p_{k-1} * \lambda_{new}.
	\end{equation}
	Through this way, we can adjust the ratio (i.e., $\lambda_{new}$ for new interactions and $\lambda_{res}$ for interactions in the reservoir) of the sampling probability of the $k^{th}$ received interaction against that of the $({k-1})^{th}$ received interaction, and thus adjust the emphasis of our approach on newer interactions with more flexibilities. After that, by iteratively performing~\Cref{eq:degrade_lambda} and assuming the sampling probability of the earliest user-item interaction is $p_1$, we can get $p_k$ as follows, 
	
	\begin{equation}
	\label{eq:freshness}
	p_k = p_{1} * \overbrace{\lambda_{new} * \lambda_{new} * \cdots * \lambda_{new}}^{k-1}  = p_{1} * (\lambda_{new})^{k-1}.
	\end{equation}
%	\begin{equation}
%	\label{eq:freshness}
%	p_k = p_1 * \lambda_{new}^{k-1}.
%	\end{equation}
	Based on~\Cref{eq:freshness}, with the size $|\mathbf{N}|$ of new data, we can infer the normalized sampling probability of the $k^{th}$ user-item interaction as below,
	\begin{equation}
	\label{eq:sample_probability}
	P(k | \lambda_{new},|\mathbf{N}|) = \frac{p_k}{\sum_{i=1}^{|\mathbf{N}|} p_i}=\frac{\lambda_{new}^{k-1} * (1-\lambda_{new})}{1-\lambda_{new}^{|\mathbf{N}|}}.
	\end{equation}
	Then, with $P(k | \lambda_{new},|\mathbf{N}|)$,  STS samples $\mathbf{SS}_{new}$ from the new data. Note that the sampling process is with replacement among individual models, which means that one user-item interaction can be possibly sampled by multiple individual models. Using the similar method, STS samples $\mathbf{SS}_{his}$ from the reservoir. Then, the final sampled training data set $\mathbf{SS}$ can be obtained by merging $\mathbf{SS}_{new}$ and $\mathbf{SS}_{his}$  as follows, 
	\begin{equation}
	\label{eq:merge}
	\mathbf{SS} = \mathbf{SS}_{new} \cup \mathbf{SS}_{his}.
	\end{equation}
	
	The aforementioned parameters $\alpha$, $\lambda_{new}$, and $\lambda_{res}$ provide STS with flexibilities to effectively handle data stream with various characteristics and receiving speeds. For example, in the scenario where the concept drift happens frequently and the new data should be more emphasized to better capture the short-term user preferences, we can increase the values of $\alpha$, $\lambda_{new}$ and  $\lambda_{res}$ to increase the proportion of sampled new data and the sampling probabilities of newer data in the sample space to better handle the data stream. Furthermore, STS has a strong generalization capability and can be easily derived to the existing sampling approaches. For example, STS can be derived to the sliding window based sampling~\cite{sliding} by setting $\alpha$ to $\frac{|\mathbf{N}|}{bs}$ and  setting both $\lambda_{new}$ and $\lambda_{res}$ to large values, e.g., 1.5, and it has a similar effect to random sampling via randomly selecting $\alpha$ from $[0, 1]$ and setting both $\lambda_{new}$ and $\lambda_{res}$ to $1$.
	
	With the representative data sampled by STS from both new data and reservoir, the individual recommendation models can be trained by AEL, which will be introduced in the following subsection.\\
	
	\vspace{1pt}
	\noindent\textbf{Time Cost Analysis:} The time cost of STS is acceptable, as the time complexities of~\Cref{eq:degrade_lambda,eq:freshness,eq:sample_probability,eq:merge} are $O(bs)$ for sampling a training batch of user-item interactions. 
	%Moreover, the discounting factor $\lambda$ can be adjusted based on various factors, including the features of workload and the parameters of the individual recommendation models, to adapt to the specific application scenario. 
	%Moreover, STS has strong generality. Specifically, it can be derived to the sliding windows based approaches~\cite{sliding} by setting $\lambda$ close to $0$, e.g., 0.1, and has similar effect to random sampling via randomly selecting $p$ and setting $\lambda$ close to $1$, e.g., 0.999999.
	\subsection{Adaptive Ensemble Learning}
	\label{ssec:ael}
	With the data sampled by STS, our proposed AEL first performs concurrent training to train multiple individual models, and then fuses the results of these trained models via an effective sequential adaptive fusion method to obtain the final recommendation results with higher accuracy. 
	%obtain the final recommendation results. Different from the existing ensemble learning methods which typically train the models one by one (i.e., the Boosting~\cite{adaboost} based ensemble learning) or average the results of multiple models for the final result (i.e., Bagging~\cite{bagging} based ensemble learning), our proposed AEL first trains the recommendation models in parallel and then fuses the results of multiple individual models with elaborately calculated fusion weights based on the recommendation accuracies in the last iteration. 
	More details about AEL are presented below.\\
	%For fast adaption to concept drifts while tracking long-term user preferences, we first perform online-offline joint training on individual recommendation models to consider both new data and historical data. Then, we employ ensemble learning to make the final recommendation based on the individual models. The two parts together form AEL, whose process is shown in~\Cref{alg:HEL}.
	%	\begin{algorithm}
	%		\footnotesize
	%		
	%		\Input{individual models: \qc, new training data: $T_N$, historical training data: $T_H$, size of positive records: $\bs$, whole historical data: $\hr$
	%		}
	%		\Output{the aggregated model: $\mathcal{M}$ }
	%		\While{\nr \rm{is not empyt}}
	%		{
	%			\If{\qc \rm{is not empty}}
	%			{
	%				$m \leftarrow \qc.\pop{}$\\
	%				$T^+ = T_N \cup T_H$ \label{line:alg_positive}\\
	%				$T^-$ = negativeSample($T^+$, $\hr$) \label{line:alg_negative}\\
	%				$T = T^+ \cup T^- $ \label{line:alg_all_data}\\
	%				$m$ = $\mathcal{T}$($m$, $F$) by using one-pass strategy \label{line:training}\\
	%				$\qc.\push{M}$
	%			}
	%		}
	%		
	%		\caption{Hybrid\_Ensemble\_Learning}
	%		
	%		\label{alg:HEL}
	%		\Return{$\mathcal{M}(\cdot) = \frac{1}{n_m}\sum \limits_{M \in \qc} M(\cdot)$} \label{line:average}\\
	%	\end{algorithm}
	
	\vspace{1pt}
	\noindent\textbf{Concurrent Training:} AEL trains multiple individual recommendation models in parallel for computation efficiency, as the individual models are independent from one another during the training process. This feature of concurrency contributes to more effectively handling the streaming data, especially when confronting the excessive amount of data in an overload scenario. Moreover, to overcome the natural absence of negative feedback in recommendations with implicit user-item interactions, AEL performs negative sampling, a technique which has been widely used in the literature~\cite{negative2017cikm, eals}, for more effective training. To guarantee the effectiveness of negative sampling, AEL utilizes the aforementioned reservoir to check if an interaction between a user and an item exists.

	For individual recommendation models, AEL can employ existing monolithic SRSs directly or adapt offline RSs to the streaming setting by incrementally updating the recommendation models with an online update mechanism, e.g., stochastic gradient descent used in~\cite{ISGD}. In this paper, AEL delivers the best performance by adapting Neural Matrix Factorization (NeuMF) proposed in~\cite{ncf} to the streaming setting. Specifically, NeuMF is a neural network based RS, which combines other two basic RSs, i.e., Generalized Matrix Factorization (GMF) and Multiple Layer Perceptron (MLP), to achieve more accurate recommendations. 
	%Let $p_u$ and $q_i$ denote the latent factors of user and item, respectively. 
	As our proposed STS-AEL achieves high recommendation accuracies when ensembling these three monolithic models, i.e., GMF, MLP, and NeuMF, we briefly introduce them in the following. 
	
	As its name indicates, GMF is a generalized matrix factorization model which enhances the original matrix factorization model with non-linear transformation for stronger modelling capabilities. Specifically, GMF improves matrix factorization with a nonlinear activation function $a_{out}^{GMF}$ as below,
	\begin{equation}
	\label{eq:gmf1}
	\phi^{\textit{GMF}} = \textbf{p}_{u}^{\textit{GMF}} \otimes \rm{\textbf{q}}_{v}^{\textit{GMF}},
	\end{equation}
	\begin{equation}
	\label{eq:gmf2}
	\hat{y}_{u,v}^{GMF} = a_{out}^{GMF}(\textbf{W}_{\textit{GMF}}^\textit{T} \phi^{\textit{GMF}} + \textbf{b}_{GMF}),
	\end{equation}
	where $\mathbf{p}_{u}$ and $\mathbf{q}_{v}$ represent the embedding of user $u$ and the embedding of item $v$, respectively, $\otimes$ denotes the element-wise multiplication, \textbf{W}$_{GMF}$ denotes the weight matrix, $\textbf{b}_{GMF}$ denotes the bias vector, and $\hat{y}_{u,v}^{GMF}$ indicates the predicted probability for an interaction between user $u$ and item $v$. In such a way, through enhancing the conventional matrix factorization with nonlinearity, GMF can achieve stronger fitting ability, and thus make more accurate predictions. 
	
	Different from GMF which learns user preferences from the interactions based on a fixed dot product between the user embedding and the item embedding, MLP aims to improve the modelling flexibilities with a multiple layer perception structure, and thus achieve higher recommendation accuracies. Specifically, MLP first concatenates the user embedding and item embedding as follows, 
	\begin{equation}
	\label{eq:mlp1}
	\phi_0^{\textit{MLP}} = \begin{bmatrix} 
	\textbf{p}_{u}^{MLP};
	\textbf{q}_{v}^{MLP}    
	\end{bmatrix},
	\end{equation}
	and then feeds the concatenated embedding to a $L$-layer perceptron for training with interactions between users and items as below,
	%\begin{equation}
	%\rm{\textbf{z}}_1 = \phi_1^{\textit{MLP}} = \begin{bmatrix} 
	%\textbf{p}_{u}  \\
	%\textbf{q}_{i}  
	%\end{bmatrix},
	%\end{equation}
	
	%\begin{equation}
	%\phi_k ^ {MLP} = a_k^{MLP}(\rm{\textbf{W}}_{k}^\textit{T}\phi_{\textit{k}-1}^{\textit{MLP}} + \textbf{b}_{k}), \label{eq:mlp_inner}
	%\end{equation}
	%\begin{equation}
	%\hat{y}_{u,v}^{MLP} = a_{L}^{MLP}(\rm{\textbf{W}}_{\textit{L}}^\textit{T} \phi_{\textit{L}-1}^{\textit{MLP}}), \label{eq:mlp_output} 
	%\end{equation}

	\begin{align}
	\phi_1 ^ {MLP} &= a_1^{MLP}(\textbf{W}_1^\textit{T}\phi_{0}^{\textit{MLP}} + \textbf{b}_1), \label{eq:mlp_input} \\
	&\mathrel{\makebox[\widthof{=}]{\vdots}} \nonumber \\
	\phi_k ^ {MLP} &= a_k^{MLP}(\textbf{W}_{k}^\textit{T}\phi_{\textit{k}-1}^{\textit{MLP}} + \textbf{b}_{k}), \label{eq:mlp_inner} \\
&\mathrel{\makebox[\widthof{=}]{\vdots}} \nonumber \\
	\hat{y}_{u,v}^{MLP} &= a_{L}^{MLP}(\textbf{W}_{\textit{L}}^\textit{T} \phi_{\textit{L}-1}^{\textit{MLP}}  + \textbf{b}_{L}), \label{eq:mlp_output} 
	\end{align}
	where $\textbf{W}_{k}$, $\textbf{b}_{k}$, and $a_k$ denote the weight matrix, bias vector, and activation function for the $k^{th}$ ($1 \leq k \leq L$) layer, respectively. MLP obtains much flexibility from the concatenated embedding and nonlinear perceptrons, and thus can capture the user preferences effectively. 
	
	To further improve the performance of the recommendations, NeuMF fuses these two recommendation models, i.e, GMF and MLP, to complement each other for better learning the user preferences towards items. Specifically, NeuMF first concatenates the features learned by GMF and MLP, and then transforms the concatenated feature by a nonlinear function $a_{out}^{NeuMF}$ as below,
	\begin{equation}
	\label{eq:neumf}
	\hat{y}_{u,v}^{NeuMF} = a_{out}^{NeuMF}(\textbf{W}_{\textit{NeuMF}}^\textit{T} \begin{bmatrix} 
	\phi^{\textit{GMF}};
	\phi_{L-1}^{\textit{MLP}}
	\end{bmatrix} + \textbf{b}_{NeuMF}),
	\end{equation}
	where \textbf{W}$_{NeuMF}$ and \textbf{b}$_{NeuMF}$ denote the weight matrix and the bias vector, respectively. Through such a process, NeuMF expects to 
	combine the advantages of both GMF and MLP, and thus deliver more accurate recommendations.
	
	For the training, following the work in~\cite{ncf,ssrm,wang2019modeling}, we employ the binary cross-entropy loss as the loss function for the training purpose as below,
	\begin{equation}
	\label{eq:loss}
	\ell(y_{u,v},\hat{y}_{u,v}) = -(y_{u,v} \text{log} \hat{y}_{u,v} + (1-y_{u,v})\text{log} (1-\hat{y}_{u,v})),
	\end{equation}
	where $y_{u,v}$ indicates if an interaction between user $u$ and item $v$ exists, and $\hat{y}_{u,v}$ is the predicted probability for this interaction. Specifically, this loss function encourages larger $\hat{y}_{u,v}$ if the interaction between user $u$ and item $v$ exists (i.e., $y = 1$), and encourages smaller $\hat{y}_{u,v}$ otherwise. With this binary cross-entropy loss function defined in~\Cref{eq:loss}, the individual recommendation models can be trained via stochastic gradient descent.\\

	\vspace{1pt}
	\noindent\textbf{Sequential Adaptive Fusing:} The proposed sequential adaptive fusion approach improves fusion performance by assigning elaborately calculated weights to multiple individual models in the streaming scenario, where the training and test are iteratively conducted with the data stream. 
	%The intuitive idea of sequential adaptive fusion is utilizing the test accuracies of each individual model regarding each user-item pair in the last iteration to guide the fusion in the current iteration. 
	Specifically, AEL contains four key steps: 1) calculate and store the prediction accuracy for each interaction and the corresponding user-item pair (i.e., the user and item related to this interaction) for each individual model in the current iteration, 2) with the prediction accuracies and the corresponding user-item pairs stored in the last iteration, estimate the confidence of each individual model to predict the interaction for the target user-item pair, 3) based on the calculated confidence, use an AdaBoost-like method~\cite{OWE,AWE} to calculate the fusion weights for all the individual models, and 4) fuse the predictions of those models with the fusion weights to obtain the ensembled prediction. More details are presented below.
	
	AEL maintains a set $\mathbf{P}$ containing tuples of prediction accuracies and the corresponding user-item pairs in the last iteration for each individual model, taking the $k^{th}$ individual model as an example, 
	\begin{equation}
	\mathbf{P}_k=\{\langle acc_{1}^k, {u_1}^k,{v_1}^k\rangle, \cdots, \langle acc_{j}^k, {u_j}^k,{v_j}^k\rangle, \cdots,\langle acc_{g}^k, {u_g}^k,{v_g}^k\rangle\}, 
	\label{eq:pk}
	\end{equation}
	where $acc_{j}^k$ ($1 \leq j \leq g$) represents the accuracy of the $k^{th}$ individual model regarding user-item pair $\langle u_j^k, v_j^k \rangle$, and $g$ is the size of $\mathbf{P}_k$. To predict the interaction between user $u$ and item $v$ by the $k^{th}$ individual model, AEL first calculates the similarity (we employ cosine similarity in this paper to achieve the best performance) between the target user-item pair $\langle u,v\rangle$ and each of the user-item pairs $\langle u', v'\rangle$ in $\mathbf{P}_k$ based on their embeddings,
	\begin{equation}
	\text{cos\_sim}_{\langle u,v\rangle,\langle u',v'\rangle}^k = \frac{[\textbf{p}_u ^{k}; \textbf{q}_v^k] \cdot [\textbf{p}_{u'} ^{k}; \textbf{q}_{v'}^k]}{\|[\textbf{p}_u ^{k}; \textbf{q}_v^k]\|_2 \|[\textbf{p}_{u'} ^{k}; \textbf{q}_{v'}^k]\|_2}, \label{eq:sim} \\
	\end{equation}
	where $[\textbf{p}_u; \textbf{q}_v]$ indicates the concatenation of embeddings $\textbf{p}_u$ and $\textbf{q}_v$, and  $\|*\|_2$ represents the L2 norm. Then, based on these similarities, AEL creates subset $\mathbf{S}_k \subseteq \mathbf{P}_k$ (taking the $k^{th}$ individual model as an example) for each individual model by extracting the top $e$ (a predetermined parameter representing the size of $\mathbf{S}_k$) tuples which have most similar user-item pairs to the target $\langle u, v\rangle$ from $\mathbf{P}_k$. Then, we can estimate the confidence of each model for the prediction of the interaction for the target $\langle u, v\rangle$. Specifically, with $\mathbf{S}_{k}$, AEL calculates the confidence $c_{u,v}^k$ of the $k^{th}$ individual model to predict the interaction for the target $\langle u,v\rangle$ as follows,
	\begin{equation}
	c_{u,v}^k = \frac{1}{|\mathbf{S}_{u,v}^k|}\sum_{\langle\textbf{p}_{u'}^k,\textbf{q}_{i'}^k,a_{u',i'}^k\rangle \in \mathbf{S}_{u,v}^k}a^k_{u',v'}. \label{eq:pp}
	\end{equation}
	For the sake of simplicity, we use $\mathbf{c}_{u,v}$ to represent the vector containing the confidence of all the individual models to predict the target $\langle u,v \rangle$, i.e.,
	\begin{equation}
	    \mathbf{c}_{u,v} = [c_{u,v}^1, \cdots,c_{u,v}^o]^T,
	    \label{eq:vector}
	\end{equation}
	where $o$ is the number of individual models. With this estimated confidence $\mathbf{c}_{u,v}$, the fusion weights for the prediction of the interaction between user $u$ and item $v$ can be calculated and normalized with an AdaBoost-like~\cite{Aboost} method as below,
	\begin{equation}
	\textbf{fw}_{u,v}' = \frac{\textbf{c}_{u,v}}{1-\textbf{c}_{u,v}}, 
	\label{eq:weight1}
	\end{equation}
	\begin{equation}
	\textbf{fw}_{u,v} = \frac{\textbf{fw}_{u,v}'}{\|\textbf{fw}_{u,v}'\|_1}, \label{eq:weight2}
	\end{equation}
    where $\|*\|_1$ represents the L1 norm and $\textbf{fw}_{u,v}$ represents the fusion weights of the individual models for $\langle u, v\rangle$. As shown in~\Cref{eq:weight1,eq:weight2}, AEL assigns higher fusion weights to the models those with more confidence (i.e., larger $\mathbf{c}_{u,v}$), for more effective fusion. Finally, AEL fuses the predictions $\mathbf{\hat{y}}$ 
	from the multiple individual recommendation models with $\textbf{fw}_{u,v}$ to get the final prediction as follows,
	\begin{equation}
	\hat{y}_{u,v}^{final} = \textbf{fw}^T_{u,v}\mathbf{\hat{y}}_{u,v}. \label{eq:fusion_for_ael}
	\end{equation} \\
	%As the sequential adaptive fusing employs the embeddings of users and items to calculate the fusion weights, thus STS-AEL can only ensemble neural network based models.\\
	\noindent \textbf{Time Cost Analysis:} The time cost of AEL mainly contains two parts, i.e., 1) the time cost of the concurrent training, and 2) the time cost of the sequential adaptive fusion. As multiple individual models are trained in parallel in the concurrent training process, the time complexity for the concurrent training is roughly equal to that of the corresponding monolithic recommendation model. As for the sequential adaptive fusion, the prediction processes of the individual models can also be parallel. Thus, compared with the monolithic recommendation model, the extra time cost introduced by AEL mainly comes from the calculation of fusion weights, which is described by  \Cref{eq:sim,eq:pp,eq:vector,eq:weight1,eq:weight2}. The time complexities of \Cref{eq:sim,eq:pp,eq:vector,eq:weight1,eq:weight2} can be easily calculated, i.e., $O(|\mathbf{{p}_{u}}|*|\mathbf{{q}_{v}}|*|\mathbf{S}_{k}|)$, $O(|\mathbf{S}_{k}|)$, $O(1)$, $O(o)$, and $O(o)$, respectively. Obviously, the extra time cost mainly depends on \Cref{eq:sim}, i.e., $O(|\mathbf{{p}_{u}}|*|\mathbf{{q}_{v}}|*|\mathbf{S}_{k}|)$, since it is the highest one. This time complexity is acceptable for the following two reasons: 1) it is constant once the parameters, i.e., the sizes of latent factors $|\mathbf{{p}_{u}}|$ and $|\mathbf{{q}_{v}}|$) and the size of $\mathbf{S}_{k}$ (i.e., a set of the interactions and corresponding recommendation accuracies from the $k^{th}$ individual model), are determined, and 2) it is not affected by the number of users or the number of items.  Although AEL has a higher time complexity than the classic Bagging~\cite{bagging} based ensemble learning, which commonly averages the results of multiple individual models for the fusion purpose with a time complexity of $O(1)$, our proposed AEL greatly improves the fusion performance by fusing the results of multiple individual models with elaborately calculated weights. As for the Boosting~\cite{adaboost} based ensemble learning methods, they typically train the individual models one by one, and thus need much more training time compared with our proposed AEL which trains the individual models in parallel.
%	Moreover, the fusion weights can be calculated in parallel with the prediction processes of individual models as they are independent from each other. Thus, the extra cost from the fusion weight calculation is acceptable. In summary, AEL introduces trivial extra time cost compared with the monolithic recommendation model.
	
	\section{Experiments}
	In this section, we present the results of the extensive experiments we conducted
	which aim to answer the following four research questions:
	\begin{enumerate}[start=1,label={\upshape\bfseries RQ\arabic*.},wide = 0pt, leftmargin = 3.2em]
		%\vspace{-1mm}
		\item How does our proposed STS-AEL perform when compared with the state-of-the-art approaches? 
		%\vspace{-1mm}
		%\item How does our proposed STS-AEL framework outperform its individual recommendation model?
		%	\vspace{-1mm}
		%\item How does STS improve the recommendation accuracydddd?
		\item How does the number of individual models ensembled by STS-AEL affect the recommendation accuracy?
		\item How does our proposed STS perform when compared with the existing sampling methods?
		%	\vspace{-1mm}
		%\item How does AEL improve the recommendation accuracydddd?
		\item How does our proposed AEL perform when compared with the existing ensemble methods?

		%\vspace{-1mm}

	\end{enumerate}

	\subsection{Experimental Settings}
	Before presenting the results and analysis of the experiments, we first introduce the experimental settings, i.e., the datasets, evaluation policy, evaluation metrics, and comparison approaches, for readers to better comprehend the experiments.\\
	
	\vspace{1pt}
	\noindent\textbf{Datasets:} In the experiments, we employ three real-world datasets, i.e., MovieLens (1M)\footnote{https://grouplens.org/datasets/movielens/1m}, Netflix\footnote{https://www.kaggle.com/netflix-inc/netflix-prize-data}, and Yelp\footnote{https://www.yelp.com/dataset/challenge}, all of which are widely used in the literature~\cite{spmf,eals}, to evaluate our proposed STS-AEL and baselines. Since the original Netflix dataset contains over 100 million interactions, which is beyond our computation capacity, we randomly select the data of 5000 users for the experiments. In addition, we follow the common practice~\cite{eals,bpr} to retain the users who have more than ten interactions on all three datasets to reduce the data sparsity. The statistics of the tuned datasets are summarized in~\Cref{table:datasets}. 
	%The three datasets we employed contain records in the form of a four-element tuple $langleuser, item, rating, timestamp>$. 
	Since this work focuses on the recommendations with implicit user-item interactions, following the common practice~\cite{ocfif,nmrn-gan,eals}, we transform the explicit data (i.e., users' ratings on items) in all three datasets into the implicit ones, where it is 1 if an explicit interaction exists and 0 otherwise. 
	\begin{table}[]
		\centering
		\begin{tabular}{|c|c|c|c|c|}
			\hline
			\textbf{Datasets} & \textbf{\#Users} & \textbf{\#Items} & \textbf{\#Interactions} & \textbf{Sparsity} \\ \hline
			MovieLens         & 6400             & 3703             & 994169                  & 95.81\%           \\ \hline
			Netflix           & 5000             & 16073            & 1010588                 & 98.74\%           \\ \hline
			Yelp              & 25677            & 25815            & 731671                  & 99.89\%           \\ \hline
		\end{tabular}
		%	\vspace{-1mm}
		\caption{Experimental datasets. The symbol \# in this table denotes the number, e.g., \#Users represents the number of users.}
		%	\vspace{-3mm}
		\label{table:datasets}
	\end{table}\\
	
	\vspace{1pt}
	\noindent \textbf{Evaluation Policy:} Similar to~\cite{eals,spmf}, we first sort the data by their receiving time, and then divide them into a training set (where the data are used for incremental training) and a test set (where the data are first used for testing and then used for incremental training) to simulate the historical data and upcoming data, respectively, in the streaming scenario. The proportion of the training set is set to 85\%, 90\%, and 95\%, respectively, and following~\cite{eals}, we report the results in the case where the training set proportion is 90\% while the results in other two cases are similar to the reported ones.  %The base training set is regarded as historical data to initialize the recommendation models, and thus to perform more stable and accurate evaluation. And for incremental evaluation and training set, the new data is first used to evaluate the current model, and then treated as part of the training data  to simulate the real-word practice of handling data streams. After the first training, the new data will be regarded as historical data.
	%Moreover, we use the number of data  received or processed in one unit time to simulate the data receiving or processing speeds. To observe the performance over different workload intensities, we use the amount of fix the data processing speed $sp_{p} = 256$ with various data receiving speed $sp_{r}$, where $sp_r < 256$ and $sp_r > 256$ indicate the underload scenario and overload scenario, respectively.\\
	Moreover, to observe the performance of our proposed STS-AEL and that of the baselines w.r.t. different workload intensities, we train all the models with a fixed number $n_{p}$ (we set $n_p = 256$ in this paper) of user-item interactions in each iteration and adjust the number $n_{r}$ of user-item interactions received in this training period to simulate the cases with different workload intensities. For the sake of simplicity, we use $n_{p}$ and $n_{r}$ to simulate the data processing speed $sp_{p}$ and data receiving speed $sp_{r}$, respectively, where  $sp_{p} > sp_{r}$ and $sp_{p} < sp_{r}$ indicate the underload scenario and overload scenario, respectively. \\
	
	\vspace{1pt}
	\noindent  \textbf{Evaluation Metrics:} We adopt the ranking-based evaluation strategy, which is widely used for the evaluation of streaming recommendations with implicit data~\cite{spmf,eals}. Specifically, for each given interaction between a target user and a target item, we randomly sample 99 items which are not interacted with this user as negative items, and rank the target item among the 100 items (i.e., the target one plus the 99 sampled ones). Then, the recommendation accuracy is evaluated by two widely used metrics: Hit Ratio (\textbf{HR}) and Normalized Discounted Cumulative Gain (\textbf{NDCG})~\cite{he2015trirank,eals,spmf}. Specifically, taking HR@10 and NDCG@10 as examples, HR@10 tests if the target item is ranked in the top 10 recommended items, while NDCG@10 considers the specific ranking position of the target item in the top 10 recommended items. \\

	\vspace{1pt}
	\noindent\textbf{Comparison Approaches:} We compare the performance of our proposed STS-AEL framework with that of nine baseline models, including one ensemble model (i.e., OCFIF) and eight monolithic models (i.e., iBPR, iGMF, iMLP, iNeuMF, iTPMF-CF, RCD, eAls, and SPMF). The brief introduction of these baselines are as follows.  
	\begin{itemize}
		%\vspace{-1mm}
		
		%\vspace{-1mm}

		\item \textit{Bayesian Personalized Ranking} (BPR) is a representative pair-wise ranking approach proposed by Rendle et al.~\cite{bpr} to optimize the matrix factorization. We adapt this work to the streaming setting, named as \textbf{iBPR}, by incrementally updating the recommendation model with new data.
		\item \textit{Neural Matrix Factorization} (NeuMF)~\cite{ncf} is an advanced matrix factorization model, which combines two recommendation models, i.e., \textit{Generalized Matrix Factorization} (GMF) and \textit{Multi-Layer Perceptron} (MLP), for higher recommendation accuracy.  We adapt these three offline recommendation models, i.e., NeuMF, GMF, and MLP, to the streaming setting, named as \textbf{iNeuMF}, \textbf{iGMF}, and \textbf{iMLP}, respectively, by feeding them with newly receiving data continuously.
		\item \textit{Time-window based probabilistic Matrix Factorization for Collaborative Filtering} (TPMF-CF)~\cite{applied_intelligence_cf_time_windows} is a representative probabilistic matrix factorization based approach which adopts the time window technique to construct a 3D user-item-time model. As the TPMF-CF is originally designed for the offline scenario, we adapt TPMF-CF to the streaming setting, named as \textbf{iTPMF-CF}, by incrementally training the recommendation model with new data. Furthermore, to keep the highlight of the TPMF-CF, we also employ a sliding window based sampling approach to improve its recommendation accuracy.
		\item \textit{Randomized block Coordinate Descent} (\textbf{RCD}) and \textit{Element-wise Alternating Least Squares} (\textbf{eAls}) are two representative approaches employed by Devooght et al.~\cite{rcd} and He et al.~\cite{eals}, respectively, to optimize the streaming matrix factorization. To be consistent with the proposed STS-AEL and other baselines, we enhance eAls and RCD with abilities of the batch process to increase their throughput.
		\item \textit{Stream-centered Probabilistic Matrix Factorization} (\textbf{SPMF})~\cite{spmf} is a state-of-the-art monolithic SRS based on the probabilistic matrix factorization, which improves the work in~\cite{ISGD}. SPMF is originally performed along with a ranking-based sampling method. However, the ranking-based sampling has excessive computation complexity since it evaluates all the user-item interactions in the reservoir and rank them by the test accuracies for each sampling process, and thus is not suitable for our evaluation policy where sampling needs to be performed frequently. Therefore, we sample data for SPMF with our proposed STS for a fair comparison.
		%	\vspace{-1mm}
		%, to distinguish them with the original models.
		%	\vspace{-1mm}
		\item \textit{Online Collaborative Filtering with Implicit Feedback} (\textbf{OCFIF})~\cite{ocfif} is the only ensemble SRS, which combines multiple matrix factorization models to deliver more accurate streaming recommendations.  It is specifically devised to ensemble matrix factorization models, and thus cannot ensemble other models for the evaluation.
		\item \textit{Stratified and Time-aware Sampling based Adaptive Ensemble Learning} (\textbf{STS-AEL}) is our proposed SRS. For the evaluation, we take each of the top three best-performing monolithic baselines (as shown in Experiment 1), i.e., iNeuMF, iMLP, and iGMF, as its individual model and set different numbers (i.e., 2, 4, 6, and 8) of individual models to compose different forms of STS-AELs. For example, STS-AEL\_8-iNeuMF indicates an STS-AEL framework ensembling 8 iNeuMF models. 
		%Since STS-AEL can only ensemble embedding based models, SPMF is not taken as its individual model. 
	\end{itemize}
	\textbf{Parameter Setting:} For baselines, We initialize them with the parameters reported in their  papers and tune them based on our experimental scenarios to achieve the best performance for a fair comparison. 
	%For OCFIF, we empirically set the number of matrix factorization models $o = 10$ and the parameter $p$ of the $i^{th}$ model as $(1-\frac{i}{o+1})/(\frac{i}{o+1})$ to deliver the best performance.
	%For neural network based approaches, i.e., iGMF, iMLP, iNeuMF, and STS-AEL frameworks ensembling these three models, we empirically set the 
	%embedding size to 8,  batch size to $256$, and neural network structure to $64 \rightarrow 32 \rightarrow 16 \rightarrow 8$ (the number indicates the hidden units in each layer) to achieve the best performance.
	For our proposed STS-AEL, we empirically set the learning rate as 0.001, and initialize the parameters in embedding layers with the Gaussian distribution  $X\sim N(0,0.25)$, inner layers with Glorot initialization~\cite{glorot_initialization}, and the output layer with LeCun initialization~\cite{lecun_initialization}.
	Besides, we adopt L2 regularization to avoid overfitting and Adaptive Moment Estimation (Adam)~\cite{adam} for the regularization. In addition, we manually adjust the parameters $\alpha$, $\lambda_{new}$, and $\lambda_{res}$ based on the data receiving speed and the characteristics of the dataset to achieve the best performance. Without loss of generality,  following \cite{spmf,ssrm}, all the baselines and the proposed STS-AEL process data stream in batch to increase the throughput.
	\subsection{Performance Comparison and Analysis}
	%	Extensive experiments are conducted to answer \textbf{RQ1} - \textbf{RQ4}.\\
	\npdecimalsign{.}
	\nprounddigits{3}
	% \usepackage{multirow}
	\begin{table*}
		\centering
		\begin{subtable}[b]{\linewidth}
			%\resizebox{\textwidth}{15mm}{
				\centering
			\footnotesize{
						\begin{tabular}{c@{~}|c@{~}|c@{   }|p{9mm}<{\centering}@{     }p{9mm}<{\centering}@{      }p{9mm}<{\centering}@{   }|p{9mm}<{\centering}@{     }p{9mm}<{\centering}@{  }p{9mm}<{\centering}@{   }} 
							\hline
%				\begin{tabular}{c|c|c|ccc|ccc} 
%					\hline
					
					\multicolumn{3}{c|}{Dataset}                                                                                     & \multicolumn{6}{c}{MovieLens}     \\ \hline
					\multicolumn{3}{c|}{Metrics}                                                                                      & \multicolumn{3}{c|}{HR@10}                                                                                   & \multicolumn{3}{c}{NDCG@10}                                                          \\ \hline
					\multicolumn{3}{c|}{Data Receiving Speeds}                                                                        & 128                                & 256                                & 512                                & 128                                & 256                                & 512                                \\ \hline
					\multirow{8}{*}{Baselines}                                               & \multirow{7}{*}{Monolithic}  & eAls    & 0.231                              & 0.231                              & 0.234                              & 0.106                              & 0.106                              & 0.108                              \\
					&                              & RCD    
					& 0.287                              & 0.297                              & 0.278                              & 0.139                              & 0.145                              & 0.138                                 \\
					&                              & iBPR   
					& 0.303                              & 0.303                              & 0.279                              & 0.147                              & 0.147                              & 0.134        \\
					&                              & iTPMF-CF  	
					& 0.408                              & 0.401                              & 0.426                              & 0.222                              & 0.216                              & 0.229        \\
					&                              & SPMF   
					& 0.472                              & 0.466                              & 0.434                              & 0.262                              & 0.259                              & 0.234                                 \\
					&                              & iGMF    
					& 0.525                			 & 0.529                              & 0.477                              & 0.295                              & 0.297                              & 0.265                                 \\
					&                              & iMLP    
					& 0.538                              & 0.539                              & 0.488                              & 0.304                              & 0.303                              & 0.272                                          \\
					&                              & iNeuMF  
					& \underline{0.551} & \underline{0.546} & \underline{0.496} & \underline{0.311} & \underline{0.307} & \underline{0.275}  \\ \cline{2-3}
					& Ensemble                     & OCFIF   
					& 0.532                              & 0.508                              & 0.467                              & 0.291                              & 0.279                              & 0.256                                  \\ \hline
					\multirow{3}{*}{\begin{tabular}[c]{@{}c@{}}Our  STS-AEL  \\Framework\end{tabular}} 
%					& \multicolumn{2}{c|}{STS-AEL\_8-SPMF}   
%					& 0.546                              & 0.550                              & 0.552                              & 0.320                              & 0.328                              & 0.333                                            \\
					& \multicolumn{2}{c|}{STS-AEL\_8-iGMF}   
					& 0.592                              & 0.584                              & 0.590                              & 0.344                              & 0.340                              & 0.344                             \\
					& \multicolumn{2}{c|}{STS-AEL\_8-iMLP}  
					& 0.591                              & 0.586                              & 0.583                              & 0.341                              & 0.340                              & 0.338                                     \\
					& \multicolumn{2}{c|}{STS-AEL\_8-iNeuMF}
					& \textbf{0.608}    & \textbf{0.607}    & \textbf{0.598}    & \textbf{0.351}    & \textbf{0.353}    & \textbf{0.346}     \\ \hline
					\multicolumn{3}{c|}{\makecell{Improvement percentage \\ over the best-performing baseline}}                                                              
					& 10.3\%                             & 11.2\%                             & 20.6\%                             & 12.9\%                             & 15.0\%                             & 25.8\%                                \\ \hline 
				\end{tabular}
			}
			\caption{\footnotesize Results on MovieLens}
					\label{table:performance_comparison:a_ml}
			\hspace{\fill}
		\end{subtable}
		
		\begin{subtable}[b]{\linewidth}
				\centering
			%\resizebox{\textwidth}{15mm}{
			\footnotesize{
						\begin{tabular}{c@{~}|c@{~}|c@{   }|p{9mm}<{\centering}@{     }p{9mm}<{\centering}@{      }p{9mm}<{\centering}@{   }|p{9mm}<{\centering}@{     }p{9mm}<{\centering}@{  }p{9mm}<{\centering}@{   }} 
					\hline
					
					\multicolumn{3}{c|}{Dataset}                                                                                     & \multicolumn{6}{c}{Netflix}     \\ \hline
					\multicolumn{3}{c|}{Metrics}                                                                                      & \multicolumn{3}{c|}{HR@10}                                                                                   & \multicolumn{3}{c}{NDCG@10}                                                          \\ \hline
					\multicolumn{3}{c|}{Data Receiving Speeds}                                                                                                      & 128                                & 256                                & 512                                & 128                                & 256                                & 512                                \\ \hline
					\multirow{8}{*}{Baselines}                                               & \multirow{7}{*}{Monolithic}  & eAls    &  0.395                              & 0.389                              & 0.362                              & 0.211                              & 0.207                              & 0.192                               \\
					&                              & RCD     & 
					0.447                              & 0.436                              & 0.435                              & 0.226                              & 0.226                              & 0.219                              \\
					&                              & iBPR    & 
					0.685                              & 0.686                              & 0.627                              & 0.396                              & 0.395                              & 0.360     \\
					&                              & iTPMF-CF  &  	
					0.514	                       &  0.5231                     &0.540                              & 0.290                              &0.298                              & 0.309        \\
					&                              & SPMF    & 
					0.699                              & 0.676                              & 0.636                              & 0.425                              & 0.410                              & 0.374                                    \\
					&                              & iGMF    & 
					0.747                              & 0.748                              & 0.577                              & 0.482                              & 0.482                              & 0.352                                 \\
					&                              & iMLP    & 
					0.787                              & 0.782                              & 0.624                              & 0.519                              & 0.510                              & 0.369                                  \\
					&                              & iNeuMF  & 
					\underline{0.801} & \underline{0.798} & \underline{0.711} & \underline{0.531} & \underline{0.529} & \underline{0.443} \\ \cline{2-3}
					& Ensemble                     & OCFIF   & 
					0.745                              & 0.734                              & 0.606                              & 0.457                              & 0.453                              & 0.357                                  \\ \hline
					\multirow{4}{*}{\begin{tabular}[c]{@{}c@{}}Our  STS-AEL  \\Framework\end{tabular}} 
%					& \multicolumn{2}{c|}{STS-AEL\_8-SPMF}   &  0.775                              & 0.771                              & 0.766                              & 0.508                              & 0.505                              & 0.498                                 \\
					& \multicolumn{2}{c|}{STS-AEL\_8-iGMF}   & 
					0.789                              & 0.783                              & 0.785                              & 0.524                              & 0.517                              & 0.518                              \\
					& \multicolumn{2}{c|}{STS-AEL\_8-iMLP}   & 
					0.813                              & 0.805                              & 0.798                              & 0.546                              & 0.535                              & 0.525                              \\
					& \multicolumn{2}{c|}{STS-AEL\_8-iNeuMF} & 
					\textbf{0.840}    & \textbf{0.830}    & \textbf{0.821}    & \textbf{0.576}    & \textbf{0.562}    & \textbf{0.552}      \\ \hline
					\multicolumn{3}{c|}{\makecell{Improvement percentage \\ over the best-performing baseline}}                                                                      &  4.80\%                              & 4.00\%                              & 15.5\%                             & 8.40\%                              & 6.20\%                              & 24.6\%                                                    \\ \hline 
				\end{tabular}
			}
			%	\vspace{-2mm}
			
			%	\vspace{-2mm}
			\caption{\footnotesize Results on Netflix}
								\label{table:performance_comparison:b_nf}
			\hspace{\fill}
		\end{subtable}
		
		\begin{subtable}[b]{\linewidth}
			\centering
			%\resizebox{\textwidth}{15mm}{
			\footnotesize{
						\begin{tabular}{c@{~}|c@{~}|c@{   }|p{9mm}<{\centering}@{     }p{9mm}<{\centering}@{      }p{9mm}<{\centering}@{   }|p{9mm}<{\centering}@{     }p{9mm}<{\centering}@{  }p{9mm}<{\centering}@{   }} 
					\hline
					\multicolumn{3}{c|}{Dataset}                                                                                     & \multicolumn{6}{c}{Yelp}     \\ \hline
					\multicolumn{3}{c|}{Metrics}                                                                                      & \multicolumn{3}{c|}{HR@10}                                                                                   & \multicolumn{3}{c}{NDCG@10}                                                          \\ \hline
					\multicolumn{3}{c|}{Data Receiving Speeds}                                                                        & 128                                & 256                                & 512                                & 128                                & 256                                & 512                                \\ \hline
					\multirow{8}{*}{Baselines}                                               & \multirow{7}{*}{Monolithic}  & eAls    &  0.287                              & 0.289                              & 0.290                              & 0.167                              & 0.167                              & 0.169                              \\
					&                              & RCD     &
					0.454                              & 0.452                              & 0.447                              & 0.260                              & 0.257                              & 0.259                              \\
					&                              & iBPR    & 
					0.307                              & 0.295                              & 0.188                              & 0.180                              & 0.172                              & 0.108                              \\
					&                              & iTPMF-CF  &  
					
					0.172	                    	   &  0.174                     & 0.186                              & 	0.089                              &	0.090                              &0.098        \\
					&                              & SPMF    &
					0.203                              & 0.203                              & 0.177                              & 0.107                              & 0.106                              & 0.091                              \\
					&                              & iGMF    & 
					0.499                              & 0.470                              & 0.396                              & 0.294                              & 0.276                              & 0.228                              \\
					&                              & iMLP    &
					
					\underline{0.573} & \underline{0.574} & \underline{0.438} & \underline{0.338} & \underline{0.338} & 0.246                              \\
					&                              & iNeuMF  & 
					
					0.566                              & 0.570                              & 0.435                              & 0.331                              & 0.334                              & \underline{0.247} \\ \cline{2-3}
					& Ensemble                     & OCFIF   & 
					0.260                              & 0.249                              & 0.203                              & 0.135                              & 0.129                              & 0.107                              \\ \hline
					\multirow{4}{*}{\begin{tabular}[c]{@{}c@{}}Our STS-AEL \\Framework\end{tabular}} 
%					& \multicolumn{2}{c|}{STS-AEL\_8-SPMF}   &  0.287                              & 0.318                              & 0.352                              & 0.160                              & 0.178                              & 0.199                              \\
					& \multicolumn{2}{c|}{STS-AEL\_8-iGMF}   & 
					0.647                              & 0.614                              & 0.600                              & 0.399                              & 0.371                              & 0.362                              \\
					& \multicolumn{2}{c|}{STS-AEL\_8-iMLP}   & 
					0.676                              & 0.671                              & 0.639                              & 0.414                              & 0.402                              & 0.378                              \\
					& \multicolumn{2}{c|}{STS-AEL\_8-iNeuMF} & 
					\textbf{0.717}    & \textbf{0.677}    & \textbf{0.665}    & \textbf{0.456}    & \textbf{0.415}    & \textbf{0.405}    \\ \hline
					\multicolumn{3}{c|}{\makecell{Improvement percentage \\ over the best-performing baseline}}                                                                      &  25.1\%                             & 17.9\%                             & 51.8\%                             & 34.9\%                             & 22.8\%                             & 64.0\%                             \\ \hline 
				\end{tabular}
			}
			%	\vspace{-2mm}

			%	\vspace{-2mm}
			\caption{\footnotesize Results on Yelp}
								\label{table:performance_comparison:c_yelp}
		\end{subtable}

		\caption{Performance comparison with baselines. Our proposed STS-AEL\_8-iNeuMF achieves the highest recommendation accuracies (marked with \textbf{bold} font) in all the cases, and its improvement percentages in terms of HR@10 and NDCG@10 over the best performing baselines (marked with \underline{underline}) are presented in the last row. Note that the data processing speed ($sp_{p}$) is 256, thus $sp_r = 128$ and $sp_r = 512$ indicate the underload scenario and overload scenario, respectively.}
		\label{table:performance_comparison}
	\end{table*}

	\textbf{Experiment 1: Performance Comparison with Baselines (for RQ1)}
	
	\noindent \textbf{Setting:} %To answer \textbf{RQ1}, we compare our proposed STS-AEL with all nine baselines. 
	In this experiment, we take each of
	%dynamically adjust discounting factor $\lambda$ for the best performance. To illustrate the performance improvement introduced by STS-AEL over individual models, we take all 
	iGMF, iMLP, and iNeuMF as the individual model of STS-AEL and set the number of individual models to eight for the evaluation. In addition, the proposed STS-AEL and nine baselines are evaluated on all three datasets w.r.t. a fixed data processing speed, i.e., $sp_{p} = 256$ and three different data receiving speed, i.e., $sp_{r} = 128$ (simulating the underload scenario), $sp_{r} = 256$ (simulating the ideal case where the data processing speed is equal to the data receiving speed), and $sp_{r} = 512$ (simulating the overload scenario) to make comparisons in different cases.
	
	\noindent\textbf{Result:} As \Cref{table:performance_comparison} shows, in all the cases, STS-AEL\_8-iNeuMF delivers the highest recommendation accuracies (marked with \textbf{bold} font), and the improvement percentages of STS-AEL\_8-iNeuMF over the best-performing baseline (marked with \underline{underline}) in each case are introduced in the last row, ranging from 4.0\% (compared with iNeuMF on Netflix w.r.t. $sp_r = 256$) to 51.8\% (compared with iMLP on Yelp w.r.t. $sp_r = 512$) with an average of 17.1\% in terms of HR@10, and ranging from 6.2\% (compared with iNeuMF on Netflix w.r.t. $sp_r = 256$) to 64.0\% (compared with iMLP on Yelp w.r.t. $sp_r = 512$) with an average of 22.9\% in terms of NDCG@10. 

	%\textbf{Result 2 (for RQ2):} Moreover,~\Cref{table:performance_comparison} indicates that, for all the cases, STS-AEL outperforms its corresponding individual models, particularly,  by an average of 17.5\% ranging from 4.0\% to 48.7\% in terms of HR@10 and by an average of 23.4\% ranging from 6.2\% to 57.1\% in terms of NDCG@10 w.r.t.\ iNeuMF, which is the best-performing baseline.\\
	% and a state-of-the-art neural network based recommendation model.\\
	%\textbf{Result 3:} Furthermore, for all the cases, STS-AEL\_8-ISGD outperforms OCFIF, the state-of-the-art ensemble SRS with ten matrix factorization models, by an average of 21.4\% in terms of HR@10 and 27.7\% in terms of NDCG@10. It indicates that STS-AEL is superior to OCFIF when ensembling similar individual models. \\
	\noindent\textbf{Analysis:} The superiority of our proposed STS-AEL can be explained in three aspects: 1) the proposed STS addresses concept drift while capturing long-term user preferences by wisely incorporating both new data and historical data through a stratified and time-aware strategy, 2) ensembling multiple individual models can not only avoid the limitations of a monolithic model by complementing one another but also contributes to mining user preferences more effectively by concurrent training, especially in the overload scenario, and 3) the proposed AEL improves fusion performance by assigning elaborately calculated fusion weights to multiple individual models for effective fusion.
	
	As ~\Cref{table:performance_comparison} shows, the only existing ensemble SRS, i.e., OCFIF, delivers lower recommendation accuracy even than some monolithic baseline models, e.g., iGMF, iMLP, and iNeuMF. This can be explained by the following three reasons: 1) OCFIF trains multiple individual models with the same data. It not only reduces the data processing speed, which affects the sufficient training of models, but also harms the diversities of individual models, which is essential for effective ensemble learning, 2) OCFIF selects only one individual model for the final prediction, which does not fully utilize all the individual models to obtain more accurate recommendations, and 3) OCFIF is specifically devised to ensemble matrix factorization models, which only capture the linear relations with the linear operation, i.e., dot product between the user embedding and the item embedding, while the aforementioned three models, i.e., iGMF, iMLP, and iNeuMF, all capture the more complex nonlinear relations with nonlinear operations, e.g., sigmoid function, and thus they can learn user preference towards items better.

	In addition, we can observe that the recommendation accuracies of our approach are visibly different on different datasets. The reason is that the proportions of interactions from the users used for testing against the total interactions from all users are different in the training set on the three datasets. Specifically, the calculated proportions for Netflix, Yelp, and MovieLens are 71.4\%, 49.9\%, and 28.8\%, respectively. Namely, in Netflix, the users used for testing have the most interactions for training and thus their preferences can be best learned. Therefore, our proposed approach achieves the highest recommendation accuracies on Netflix while achieves the lowest recommendation accuracies on MovieLens. Note that the recommendation accuracies of some baselines (e.g., iTPMF-CF) on Yelp are not always higher than those on MovieLens, as they cannot sufficiently learn the user preferences from the more interactions on Yelp due to their limited modelling capability.
	Moreover, the difference in the aforementioned proportions on three datasets also helps explain why the improvement percentage of our proposed STS-AEL over the best-performing baselines is the lowest on Netflix while the highest on Yelp. That is, on Netflix, the aforementioned high proportion (71.4\%) of interactions from users used for testing help all recommendation models deliver high recommendation accuracies, and thus further improvement is hard to achieve. As for Yelp where the aforementioned proportion is 49.9\%, compared with MovieLens where the proportion is 28.8\%, our proposed STS-AEL can better exploit these more interactions for higher improvements with its stronger modelling capability than that of baselines.
	
%	Therefore, we obtain the conclusion that the possible reason for the inconsistent performance on different datasets is that the users in the test sets have different proportions of interactions in the training sets on different datasets. 
	%Furthermore, we can observe that STS-AEL delivers more improvement with the weaker individual model (e.g., ISGD) and in the underload scenario (e.g., $sp_r = 128$), because weker individual model has larger improvement space and STS-AEL can employ AEL to further improve performance
	
	\noindent\textbf{Summary:} Our proposed STS-AEL significantly outperforms all the baseline models in all the cases, including the underload scenario ($sp_r = 128$) (\textbf{CH2}) and overload scenario ($sp_r = 512$) (\textbf{CH3}).
	
	In the following three experiments, we illustrate the performance of STS by setting the types of individual models to the top three best-performing monolithic models in~\Cref{table:performance_comparison}, i.e., iNeuMF, iMLP, and iGMF, respectively.\\

	\vspace{1pt}
	\noindent\textbf{Experiment 2: Impact of the Number of Individual Models (for RQ2)}
\begin{figure}[]
	\centering
	\begin{subfigure}[b]{.8\textwidth}
		\centering
		\includegraphics[width=\textwidth]{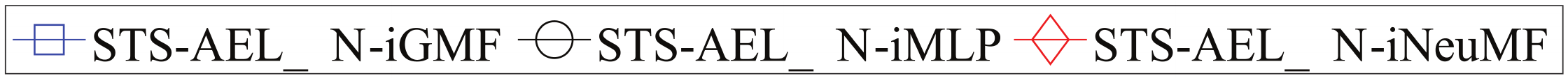}
		%\vspace{-4mm}
	\end{subfigure}%
	
	\begin{subfigure}[b]{0.327\textwidth}
		\centering
		\includegraphics[width=\textwidth]{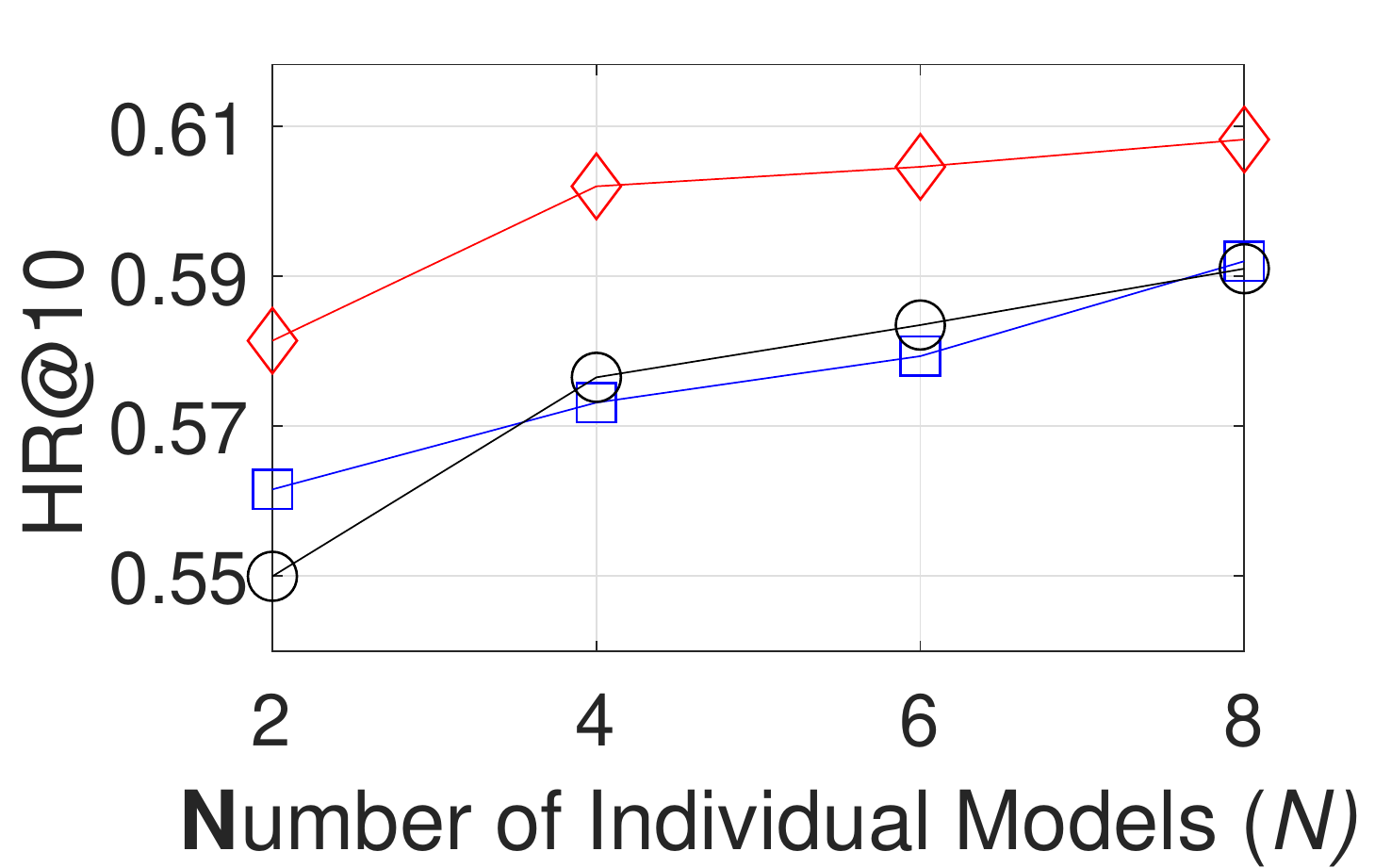}
		\caption{HR@10 on MovieLens}{w.r.t. $sp_{r} = 128$}
		\label{fig:experiment_number_hr_ml-1m}
	\end{subfigure}%
	\begin{subfigure}[b]{.327\textwidth}
		\centering
		\includegraphics[width=\textwidth]{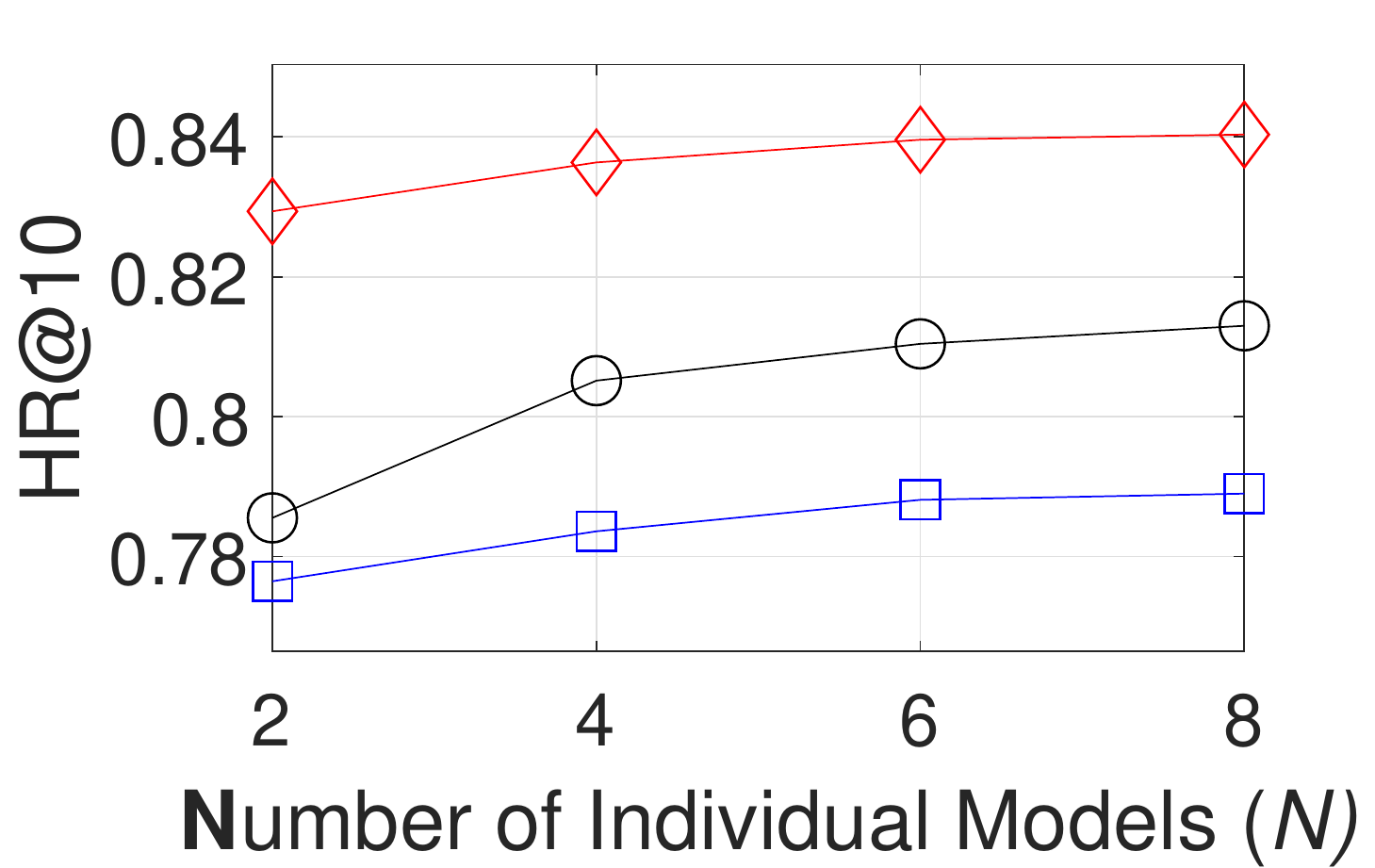}
		\caption{HR@10 on Netflix}{w.r.t. $sp_{r} = 128$}
		\label{fig:experiment_number_hr_netflix}
	\end{subfigure}%
	\begin{subfigure}[b]{.327\textwidth}
		\centering
		\includegraphics[width=\textwidth]{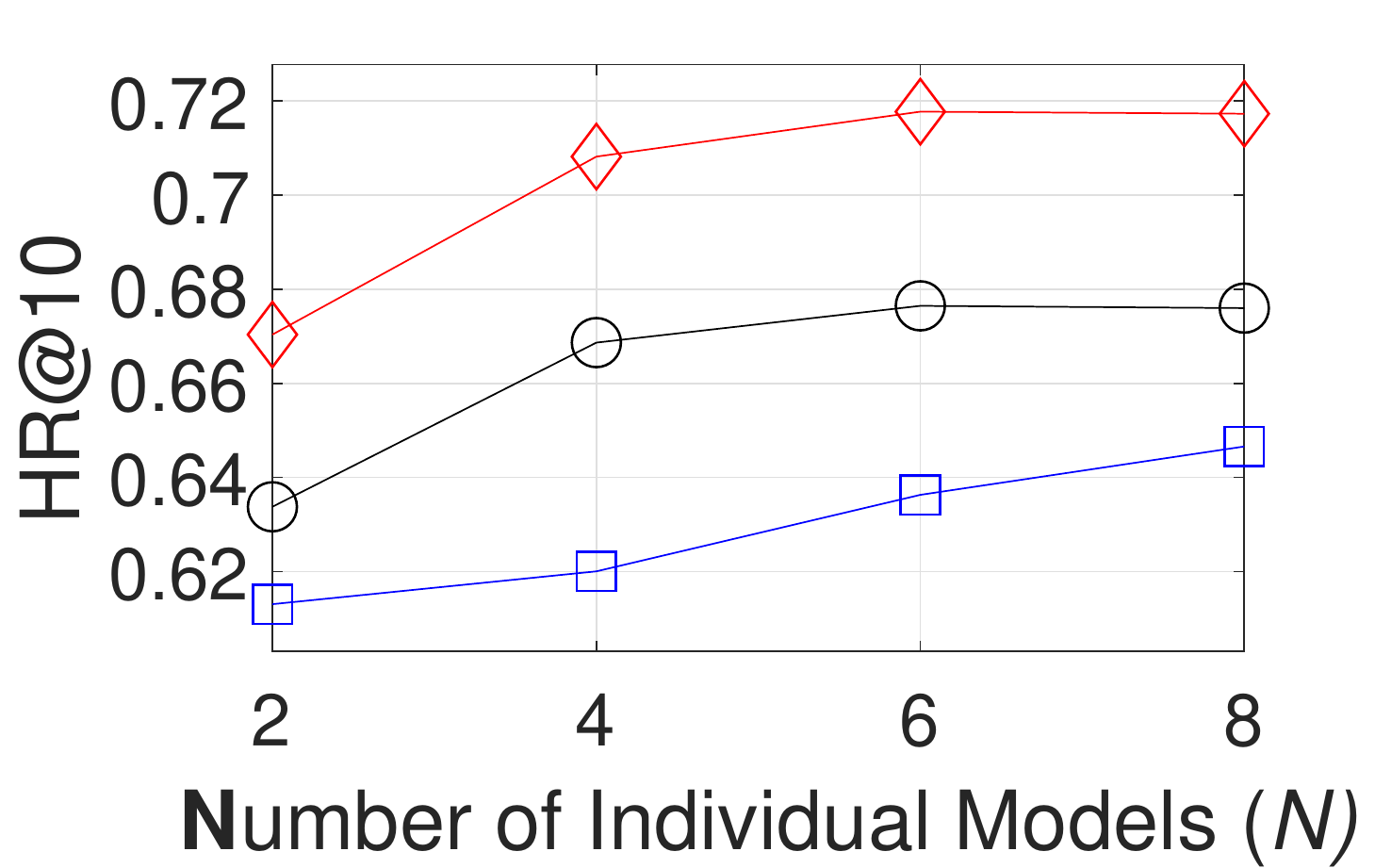}
		\caption{HR@10 on Yelp}{w.r.t. $sp_{r} = 128$}
		\label{fig:experiment_number_hr_yelp}
	\end{subfigure}%
	
	\begin{subfigure}[b]{.327\textwidth}
		\centering
		\includegraphics[width=\textwidth]{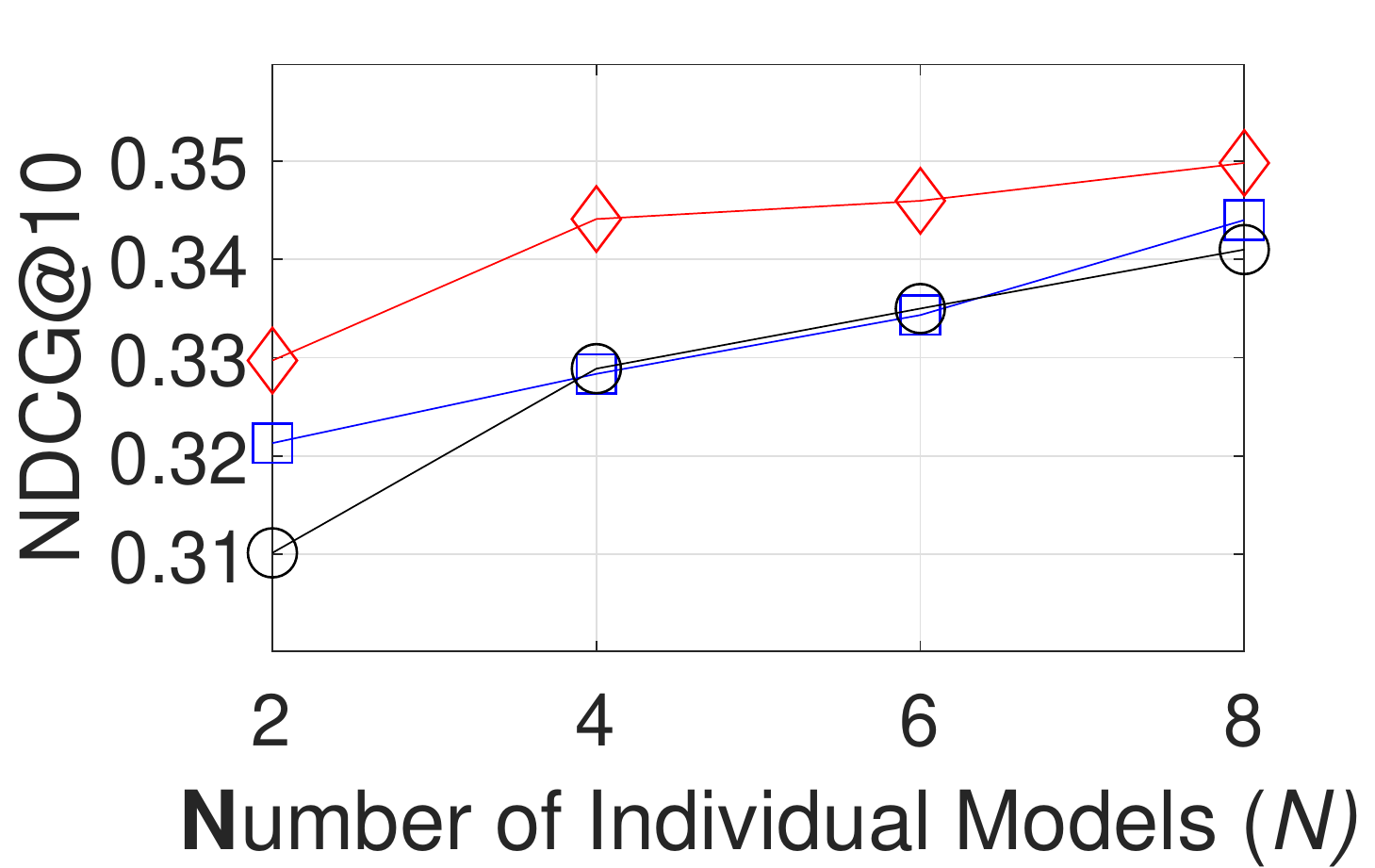}
		\caption{NDCG@10 on MovieLens}{w.r.t. $sp_{r} = 128$}
		\label{fig:experiment_number_ndcg_ml-1m}
	\end{subfigure}
	\begin{subfigure}[b]{.327\textwidth}
		\centering
		\includegraphics[width=\textwidth]{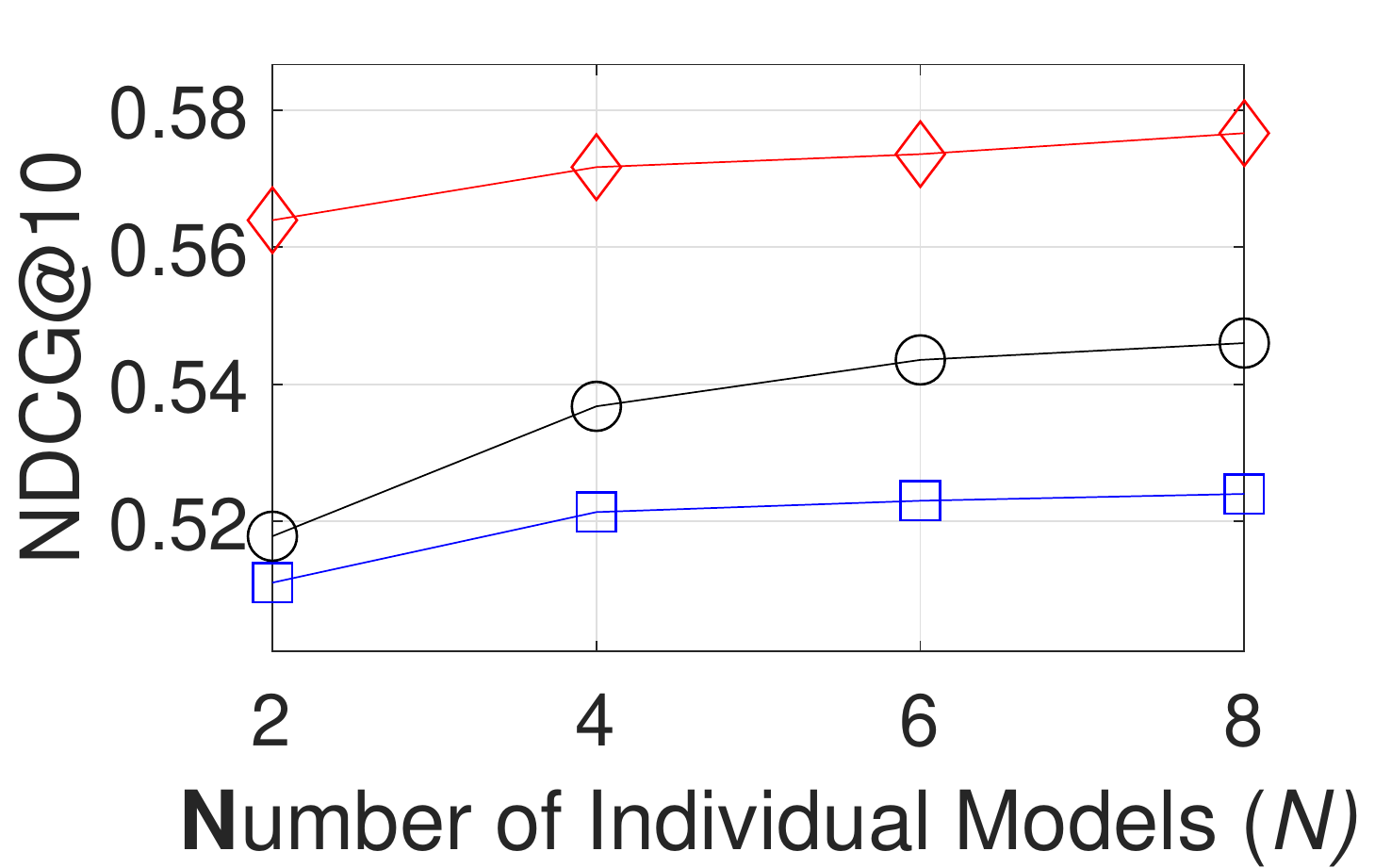}
		\caption{NDCG@10 on Netflix}{w.r.t. $sp_{r} = 128$}
		\label{fig:experiment_number_ndcg_netflix}
	\end{subfigure}
	\begin{subfigure}[b]{.327\textwidth}
		\centering
		\includegraphics[width=\textwidth]{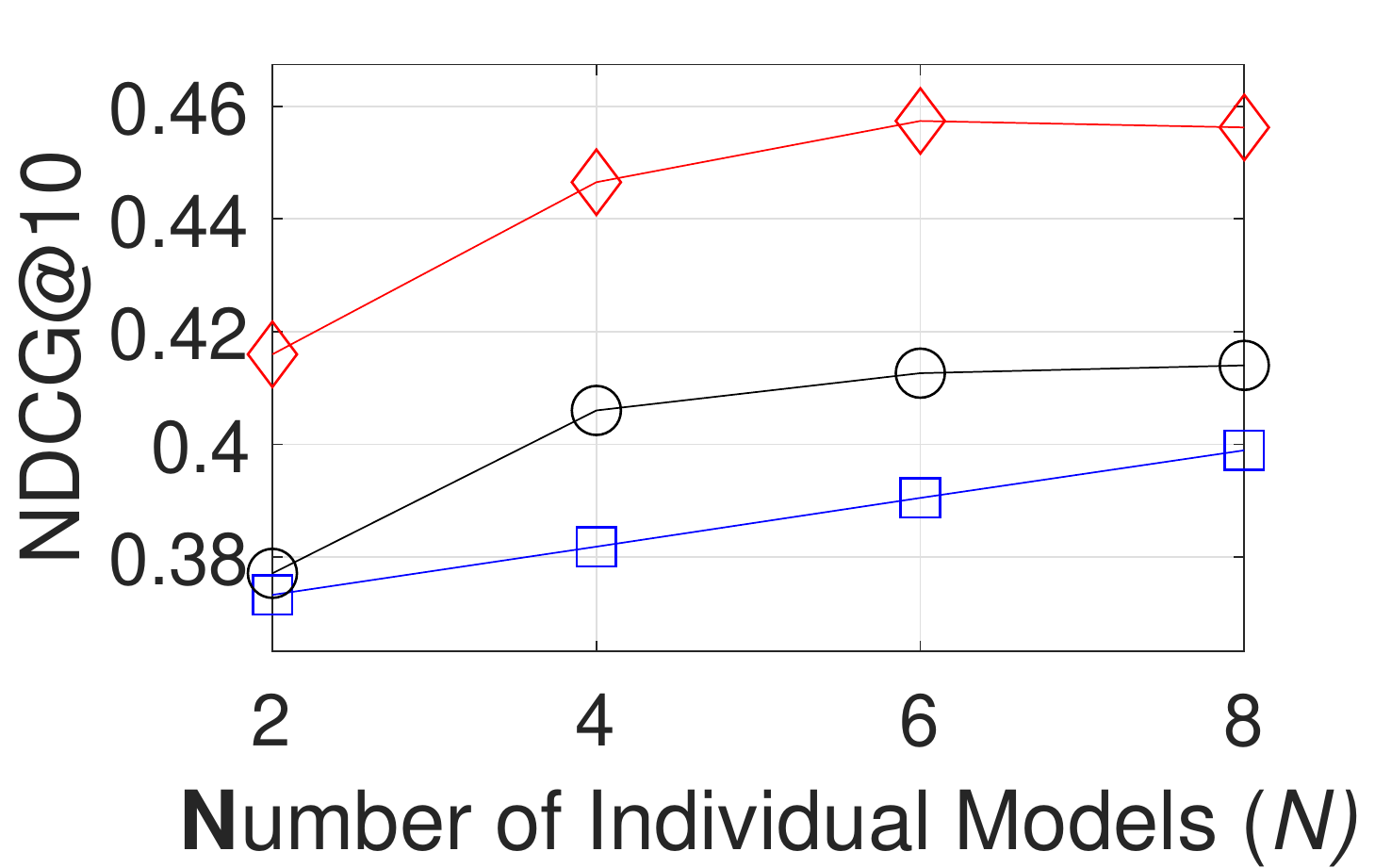}
		\caption{NDCG@10 on Yelp}{w.r.t. $sp_{r} = 128$}
		\label{fig:experiment_number_ndcg_yelp}
	\end{subfigure}
	
	\begin{subfigure}[b]{0.327\textwidth}
		\centering
		\includegraphics[width=\textwidth]{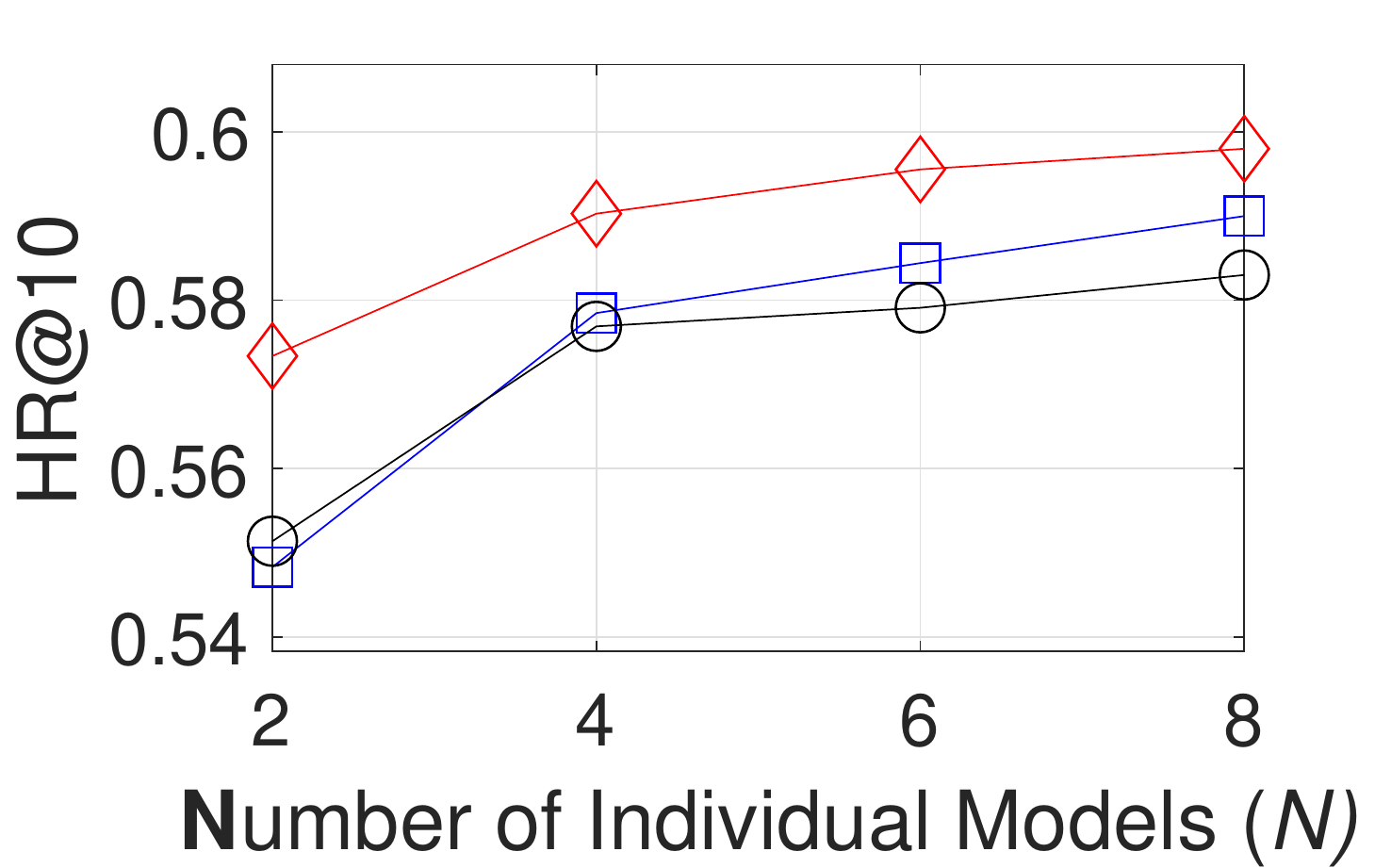}
		\caption{HR@10 on MovieLens}{w.r.t. $sp_{r} = 512$}
		\label{fig:experiment_number_hr_ml-1m_overload}
	\end{subfigure}%
	\begin{subfigure}[b]{.327\textwidth}
		\centering
		\includegraphics[width=\textwidth]{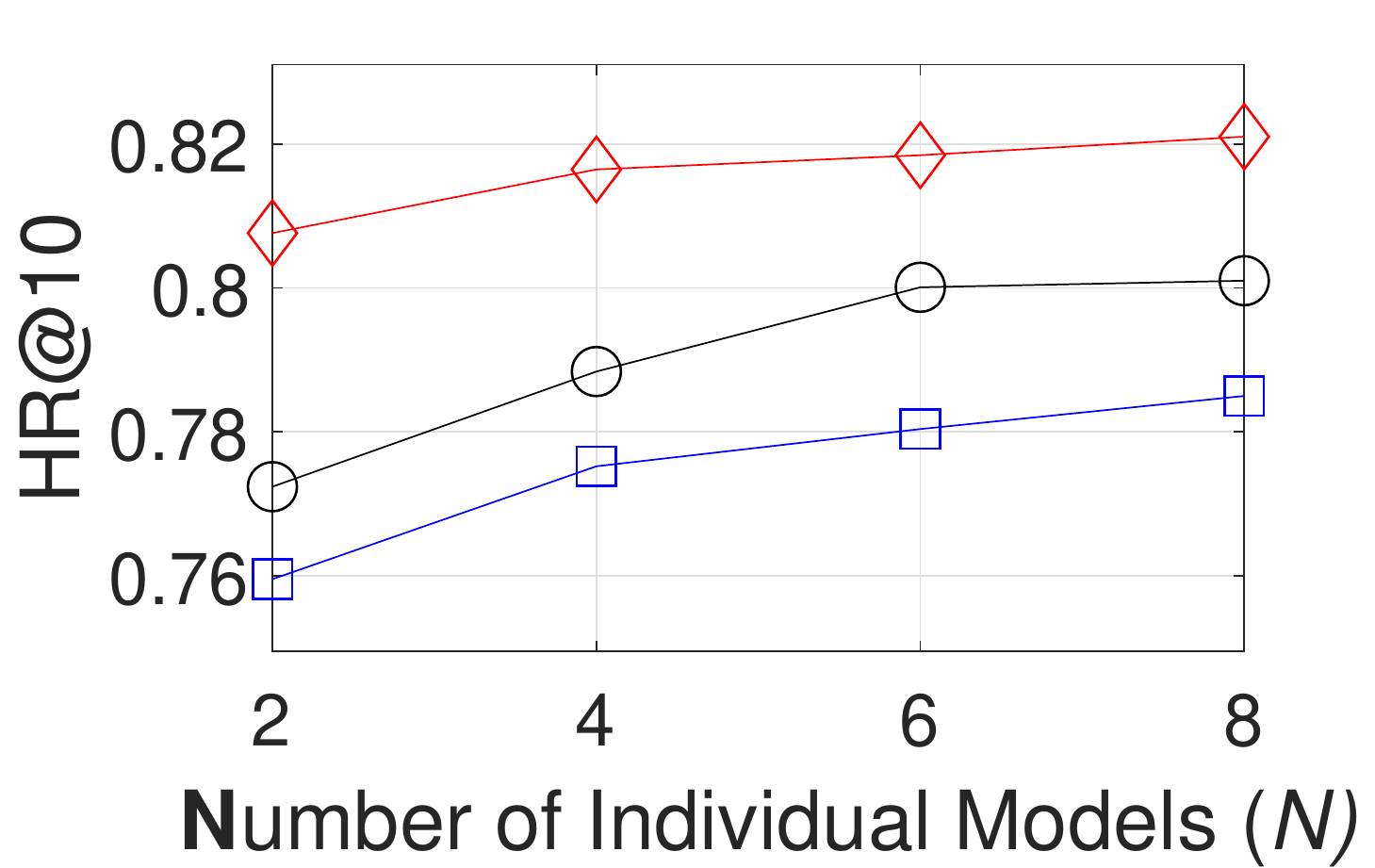}
		\caption{HR@10 on Netflix}{w.r.t. $sp_{r} = 512$}
		\label{fig:experiment_number_hr_netflix_overload}
	\end{subfigure}%
	\begin{subfigure}[b]{.327\textwidth}
		\centering
		\includegraphics[width=\textwidth]{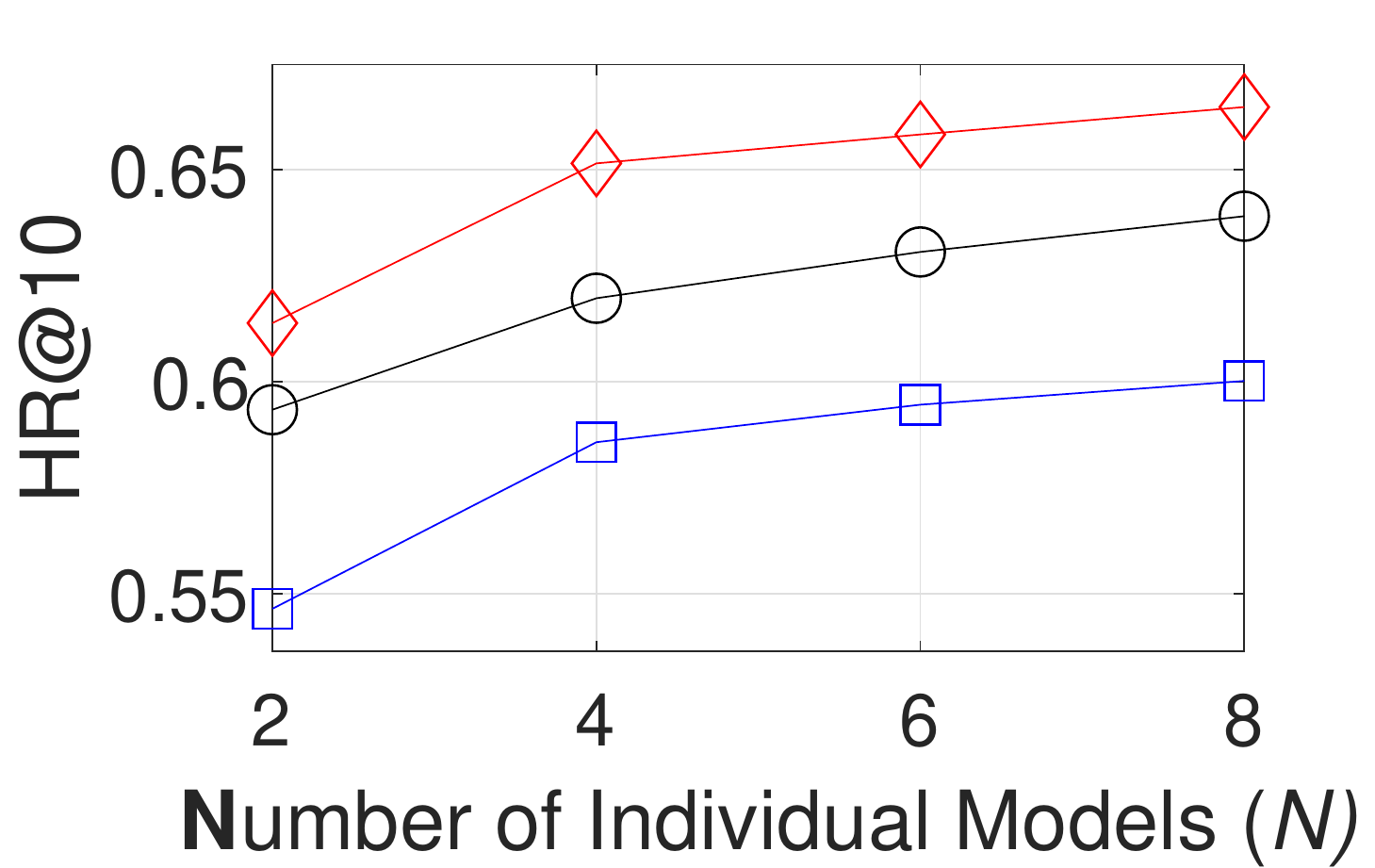}
		\caption{HR@10 on Yelp}{w.r.t. $sp_{r} = 512$}
		\label{fig:experiment_number_hr_yelp_overload}
	\end{subfigure}%
	
	\begin{subfigure}[b]{.327\textwidth}
		\centering
		\includegraphics[width=\textwidth]{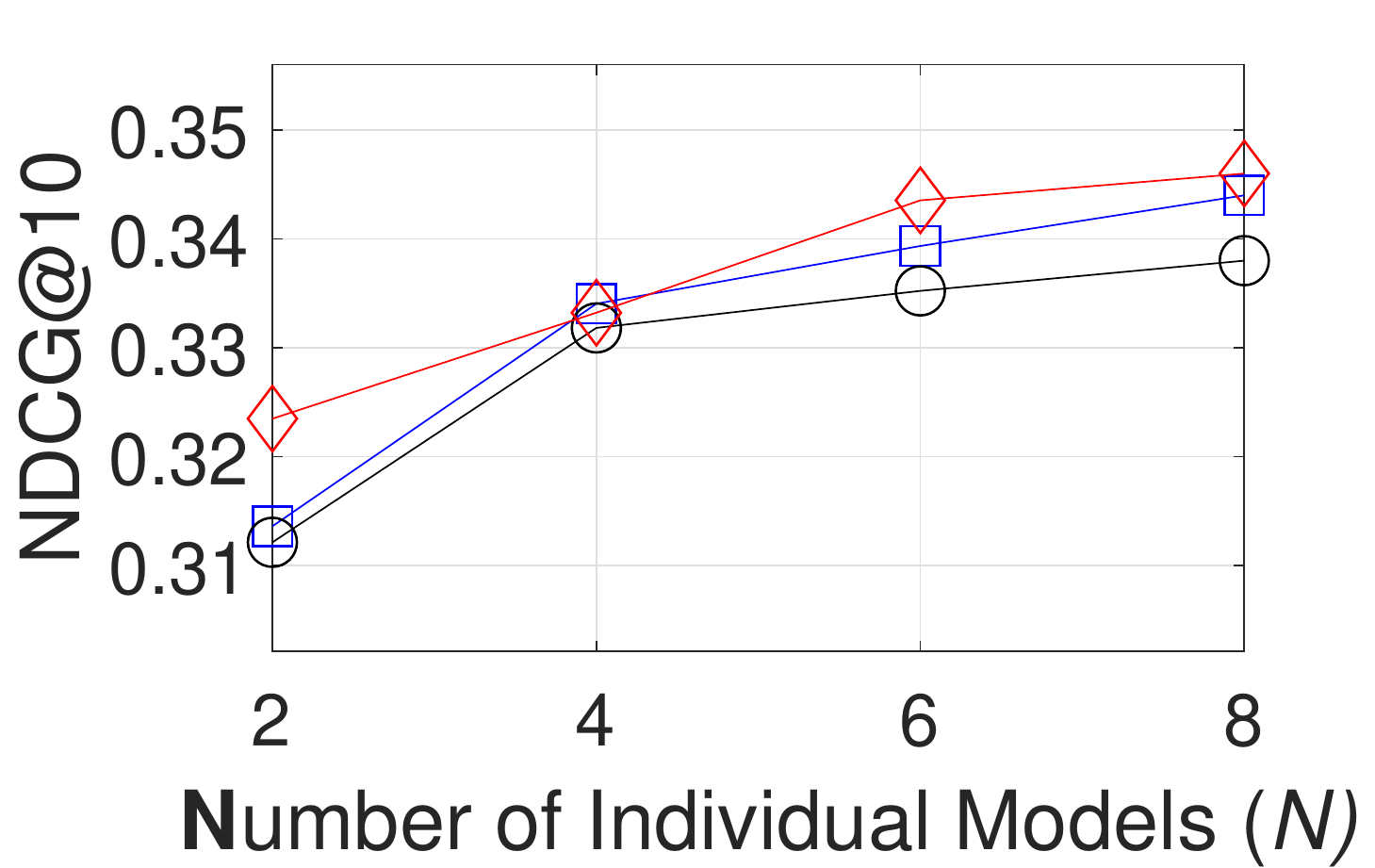}
		\caption{NDCG@10 on MovieLens}{w.r.t. $sp_{r} = 512$}
		\label{fig:experiment_number_ndcg_ml-1m_overload}
	\end{subfigure}
	\begin{subfigure}[b]{.327\textwidth}
		\centering
		\includegraphics[width=\textwidth]{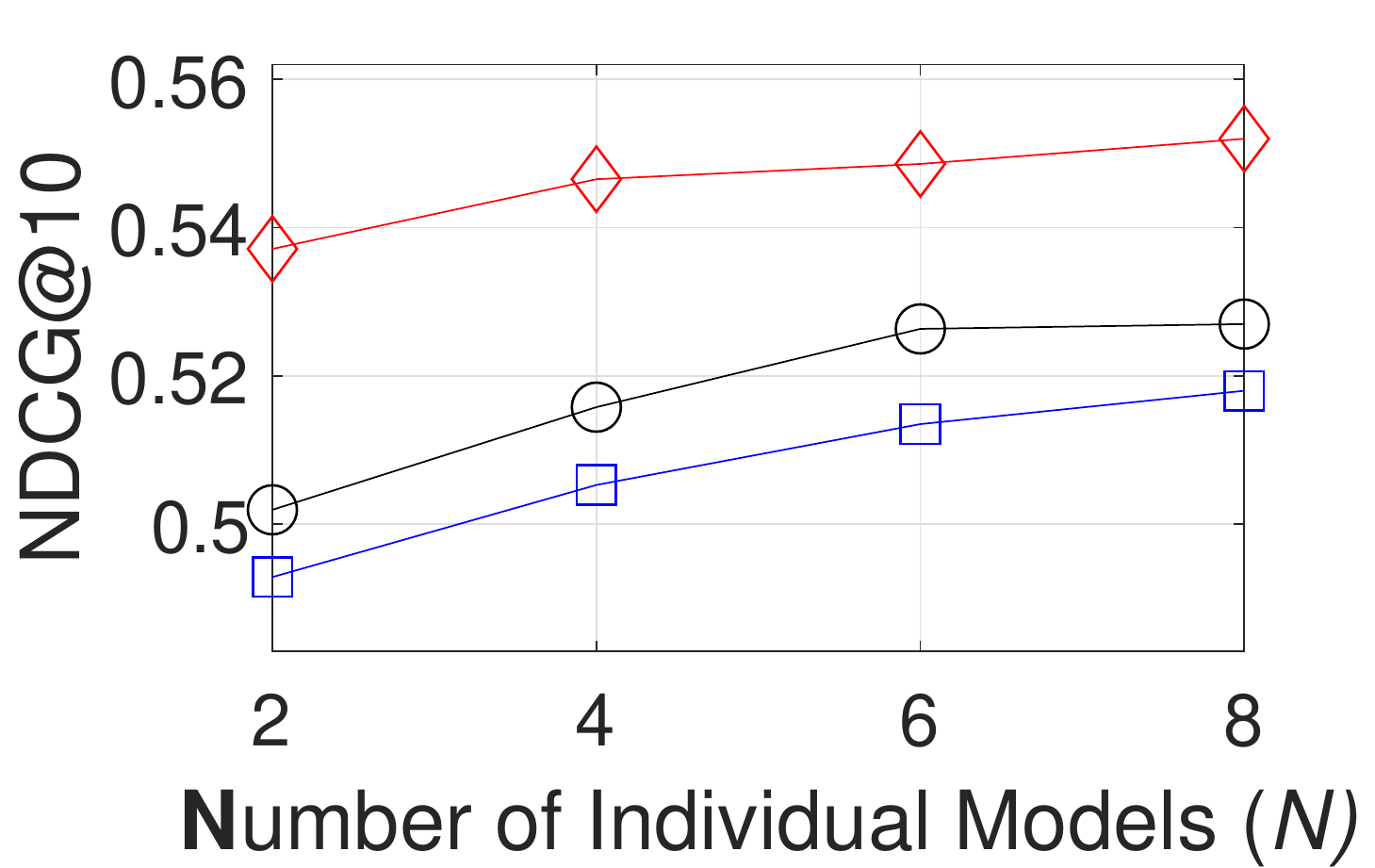}
		\caption{NDCG@10 on Netflix}{w.r.t. $sp_{r} = 512$}
		\label{fig:experiment_number_ndcg_netflix_overload}
	\end{subfigure}
	\begin{subfigure}[b]{.327\textwidth}
		\centering
		\includegraphics[width=\textwidth]{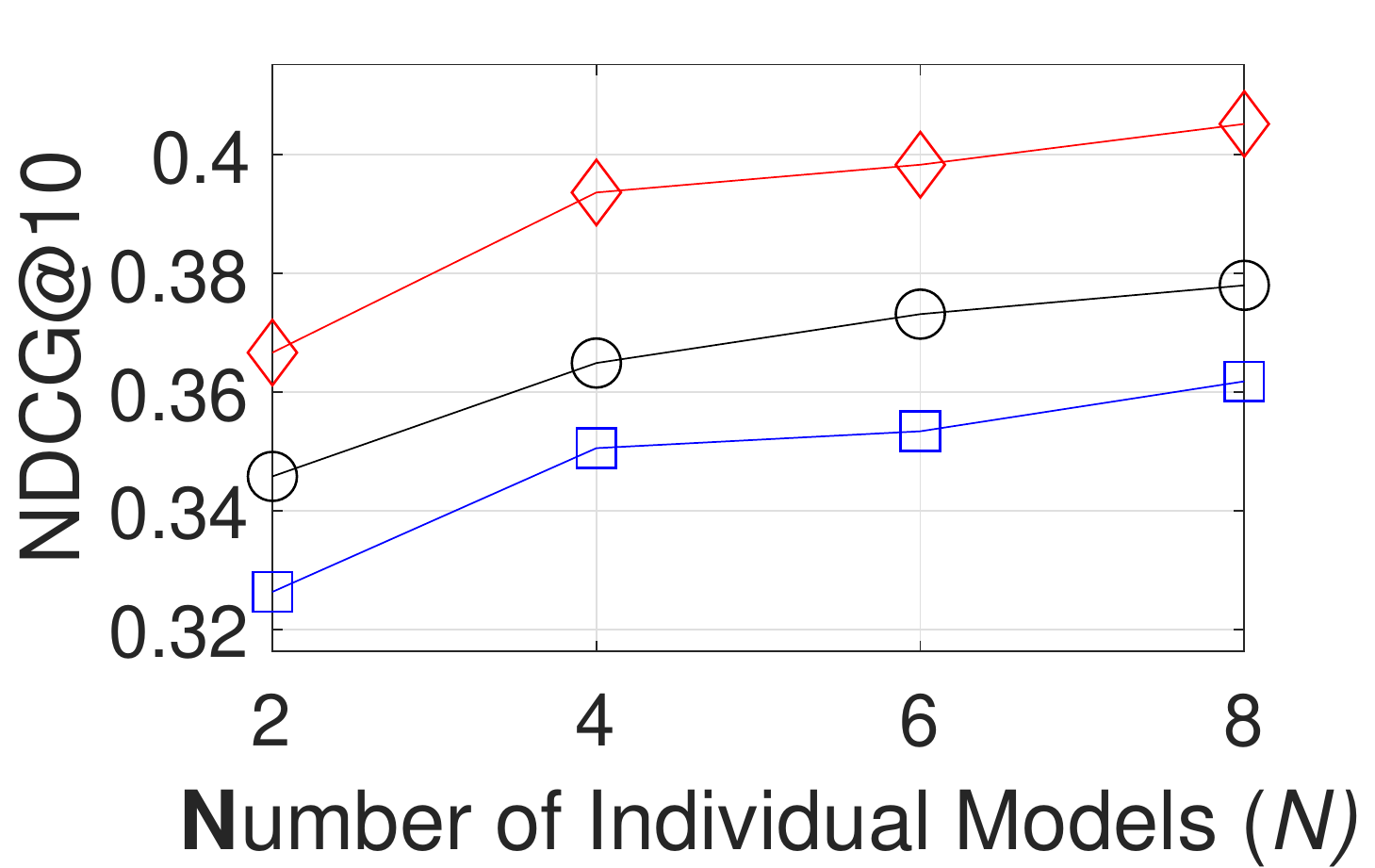}
		\caption{NDCG@10 on Yelp}{w.r.t. $sp_{r} = 512$}
		\label{fig:experiment_number_ndcg_yelp_overload}
	\end{subfigure}
	
	%	\vspace{-5mm}
	\caption{Impact of the number of individual models. To observe the impact of the number of individual models, we allow STS-AEL ensembling different numbers (i.e., 2, 4, 6, and 8) of individual models. For example, STS-AEL\_8 indicates that STS-AEL ensembles 8 individual models. As shown in Figs. 3(a) to 3(l), STS-AEL delivers higher recommendation accuracy with more individual models on all three datasets in both underload scenario (i.e., $sp_{r} = 128$ ) and overload scenario (i.e., $sp_{r} = 512$).}
	%	\vspace{-3mm}
	\label{fig:experiment_number}
\end{figure}

\vspace{1pt}
\noindent \textbf{Setting:} To answer \textbf{RQ2}, we allow our proposed STS-AEL ensembling different numbers (i.e., 2, 4, 6, and 8) of individual models for comparison. In this experiment, we report the results on all three datasets in both underload scenario (i.e., the cases where $sp_{r} = 128$) and overload scenario (i.e., the cases where $sp_{r} = 512$). 

\noindent\textbf{Result:} As shown in~\Cref{fig:experiment_number}, on all three datasets in both underload scenario and overload scenario, our proposed STS-AEL delivers higher recommendation accuracy when ensembling more individual models w.r.t. all three types of individual models, i.e., iNeuMF, iMLP, and iGMF. 

\noindent\textbf{Analysis:} The improvements when ensembling more individual models can be explained in two aspects: 1) more individual models can better complement one another through the fusion process when making recommendations, and 2) with the concurrent training, the streaming data can be better mined with more individual models, both when complemented by the historical data in the underload scenario and when confronting the excessive amount of data in the overload scenario. In addition, It can be observed that the improvements generally become smaller as the number of individual models increases. The possible reason is that after having enough individual models to effectively complement one another and to sufficiently mine streaming data in parallel, getting more individual models will no longer increase the recommendation accuracy much.

Since STS-AEL performs the best when ensembling 8 individual models, as shown in~\Cref{fig:experiment_number}, we set the number of individual models to 8 in the following experiments to answer \textbf{RQ3} and \textbf{RQ4}.\\
	%\indent In the following experiments, due to the space limitation, we report the results on MovieLens dataset in the underload scenario only to answer \textbf{RQ2} - \textbf{RQ4}, while the results in other cases are similar to the reported ones.
	% and outperforms its individual models in all the cases. 
	
	%	\begin{figure*}[t]
	%		\centering
	%		\subcaptionbox{Fig1}[.3\textwidth][c]{%
	%			\includegraphics[width=.3\textwidth]{sample_hr-eps-converted-to.pdf}}\quad
	%		\subcaptionbox{Fig1}[.3\textwidth][c]{%
	%			\includegraphics[width=.3\textwidth]{sample_hr-eps-converted-to.pdf}}\quad
	%		\subcaptionbox{Fig1}[.3\textwidth][c]{%
	%			\includegraphics[width=.3\textwidth]{sample_hr-eps-converted-to.pdf}}\quad
	%
	%
	%		\bigskip
	%		
	%		\subcaptionbox{Fig4}[.3\linewidth][c]{%
	%			\includegraphics[width=.2\linewidth]{sample_hr-eps-converted-to.pdf}}\quad
	%		\subcaptionbox{Fig5}[.3\linewidth][c]{%
	%			\includegraphics[width=.2\linewidth]{sample_hr-eps-converted-to.pdf}}\quad
	%		\subcaptionbox{Fig6}[.3\linewidth][c]{%
	%			\includegraphics[width=.2\linewidth]{sample_hr-eps-converted-to.pdf}}
	%		\caption{This is a figure}
	%	\end{figure*}
	
	\vspace{1pt}
	\noindent\textbf{Experiment 3: Superiority of STS (for RQ3)}
	\begin{figure}[]
		\centering
		\begin{subfigure}[b]{.9\textwidth}
			\centering
			\includegraphics[width=\textwidth]{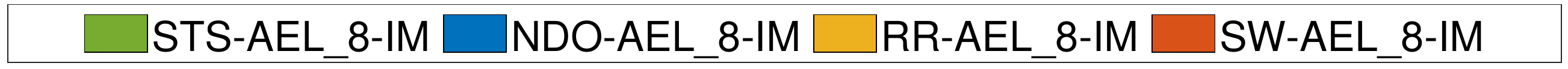}
			%\vspace{-4mm}
		\end{subfigure}%
		
		\begin{subfigure}[b]{0.333\textwidth}
			\centering
			\includegraphics[width=\textwidth]{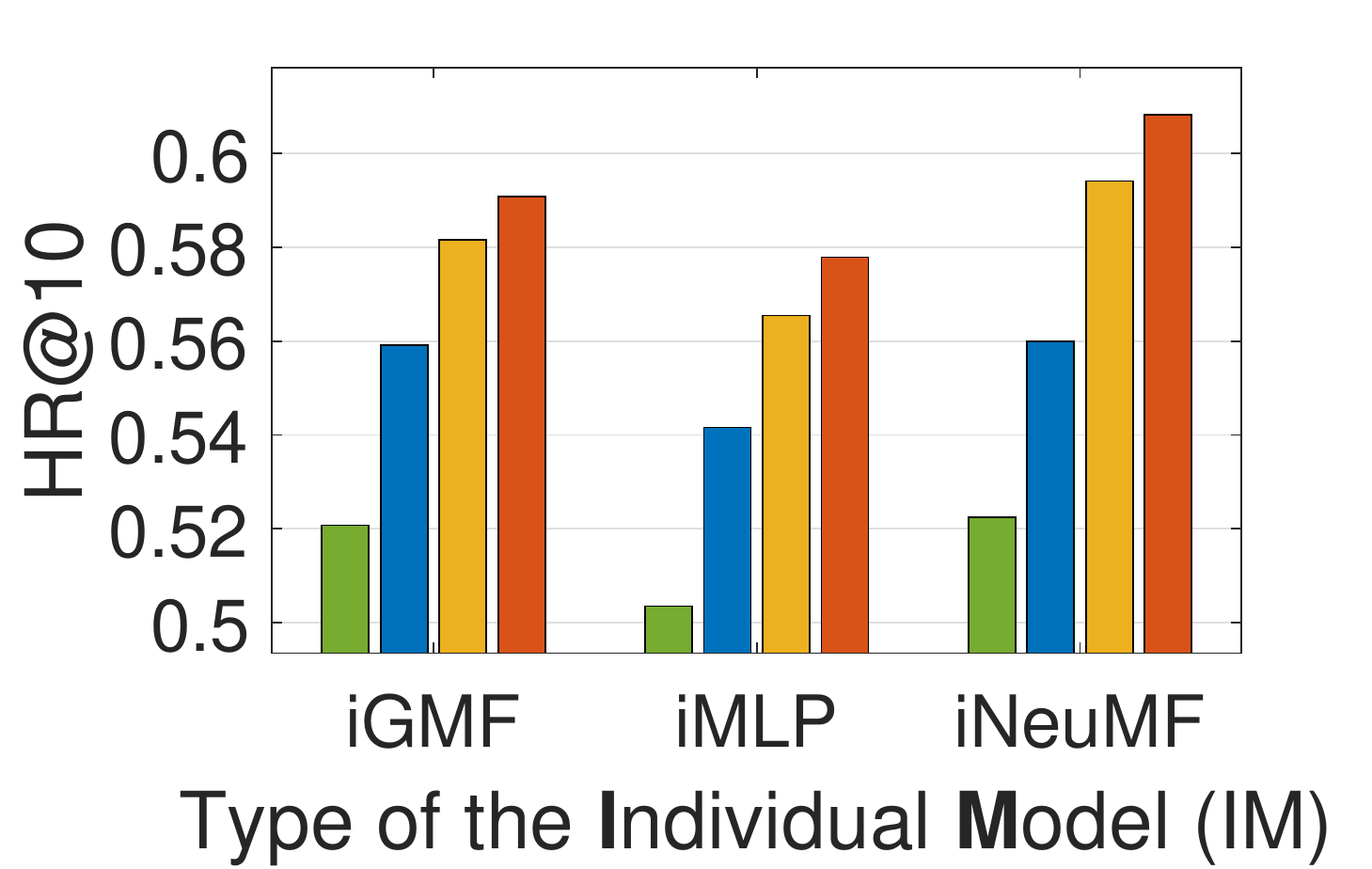}
			\caption{HR@10 on MovieLens}
			\label{fig:experiment_sample_hr_ml-1m}
		\end{subfigure}%
		\begin{subfigure}[b]{.333\textwidth}
			\centering
			\includegraphics[width=\textwidth]{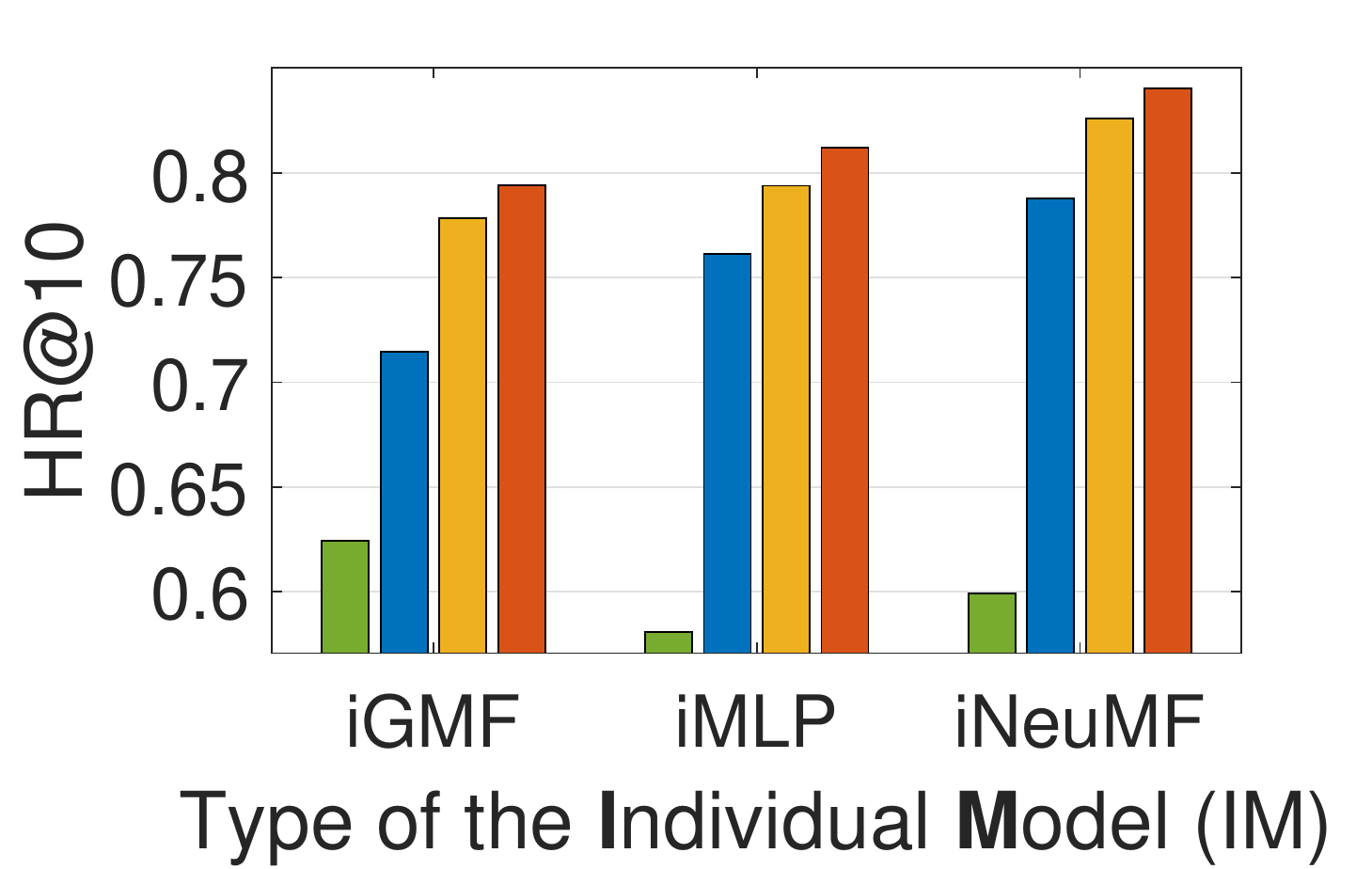}
			\caption{HR@10 on Netflix}
			\label{fig:experiment_sample_hr_netflix}
		\end{subfigure}%
		\begin{subfigure}[b]{.333\textwidth}
			\centering
			\includegraphics[width=\textwidth]{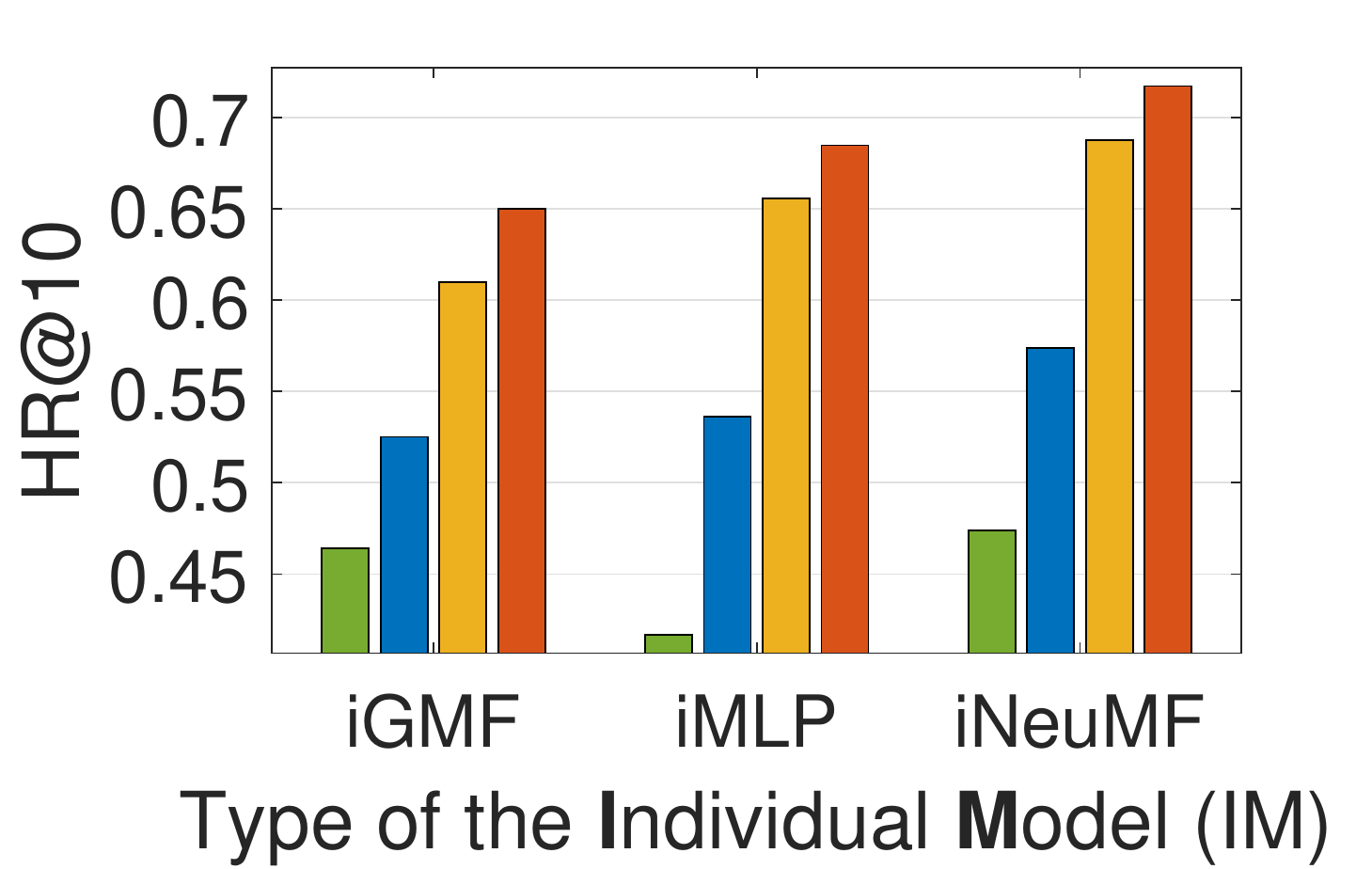}
			\caption{HR@10 on Yelp}
			\label{fig:experiment_sample_hr_yelp}
		\end{subfigure}%
		
		\begin{subfigure}[b]{.327\textwidth}
			\centering
			\includegraphics[width=\textwidth]{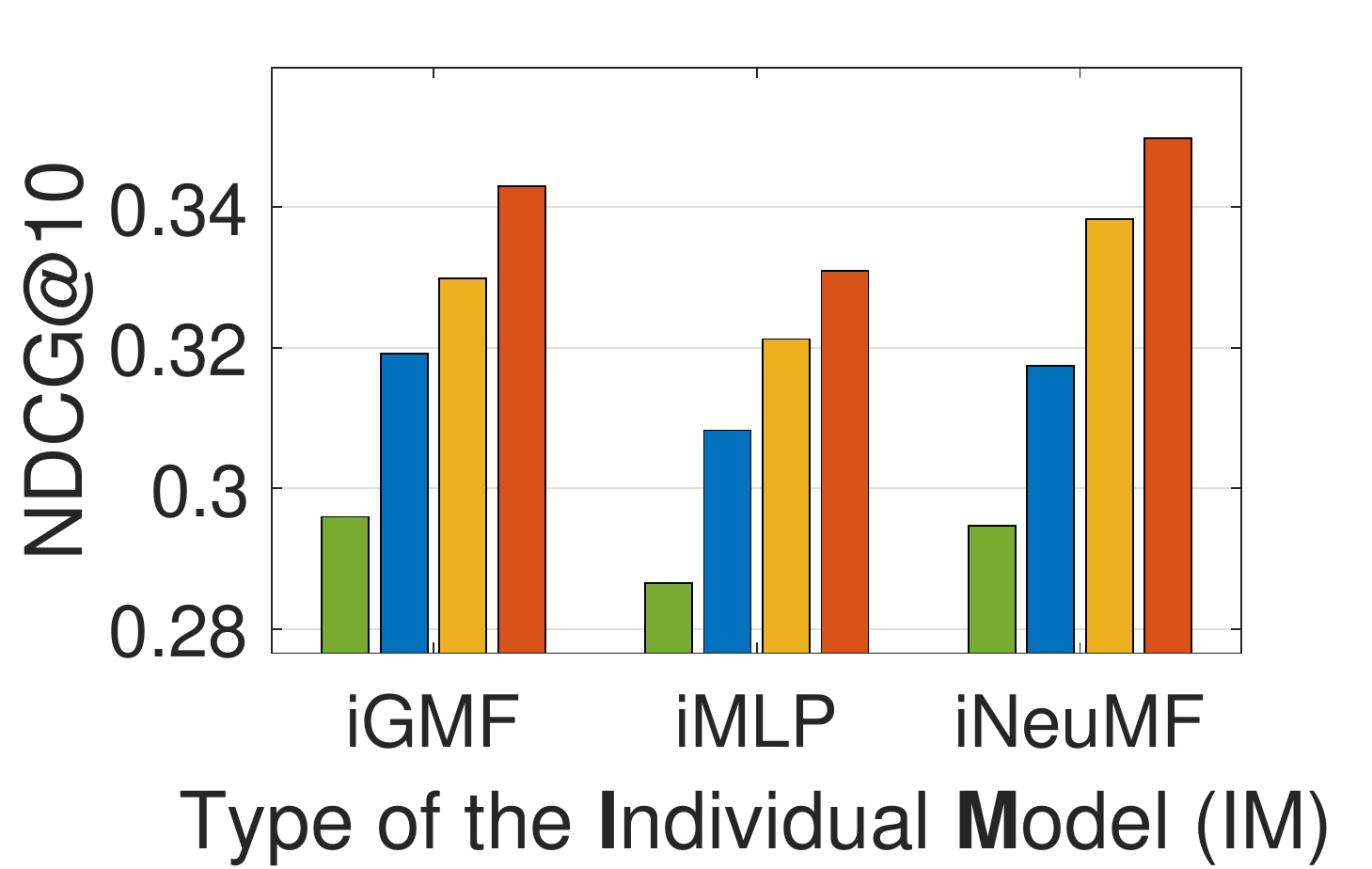}
			\caption{NDCG@10 on MovieLens}
			\label{fig:experiment_sample_ndcg_ml-1m}
		\end{subfigure}
		\begin{subfigure}[b]{.327\textwidth}
			\centering
			\includegraphics[width=\textwidth]{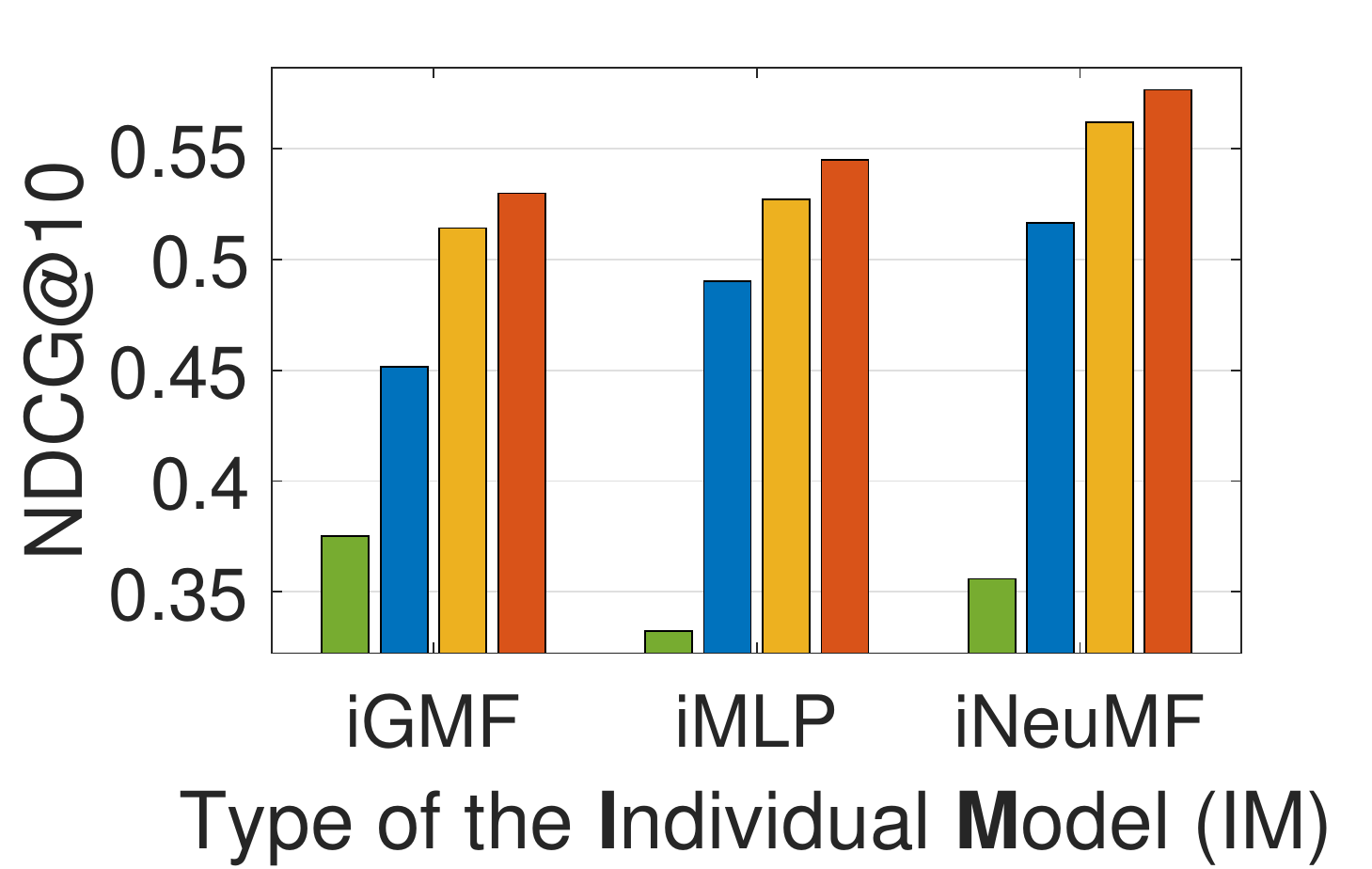}
			\caption{NDCG@10 on Netflix}
			\label{fig:experiment_sample_ndcg_netflix}
		\end{subfigure}
		\begin{subfigure}[b]{.327\textwidth}
			\centering
			\includegraphics[width=\textwidth]{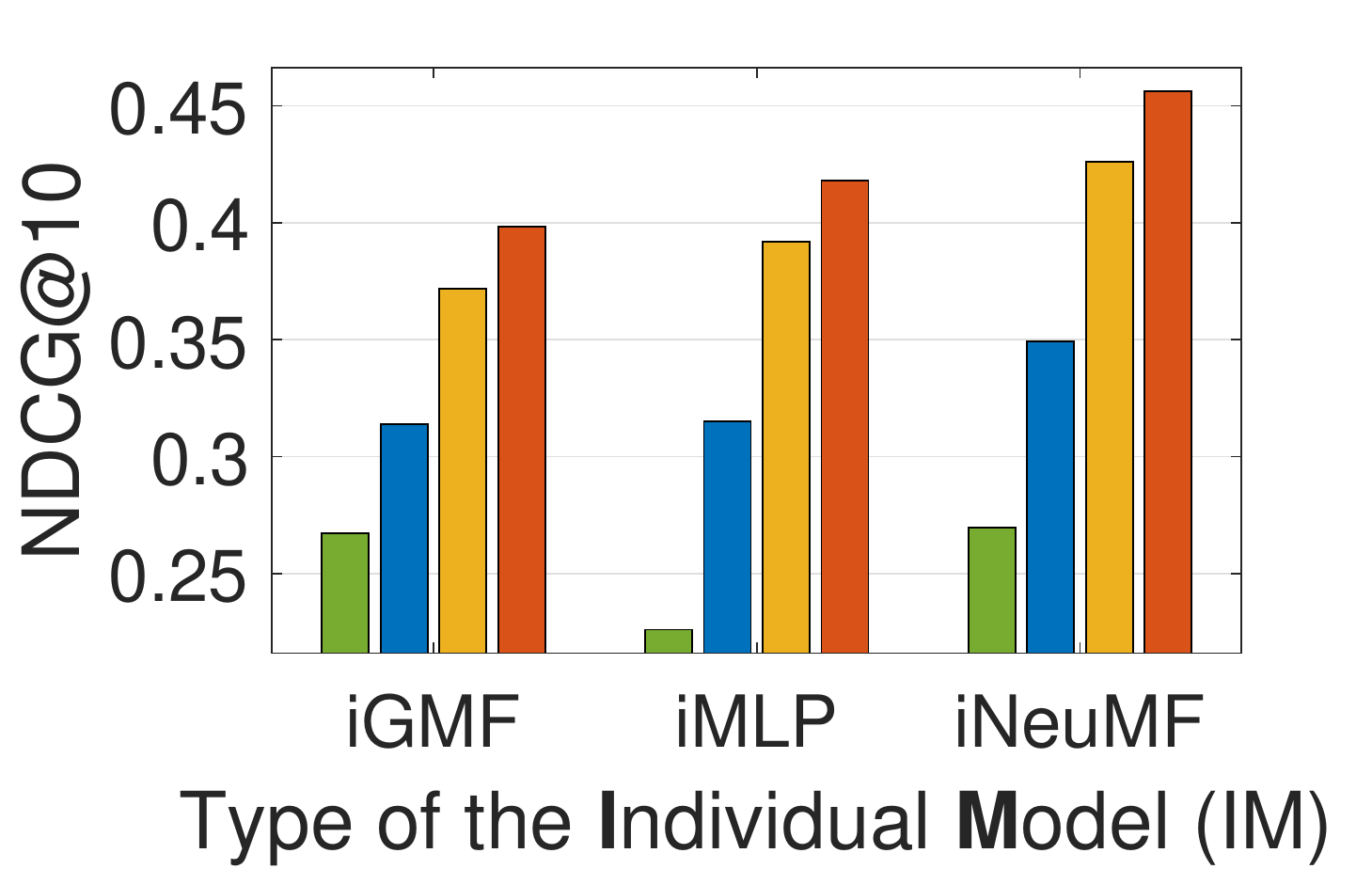}
			\caption{NDCG@10 on Yelp}
			\label{fig:experiment_sample_ndcg_yelp}
		\end{subfigure}
		
		%	\vspace{-5mm}
		\caption{The superiority of STS. To verify the superiority of our proposed STS, we replace STS with existing sampling methods for the comparison. For example, NDO-AEL indicates that NDO is employed for sampling the training data. As shown in Figs. 4(a) to 4(f), STS consistently outperforms other sampling methods in all the cases.}
		%	\vspace{-3mm}
		\label{fig:experiment_sample}
	\end{figure}
	
	\vspace{1pt}
	\noindent \textbf{Setting:} To answer \textbf{RQ3}, we replace the proposed STS with three representative sampling methods for comparisons, i.e., New Data Only (NDO)~\cite{ocfif} which only uses the newly received data for the training purpose, Reservoir-enhanced Random sampling (RR)~\cite{rmfx} which conducts random sampling with a reservoir, and Sliding Window (SW)~\cite{OWE} which uses the latest $k$ (a predetermined parameter) user-item interactions received for the training purpose. Similar to the naming scheme in the preceding experiments, we use *-AEL (e.g., NDO-AEL) to indicate which fusion method (e.g., NDO) is employed for the sampling purpose. In this experiment, we report the results on all three datasets in the underload scenario (i.e., $sp_{r} = 128$) to show the superiority of STS while the results in the overload scenario (i.e., $sp_{r} = 512$) are similar. 
	
	\noindent\textbf{Result:} As~\Cref{fig:experiment_sample} indicates, our proposed STS consistently outperforms all the other sampling methods on all three datasets. Taking the individual model of iNeuMF as an example, with which our proposed approach achieves the best overall performance, the improvements of our proposed STS over the best-performing baseline, i.e., NDO, range from 1.7\% (on Netflix) to 4.3\% (on Yelp) with an average of 2.8\% in terms of HR@10, and range from 2.6\% (on Netflix) to 7.1\% (on Yelp) with an average of 4.3\% in terms of NDCG@10. 
	
	\noindent\textbf{Analysis:} The superiority of STS can be explained by elaborately incorporating both new data and historical data in a time-aware manner while guaranteeing the proportion of new data. Thus, STS can well address concept drift while capturing long-term user preferences (\textbf{CH1}). Besides, it can be observed that NDO, which takes new data only to train the recommendation model, outperforms two other baselines, i.e., RR and SW, which both take the historical data into consideration. This indicates that improperly incorporating historical data may reduce the recommendation accuracy with our experimental settings. It is possibly caused by the insufficient emphasis on new data, which hinders effectively addressing concept drift, and ineffectual sampling for historical data, which impedes effectively capturing long-term user preferences. \\
	%\textbf{Summary:} STS-AEL delivers higher recommendation accuracy when ensembling more individual recommendation models and STS outperforms other approaches. 
	%	Since Experiment 2 indicates that STS-AEL delivers the best overall performance with $\lambda = 0.999$, due to the space limitation, we fix $\lambda = 0.999$ in the following sections to answer \textbf{RQ3} and \textbf{RQ4}.
	%\vspace{-50mm}
	
	\vspace{1pt}
	\noindent\textbf{Experiment 4: Superiority of AEL (for RQ4)}
	
	\begin{figure}[]
			\centering
		\begin{subfigure}[b]{.9\textwidth}
			\centering
			\includegraphics[width=\textwidth]{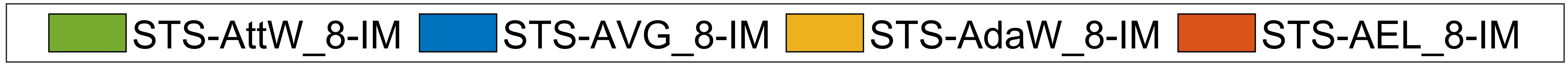}
			%		\vspace{-4.4mm}
		\end{subfigure}%
		
		\begin{subfigure}[b]{.332\textwidth}
			\centering
			\includegraphics[width=\textwidth]{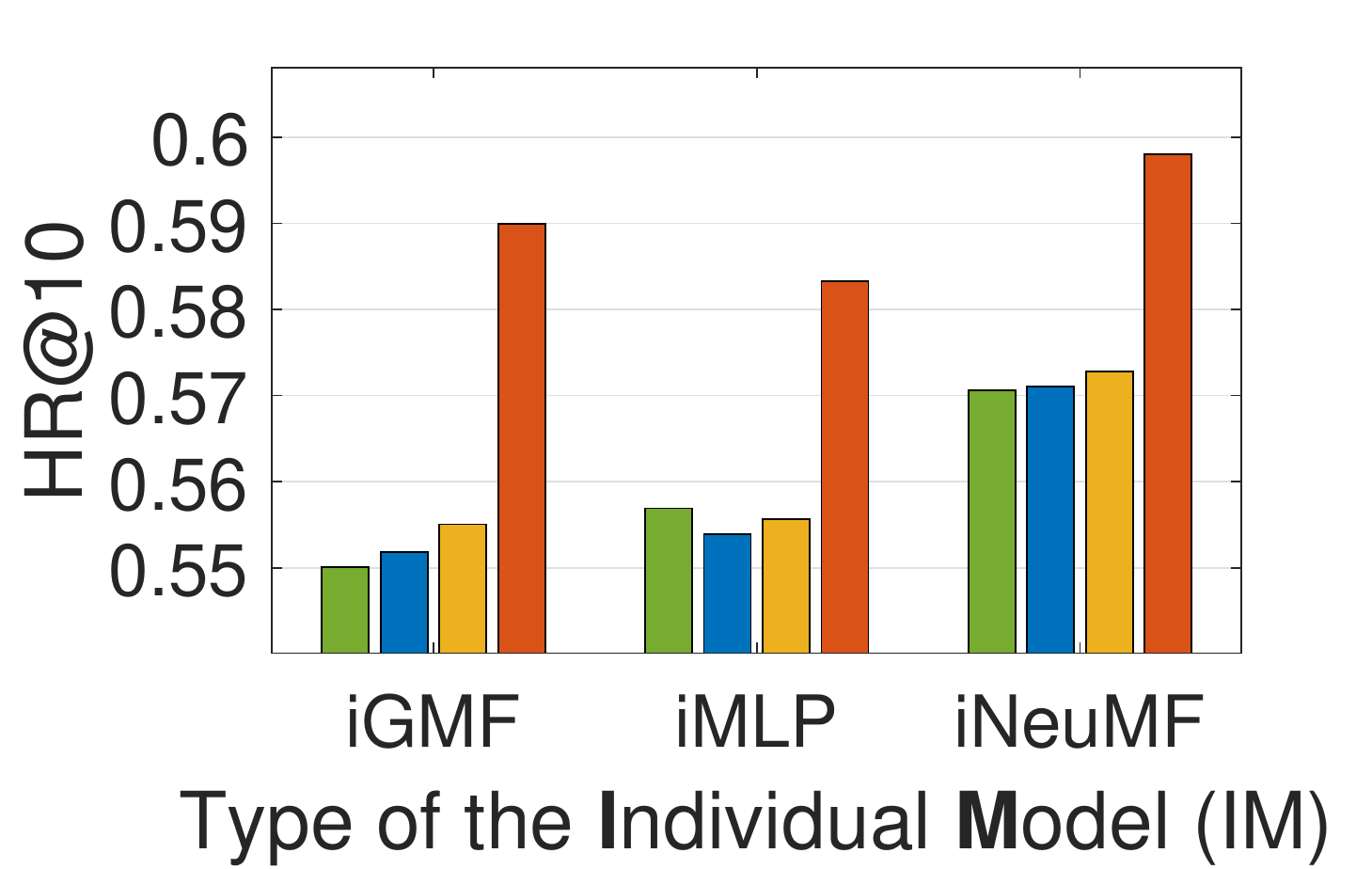}
			% \vspace{-4mm}
			\caption{HR@10 on MovieLens}
			\label{fig:experiment_fusion_hr_ml-1m}
		\end{subfigure}%
		\begin{subfigure}[b]{.332\textwidth}
			\centering
			\includegraphics[width=\textwidth]{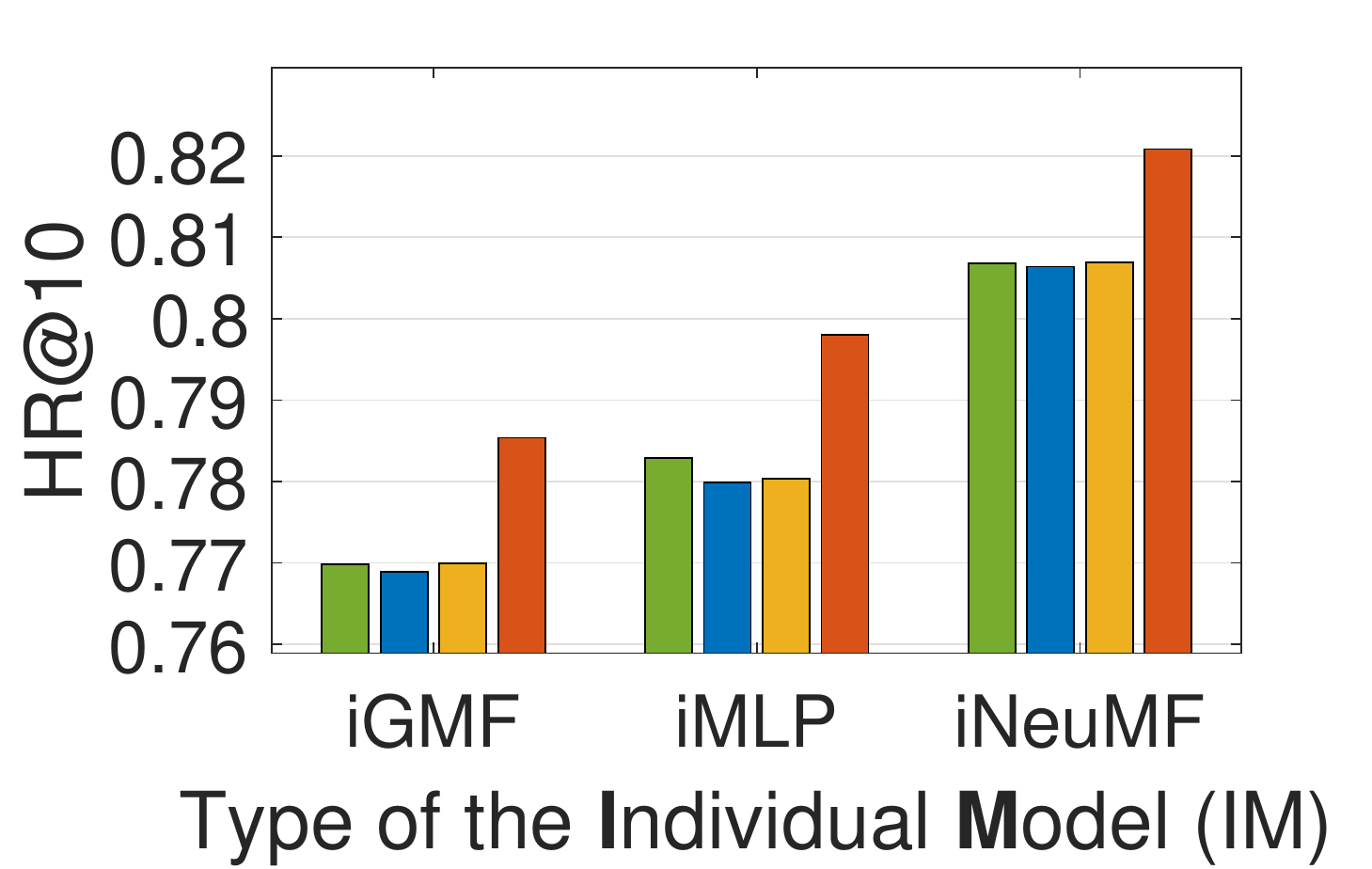}
			% \vspace{-4mm}
			\caption{HR@10 on Netflix}
			\label{fig:experiment_fusion_hr_netflix}
		\end{subfigure}%
		\begin{subfigure}[b]{.332\textwidth}
			\centering
			\includegraphics[width=\textwidth]{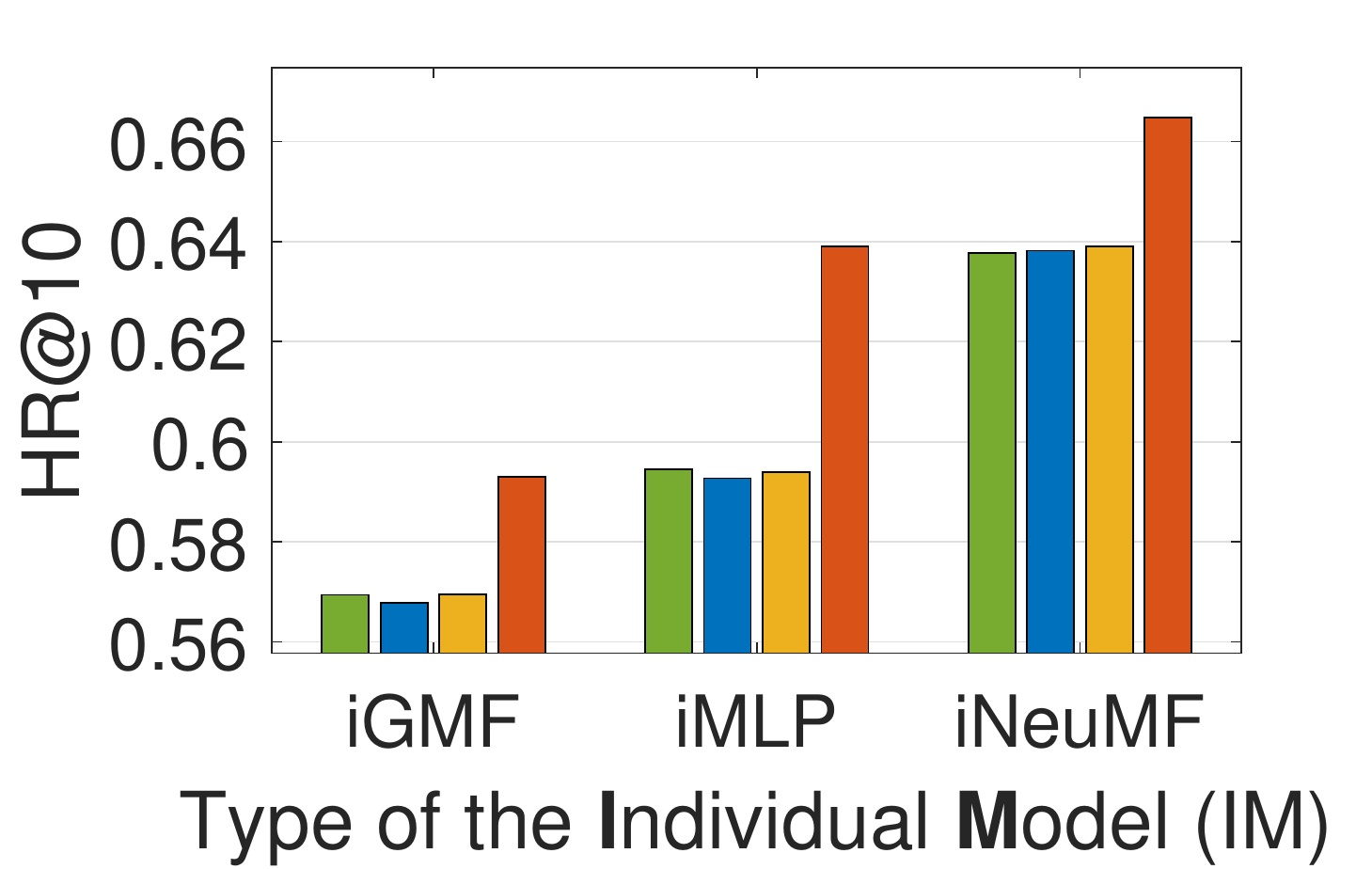}
			% \vspace{-4mm}
			\caption{HR@10 on Yelp}
			\label{fig:experiment_fusion_hr_yelp}
		\end{subfigure}%
		
		\begin{subfigure}[b]{.327\textwidth}
			\centering
			\includegraphics[width=\textwidth]{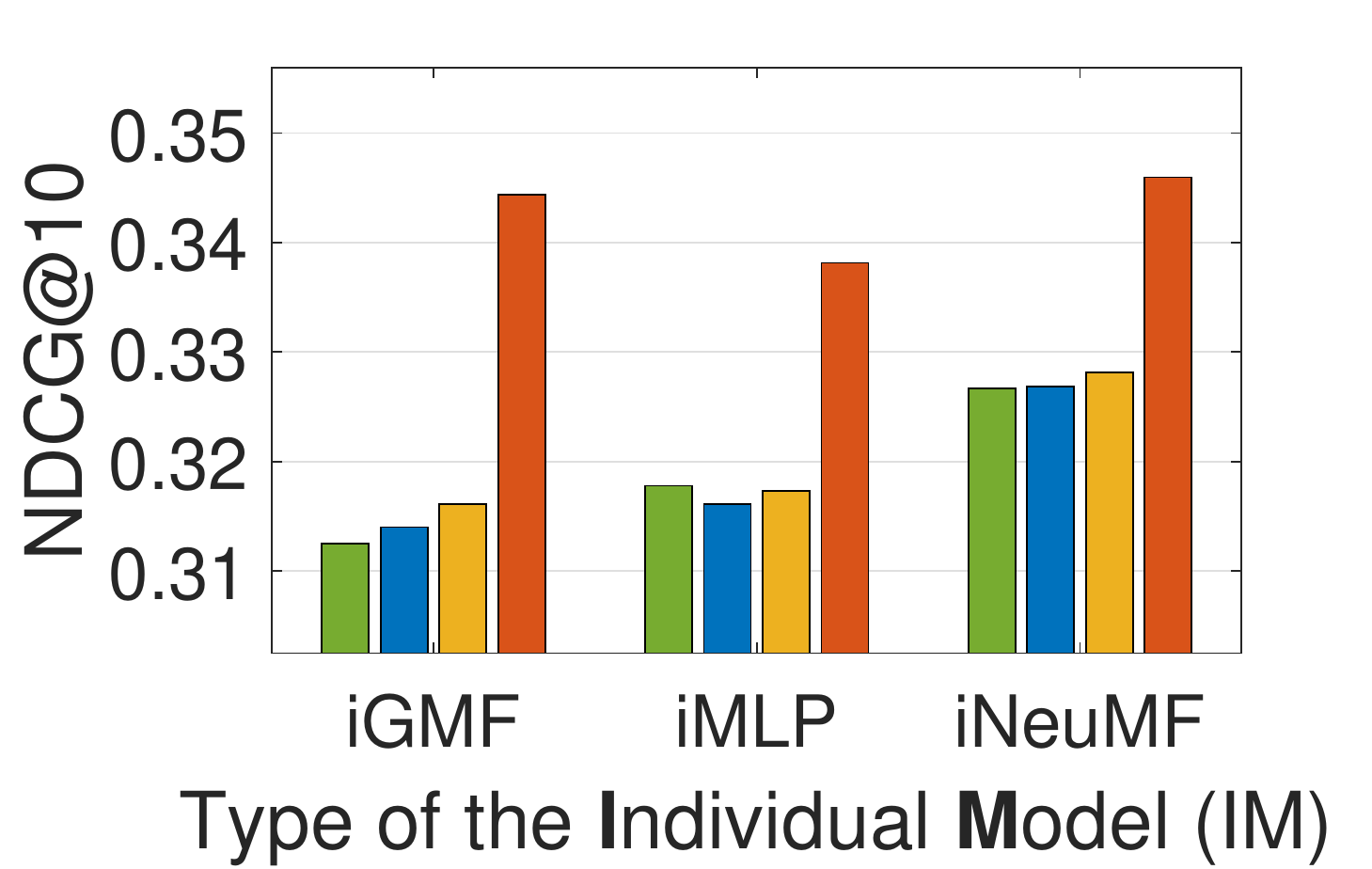}
			%		\vspace{-5mm}
			\caption{NDCG@10 on MovieLens}
			
			\label{fig:experiment_fusion_ndcg_netflix}
		\end{subfigure}
		\begin{subfigure}[b]{.327\textwidth}
			\centering
			\includegraphics[width=\textwidth]{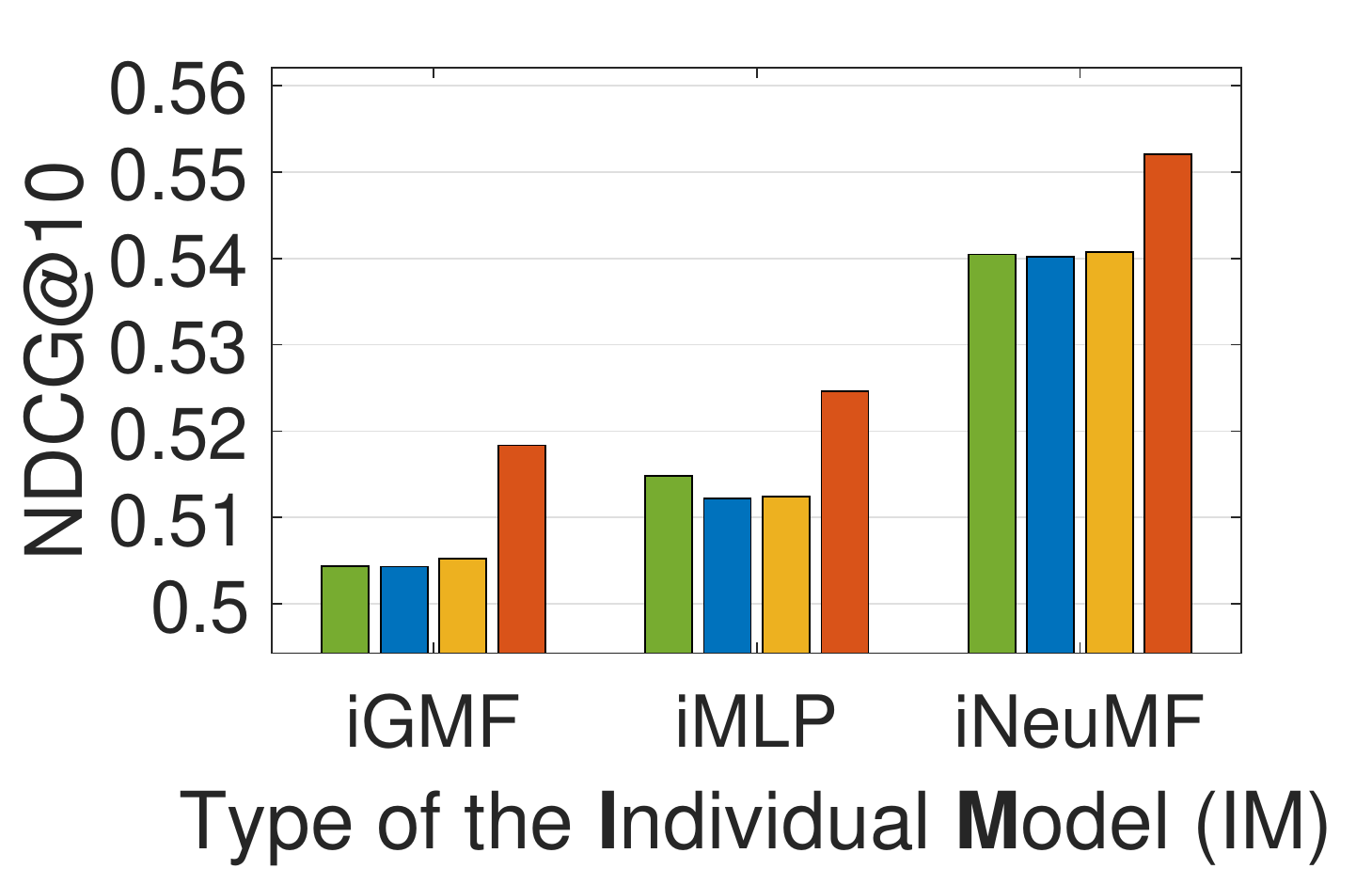}
			%		\vspace{-5mm}
			\caption{NDCG@10 on Netflix}
			\label{fig:experiment_fusion_ndcg_ml-1m}
		\end{subfigure}
		\begin{subfigure}[b]{.327\textwidth}
			\centering
			\includegraphics[width=\textwidth]{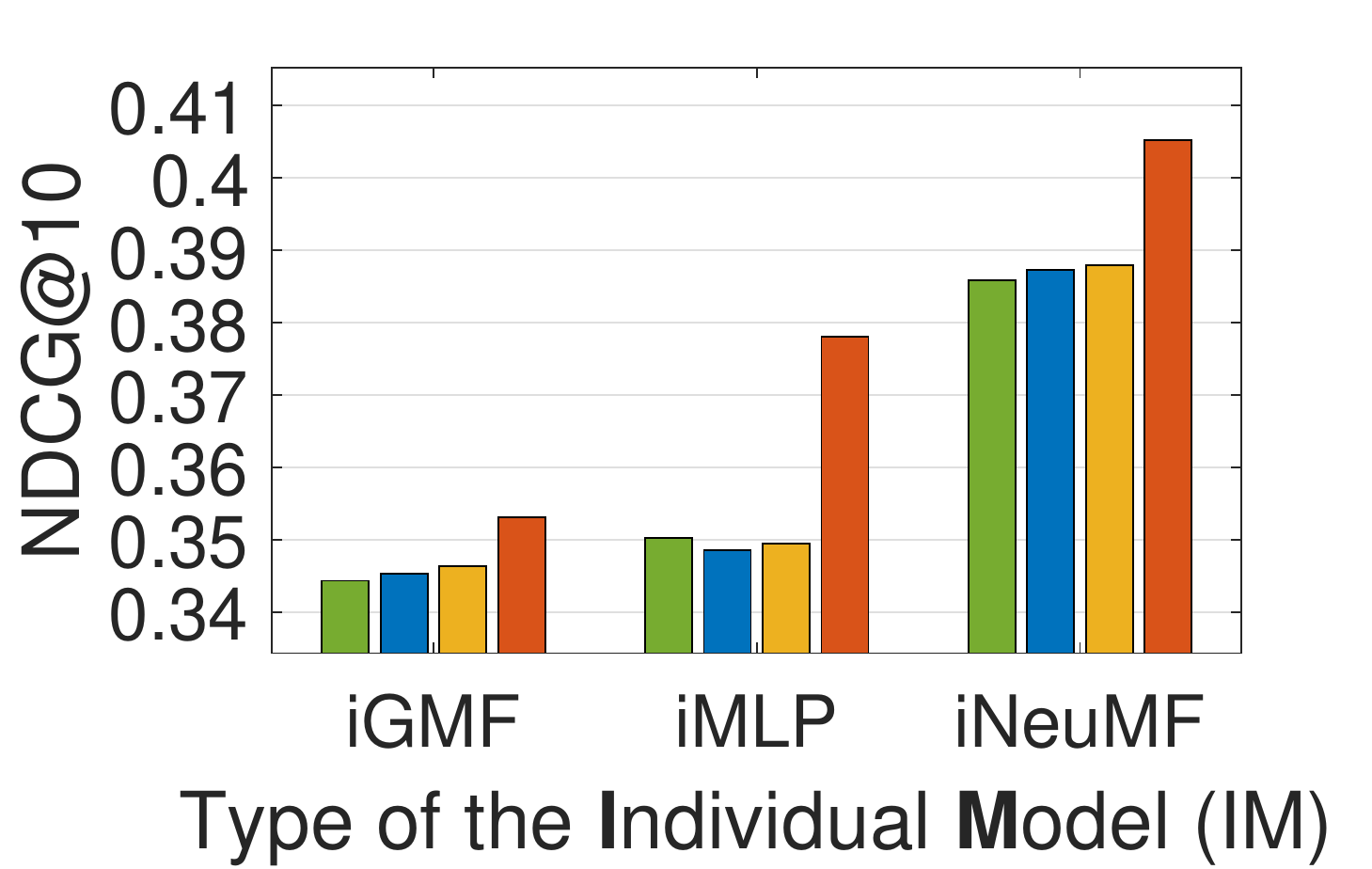}
			%		\vspace{-5mm}
			\caption{NDCG@10 on Yelp}
			\label{fig:experiment_fusion_ndcg_yelp}
		\end{subfigure}
		%	\vspace{-5mm}
		\caption{Superiority of AEL. To verify the superiority of our proposed AEL, we replace the sequential adaptive fusion in AEL with existing fusion methods for the comparison. For example, STS-AttW indicates that AttW is employed for the fusion purpose. As shown in Figs. 5(a) to 5(f),  our proposed AEL significantly outperforms all the baselines in all the cases.}
		%	\vspace{-3mm}
		\label{fig:experiment_fusion}
	\end{figure}
	
	\noindent \textbf{Setting:} To answer \textbf{RQ4}, we replace the sequential adaptive fusion in AEL with three representative fusion methods for comparisons, i.e., 1) Attentive Weighting (AttW)~\cite{schwab2019granger} which adopts an attention mechanism for the fusion, 2) AVeraGing (AVG)~\cite{oza2005online} which simply averages the results of the individual models, and 3) AdaBoost-like Weighting (AdaW)~\cite{Aboost} which considers the previous recommendation accuracy of individual models when conducting the fusion process. Similar to the naming scheme in the preceding experiments, we use STS-* (e.g., STS-AVG) to indicate which sampling method (e.g., AVG) is employed for the fusion purpose. In this experiment, we report the results on all three datasets in the overload scenario (i.e. $sp_r = 512$) while the results in the underload scenario (i.e. $sp_r = 128$) are similar. 
	
	\noindent\textbf{Result:} As~\Cref{fig:experiment_fusion} indicates, our proposed AEL significantly outperforms all the baselines on all three datasets. Taking the individual model of iNeuMF as an example, with which the ensembling approach achieves the best overall performance, the improvements over the best-performing baseline, i.e., AdaW, range from 1.7\% (on Netflix) to 6.6\% (on Yelp) with an average of 4.2\% in terms of HR@10, and range from 2.2\% (on Netflix) to 6.6\% (on Yelp) with an average of 4.7\% in terms of NDCG@10.
	
	\noindent\textbf{Analysis:} The superiority of AEL mainly comes from the elaborately calculated fusion weights, which leads to effective fusion for higher recommendation accuracy. Specifically, when calculating the fusion weights, AEL not only considers the previous recommendation accuracies of individual models but also takes the specific characteristic of the target user-item pair into account. Besides, it can be observed that AttW does not deliver good performance in this experiment, and the possible reason is that AttW is originally devised for the mixture of expert model~\cite{schwab2019granger} and is not suitable for our proposed framework.

	\section{Conclusion}
	In this paper, we have proposed a \textbf{S}tratified and \textbf{T}ime-aware \textbf{S}ampling based \textbf{A}daptive \textbf{E}nsemble \textbf{L}earning framework, called STS-AEL, for accurate streaming recommendations. Our proposed STS-AEL addresses concept drift while capturing long-term user preferences  (\textbf{CH1}) through wisely sampling the new data and historical data through a stratified and time-aware manner. Moreover, the incorporation of the sampled historical data also benefits addressing the underload problem (\textbf{CH2}). Furthermore, STS-AEL addresses the overload problem (\textbf{CH3}) by training the multiple individual models concurrently and fuse the results of these trained models with a sequential adaptive fusion approach. The extensive experiments show that the proposed STS-AEL significantly outperforms the state-of-the-art approaches. In addition, the effectiveness of the two main components, i.e., Stratified and Time-aware Sampling (STS) and Adaptive Ensemble Learning (AEL), have also been explicitly verified by the experiments. 
	
	In the future, we will study the strategy of simultaneously ensembling various types of individual models for higher accuracy of streaming recommendations.


	%\begin{acknowledgements}
	%\end{acknowledgements}



\end{document}